\documentclass[12pt,aps,nofootinbib]{revtex4}
\usepackage{graphicx}
\usepackage{bbm}
\usepackage{amssymb}
\usepackage{amsmath}

\usepackage{amsfonts}
\usepackage{eucal}
\usepackage{epsfig}
\usepackage{feynmf}
\usepackage{citesort}

\newcommand{\myvec}[1]{\mathbf{#1}}
\newcommand{\gvec}[1]{\mbox{\boldmath $#1$}}
\newcommand{\kvec}{\myvec{k}}

\newcommand{\kspaceint}[2]{\int\!\frac{d^{#1}\!{#2}}{(2\pi)^{#1}}\;}

\newcommand{\intDk}{\kspaceint{D}{k}}
\newcommand{\intDkp}{\kspaceint{D}{k^{\prime}}}

\def\Av{{\myvec{A}}}
\def\Lv{{\gvec{\lambda}}}
\def\Ab{{\mathbf{A}}}
\def\Bb{{\mathbf{B}}}
\def\Db{{\mathbf{D}}}
\def\Eb{{\mathbf{E}}}
\def\Kb{{\mathbf{K}}}
\def\Lb{{\gvec{\lambda}}}
\def\Zb{{\gvec{\zeta}}}
\def\Dm{{\mathcal{D}}}
\begin{document}
\rightline{HD-THEP-08-26}
\title{B\"odeker's Effective Theory: From Langevin Dynamics to
Dyson--Schwinger Equations}

\author{Claus Zahlten, Andres Hernandez and Michael G.~Schmidt}

\thanks{C.Zahlten@gmx.de}
\thanks{A.Hernandez@thphys.uni-heidelberg.de}
\thanks{M.G.Schmidt@thphys.uni-heidelberg.de}

\affiliation{Institut f\"ur Theoretische Physik, \
Universit\"at Heidelberg, Philosophenweg 16, \
D-69120 Heidelberg, Germany}

\date{\today}
\begin{abstract}
The dynamics of weakly coupled, non-abelian gauge fields
at high temperature is non-perturbative if the characteristic
momentum scale is of order $|\myvec{k}|\sim g^2 T$.
Such a situation is typical for the processes of electroweak
baryon number violation in the early Universe. B\"odeker has 
derived an effective theory that describes the dynamics 
of the soft field modes 
by means of a \textsc{Langevin} equation. This effective theory
has been used for lattice calculations so far
\cite{Moore:SphaleronLLO,Moore:SphaleronSym:00}. In this work
we provide a complementary, more analytic approach based on 
\textsc{Dyson--Schwinger} equations. Using methods known from
stochastic quantisation, we recast B\"odeker's \textsc{Langevin} 
equation in the form of a field theoretic path integral. 
We introduce gauge ghosts in order to help control possible gauge
artefacts that might appear after truncation, and which leads to
a BRST symmetric formulation and to corresponding Ward
identities. A second set of Ward identities, reflecting the origin
of the theory in a stochastic differential equation, is also
obtained. Finally \textsc{Dyson--Schwinger} equations are derived.
\end{abstract}

\maketitle


\section{Introduction}
Lattice calculations are an ideal tool to extract with a minimum of 
theoretical prejudice a specific piece of information from a given
theory. However, in a sense, they are kind of a `black box' that give
the answer but hide the way \emph{how} that answer comes about.

The aim of this work is to provide a complementary, more analytic 
approach to the non-perturbative physics encoded in B\"odeker's 
effective theory \cite{Boedeker:1}. The emphasis thereby lays not
primarily on the accuracy of the results where it is hardly possible
to beat the lattice calculations. Our aim is to provide a tool for a 
deeper understanding of what is really going on in the 
non-perturbative sector of hot non-abelian gauge theory and
during creation of baryon number. In particular, it allows for an analytic 
study of the sphaleron rate \cite{Shaposhnikov,Raby}
\begin{equation}\label{sphaleron}
\Gamma \equiv \lim_{V \rightarrow \infty} \lim_{t \rightarrow \infty} 
	\frac{ \langle ( N_{\rm CS}(t) - N_{\rm CS}(0) )^2 \rangle}
	{Vt} \,
\end{equation}
Such a deeper understanding is not only important for baryogenesis.
Magnetic screening and 
the corresponding identification of a magnetic mass are of quite 
general theoretical interest with applications also in the field of the quark--gluon
plasma \cite{AlexanianNair:95,JackiwPi:95,Cornwall:98,Nair:ThreeIdeas}.

We base our analysis on B\"odeker's effective theory, despite the fact
that B\"odeker has also derived a generalised \textsc{Boltzmann--Langevin}
equation which is valid to all orders in $[\log(1/g)]^{-1}$
\cite{Boedeker:3}, and of which B\"odeker's effective theory is merely
the leading logarithmic approximation.
We choose this approximation because the more general
\textsc{Boltzmann--Langevin} equation not only is far more complicated,
but is also not renormalisable by power counting \cite{Boedeker:4}.
The effective theory on the other hand is ultraviolet finite, and is
known to still be valid at next-to-leading logarithmic order provided
one uses the next-to-leading logarithmic order colour conductivity 
$\sigma$ \cite{ArnoldYaffe:Conductivity}.

The key idea of this work is rather simple. B\"odeker's effective 
theory has the form of a \textsc{Langevin} equation. It is well-known 
from stochastic quantisation that a \textsc{Langevin} equation can be 
recast in the form of a path integral 
\cite{ZinnJustin,DamgaardHueffel:SQ,ArnoldYaffe:Beyond}.
This path integral then can be reinterpreted as the functional integral 
formulation of an Euclidean quantum field theory with some `strange'
action. In this way, one gains access to all the powerful methods
developed in QFT. Specifically, it is possible to derive the 
\textsc{Dyson--Schwinger} equations of the theory, offering 
an approach to the non-perturbative sector that is independent from,
and complementary to, the existing lattice studies.

On the way to this goal a couple of obstacles have to be
overcome. These are mostly related to the peculiar role played by 
gauge invariance in the context of stochastic quantisation and 
B\"odeker's effective theory. A thorough understanding of this role
proves to be essential in pursuing our aim.

The outline of this work is as follows. Section
\ref{PathIntegralFormulation} is devoted to the transcription of
B\"odeker's theory in path integral form. From this path integral
one could proceed to derive the \textsc{Dyson--Schwinger} equations,
and in principle, could gain access to the non-perturbative
sector of the theory. At the end of the day however, one will be
forced to rely on a certain truncation scheme to extract any concrete
results from the equations. This truncation may introduce a possible
gauge dependence and thus may render the results worthless. To keep
control over the gauge dependence, it is therefore necessary to
generalise B\"odeker's equation from $A_0=0$ gauge to a more general
class of gauges before applying the formalism. 

Gauge fixing in a stochastic differential equation is quite delicate. 
One has to make sure not to destroy the \textsc{Markovian} nature of 
the equation. Applying methods developed in the context of stochastic 
quantisation~\cite{ZinnJustinZwanziger:WardIdforSQ}, we introduce 
a gauge fixing term into B\"odeker's equation thereby achieving 
the desired upgrade to a general class of flow gauges.

In Section \ref{WardIDs} we argue that any physically reasonable
truncation of the \textsc{Dyson--Schwinger} equations requires the 
introduction of \emph{gauge} ghosts. 
In the full, untruncated theory gauge ghosts are not necessary, 
which is generally true in stochastic quantisation 
\cite{DamgaardHueffel:SQ,ZinnJustinZwanziger:WardIdforSQ}. 
As was shown in Ref.~\cite{ZinnJustinZwanziger:WardIdforSQ},
gauge ghosts \emph{can} be introduced in stochastic quantisation in 
order to establish a gauge BRST symmetric formulation. 

We carry out this program in the case of B\"odeker's theory and derive
the \textsc{Ward--Takahashi} identities corresponding to the gauge BRST
symmetry of the action. These should be respected by the truncations to
be used.

The gauge Ward identities are not the only restrictions to be observed.
A second class of Ward identities exist, that are related to the 
characteristic structure of the theory reflecting its origin in a 
stochastic differential equation. This characteristic structure can as well
be expressed in the form of a BRST symmetry by introducing another kind of
ghost fields, referred to as equation of motion (EOM) ghosts in this work.
Introducing the gauge ghosts, however, destroys this second stochastic BRST
symmetry. Nevertheless, it does not change the physical contents of the
theory. The stochastic BRST symmetry is only an elegant way to express this
structure. By directly referring to the underlying physics, it is still
possible to derive the corresponding stochastic Ward identities. They
provide a second set of restrictions to be imposed on the truncations.

In Section \ref{DSE} we derive the \textsc{Dyson--Schwinger} equations of
B\"odeker's effective theory. In combination with the gauge and stochastic
Ward identities of Section \ref{WardIDs}, this constitutes an independent
approach to the non-perturbative dynamics of the soft, non-abelian gauge
fields encoded in B\"odeker's effective theory. 

In Section \ref{Outlook} we summarise and discuss our results. Appendix
\ref{Jacobian} shows the explicit calculation of some of the Jacobians
encountered in this work. We have included this Appendix in order to make
our presentation more self-contained. The \textsc{Feynman} rules
corresponding to our field theoretic transcription of B\"odeker's effective
theory are listed in Appendix \ref{FeynmanRules}. Finally, in Appendix 
\ref{NPointFunctions}, we present explicit identities for the lower
n-point function following from the general Ward identities.


\section{Path Integral Formulation of B\"{o}deker's Theory}
\label{PathIntegralFormulation}

\subsection{Transcription to a Path Integral in $A_0=0$ Gauge}
\label{sec11}

According to B\"{o}deker's effective theory the dynamics of the soft modes
of the gauge field is described to leading logarithmic order by the
\textsc{Langevin} equation \cite{Boedeker:1}
\begin{equation}
  \label{eq:Bodeker}
  \Db^{ab}\!\times\Bb^b+\sigma\dot{\Ab}^{\!a} = \Zb^a
\end{equation}
which is written in $A_0=0$ gauge and where $\Zb$ is a gaussian white
noise stochastic force. The stochastic force field incorporates the
influence of higher momentum modes and has the correlator
\begin{equation}
  \label{eq:StCorrelator}
  \left\langle
    \Zb^{ai}(t,\myvec{x})
    \Zb^{bj}(t^{\prime},\myvec{x}^{\prime})
  \right\rangle
  = 2\sigma T \,\delta^{ij} \delta^{ab} \,\delta(t-t^{\prime}) \,
    \delta^{D-1}(\myvec{x} - \myvec{x}^{\prime})
\end{equation}
reflecting its gaussian white noise character.
Here and in the following, the number of spacial dimensions is
$D-1 = 3$, however, we leave $D$ unspecified to allow for dimensional
regularisation later. The only physical parameters entering 
Eqs.~(\ref{eq:Bodeker}) and (\ref{eq:StCorrelator}), and therefore
the effective theory, are the temperature $T$, the colour
conductivity $\sigma$, and the self coupling of the gauge field hidden
in the definition of the covariant derivative
$\Db^{ab} = \delta^{ab}\nabla -gf^{abc}{\Ab}^{\!c}$.

The procedure of reformulating a \textsc{Langevin} equation like
Eq.~(\ref{eq:Bodeker}) in the form of a field theoretic path integral
is well-known \cite{ZinnJustin,DamgaardHueffel:SQ,ArnoldYaffe:Beyond}:
According to Eq.~(\ref{eq:Bodeker}), the gauge field evolves, starting
from certain initial conditions, under the influence of the stochastic
force. An arbitrary observable of the theory then is defined by some
functional of the gauge field $F[\Ab]$ and given by the expectation
value of that functional with respect to the possible realisations of
the stochastic force
\begin{equation}
  \label{eq:PathIntegral1}
  \langle F[\Ab] \rangle
  = \!\int\!\Dm\Zb\, F[\Ab^{\!s}[\Zb]] \varrho[\Zb]
  = \!\int\!\Dm\Zb\, F[\Ab^{\!s}[\Zb]]
        \exp\left\{ -\frac{1}{4\sigma T}\!\int\!\!dt\,d^{D-1}\!x\;
          \Zb^a(t,\myvec{x}) \cdot \Zb^a(t,\myvec{x})
        \right\}
\end{equation}
Here we have denoted by $\Ab^{\!s}[\Zb]$ the solution of
Eq.~(\ref{eq:Bodeker}) for a specific choice of the stochastic
force and the given initial conditions.

To proceed and recast the effective theory of the gauge field into a
form resembling the path integral formulation of an `ordinary' quantum
field theory we would rather like to have a path integral running over
the gauge field than running over the stochastic force.
This can be achieved by inserting unity in an appropriate way. In fact,
one has
\begin{equation}
  \label{eq:1byA}
  1 = \int\!\mathcal{D}\Eb\; \delta(\Eb - \Zb)
    = \int\limits_{\mathrm{(i.c.)}}\!\!\mathcal{D}\!\Ab\;
      \mathrm{Det}\!\left(\frac{\delta \Eb[\Ab]}{\delta\Ab}\right)
      \delta(\Eb[\Ab] - \Zb)
\end{equation}
where we choose the functional $\Eb[\Ab]$ as the left-hand side
of Eq.~(\ref{eq:Bodeker})
\begin{equation}
  \label{eq:E}
  \Eb^a[\Ab] = \Db^{ab}\!\times\Bb^b+\sigma\dot{\Ab}^{\!a}
\end{equation}
The invertibility of $\Eb[\Ab]$ is essential to justify the change of
variables in Eq.~(\ref{eq:1byA}). It follows from the parabolic nature
of the expression and from the restriction to those gauge field
configurations in the second path integral satisfying the initial
conditions.

Because Eq.~(\ref{eq:1byA}) holds independently of $\Zb$, it can be
inserted into the path integral~(\ref{eq:PathIntegral1}). The delta
function then assures that only those gauge field configurations
contribute to the integral that obey $\Eb[\Ab] = \Zb$. Due to our
choice of $\Eb[\Ab]$, however, this is identical to the condition
$\Ab = \Ab^{\!s}[\Zb]$. Thus, after inserting the delta function we
may replace $\Ab^{\!s}[\Zb]$ in the path integral simply by the
integration variable $\Ab$, and we are left with
\begin{equation}
  \label{eq:PathIntegral2}
  \langle F[\Ab] \rangle
  = \!\int\!\Dm\Zb\,\varrho[\Zb]
    \int\limits_{\mathrm{(i.c.)}}\!\!\mathcal{D}\!\Ab\;
    \mathrm{Det}\!\left(\frac{\delta \Eb[\Ab]}{\delta\Ab}\right)
    \delta(\Eb[\Ab] - \Zb) \,F[\Ab]
\end{equation}
Moreover, the restriction to field configurations obeying a specific set
of initial conditions can be dropped if these initial conditions are
specified at $t=-\infty$. This is a consequence of their transversal
component always being damped and the fact that any longitudinal
contribution drops out whenever a gauge invariant observable is calculated.
In case of a gauge variant quantity, however, a damping of the longitudinal
component can be achieved by introducing an additional gauge fixing
term into the \textsc{Langevin} equation \cite{DamgaardHueffel:SQ}.
This will be necessary anyway in the following section in order to
generalise from $A_0 = 0$ gauge. Henceforth, we will therefore drop the
restriction on the path integration in Eq.~(\ref{eq:PathIntegral2}).

At this point, one has two choices. One possibility is to proceed by doing
the $\Zb$ integral with the help of the delta function. This results in a
theory containing only the gauge field (and perhaps some additional ghost
fields to be introduced later), however, at the expense of rather
complicated interactions: the functional $\Eb[\Ab]$ shows up as
argument of the gaussian probability distribution, and since $\Eb[\Ab]$
contains terms up to $\Ab^{\!3}$, the action would inherit vertices of up to
sixth order.

To avoid this situation, we instead choose to introduce an additional
auxiliary field $\Lb$ to represent the delta function
\begin{equation}
  \delta(\Eb[\Ab] - \Zb) =
  \!\int\!\Dm\!\Lb\;
  \exp\left\{ i\!\int\!\!dt\,d^{D-1}\!x\;
    \Lb^{\!a}\!\cdot\!(\Eb^{a}[\Ab] - \Zb^{a})
  \right\}
\end{equation}
In this way, one can still perform the $\Zb$ integral that becomes
gaussian, thereby eliminating the stochastic force field from the
theory. One obtains
\begin{equation}
  \label{eq:PathIntegral3}
  \langle F[\Ab] \rangle =
  \!\int\!\Dm\!\Ab\Dm\!\Lb\;
  \mathrm{Det}\!\left(\frac{\delta \Eb[\Ab]}{\delta\Ab}\right)
  \!F[\Ab]\,e^{-S[\Ab,\gvec{\scriptstyle\lambda}]}
\end{equation}
with
\begin{equation} \label{eq:S[A,Lambda]}
  S[\myvec{A},\gvec{\lambda}] =
  \int\!\!dx\,
  \Bigl[
    \sigma T \,\gvec{\lambda}^a\!\cdot\!\gvec{\lambda}^a
    -i \gvec{\lambda}^a \!\cdot \Eb^a[\Ab]
  \Bigr]
\end{equation}
The determinant in Eq.~(\ref{eq:PathIntegral3}) need not be taken into
account since it can be shown to be a constant in dimensional regularisation
(see Appendix \ref{Jacobian} for an explicit calculation). We could,
nevertheless, introduce a ghost representation of the determinant referring
to the corresponding ghost fields as equation of motion (EOM) ghosts in the
following. As a benefit of doing so the action (\ref{eq:S[A,Lambda]})
would be endowed with a BRST symmetry, allowing to easily obtain a kind
of Ward identities (so-called stochastic Ward identities) reflecting the
origin of the theory in a stochastic differential equation. Since it is
desirable to obtain as many non-perturbative identities as possible in
order to find a judicious ansatz for the truncation of the
\textsc{Dyson--Schwinger} equations, introducing EOM ghosts, at first,
seems the natural way to proceed. 

However there is another type of Ward identities related to gauge invariance.
Unfortunately, the gauge ghosts to be introduced to obtain these gauge Ward
identities will break the stochastic BRST symmetries. So, instead of
introducing EOM ghosts now, we will later introduce gauge ghosts in order to
obtain the gauge Ward identities. The stochastic Ward identities will be
derived without the help of a BRST symmetry by directly referring to the
fundamental structure of the theory that reflects its origin in a stochastic
differential equation.

For now, absorbing the constant determinant in the measure, we are left with
\begin{equation}
  \label{eq:PathIntegral4}
  \langle F[\Ab] \rangle =
  \!\int\!\Dm\!\Ab\Dm\!\Lb\;
  \!F[\Ab]\,e^{-S[\Ab,\gvec{\scriptstyle\lambda}]}
\end{equation}
where the action $S$ is given by Eq.~(\ref{eq:S[A,Lambda]}).

\subsection{Upgrading to $\kappa$ Gauge}
\label{se:UpgradeKappaGauge}

B\"odeker's theory is written in $A_0=0$ gauge, and so is our transcription
as field theoretic path integral so far. At the end of the day, however,
we will be forced to use an approximation to solve the non-perturbative
equations obtained, e.g.~\textsc{Dyson--Schwinger} equations, and this
approximation might introduce gauge artefacts into the calculation. In
order to allow some control over the gauge dependence of the results, we 
need to base our derivations on a reformulation of B\"odeker's
equation in a more general gauge.

In \cite{ZinnJustinZwanziger:WardIdforSQ}, Zinn-Justin and Zwanziger have
shown that adding a term to Eq.~(\ref{eq:Bodeker}) that is tangent to the
gauge orbit 
\begin{equation} \label{eq:Boedeker+v}
  \myvec{D}^{ab} \!\times \myvec{B}^b
  + \sigma (\dot{\myvec{A}}^{\!a} + \myvec{D}^{ab} v^b[\myvec{A}])
  = \gvec{\zeta}^a
\end{equation}
has no effect on expectation values of gauge-invariant objects of the
form $F[\myvec{A}]$. This is not the most general modification of
Eq.~(\ref{eq:Bodeker}) which leaves expectation values of gauge invariant
objects unchanged \cite{Hueffel}, but it suffices for our purposes. As
long as $v^a[\myvec{A}]$ contains no time derivatives, the added term has
no effect in calculations of gauge invariant objects. 

We can reformulate this fact in a different way: Since the non-abelian
electric field is given by 
$\myvec{E}^a = -\dot{\myvec{A}}^{\!a} -\myvec{D}^{ab} A^{b0}$,
one may rewrite B\"odeker's equation in the compact form
\begin{equation} \label{eq:BoedekerWithEandB}
  \myvec{D}^{ab} \!\times \myvec{B}^b 
  - \sigma\myvec{E}^a
  = \gvec{\zeta}^a
\end{equation}
which then may be interpreted in any of the so-called flow gauges
$A^{a0} = v^a[\myvec{A}]$ with no time derivatives allowed inside
the functional $v^a[\myvec{A}]$.

The restriction that $v^a[\myvec{A}]$ does not contain time derivatives
plays a more substantial role in our context than in the context of
stochastic quantisation which was the object of Zinn-Justin and Zwanziger:
In stochastic quantisation the time variable describes a \emph{fictitious}
time that is introduced only as a device to reinterpret a given Euclidean
quantum field theory as the limit of a stochastic process for large values
of the fictitious time \cite{DamgaardHueffel:SQ}. Absence of time
derivatives in stochastic quantisation therefore means absence of
derivatives with respect to fictitious time and does not pose any
restrictions to usual time derivatives. In our context, on the contrary,
time is the real, physical time and the restrictions above narrow down
the class of possible gauges leading to a well defined \textsc{Langevin}
equation. 

Moreover, because of the different role of the time variable, we also have
a component of the gauge field that is associated with the $t$ variable of
the \textsc{Langevin} equation. In stochastic quantisation this is not the
case because $t$ is fictitious and the time associated with $A_0$ is just
the zero component of the Euclidean $\myvec{x}$ vector. To cope with this
different structure, to some extent will demand a generalisation of the
proof of Zinn-Justin and Zwanziger.

In effect, we not only have to prove that gauge invariant objects of the form
$F[\myvec{A}]$ are left invariant by the introduction of the term
$v^a[\myvec{A}]$, as was shown in \cite{ZinnJustinZwanziger:WardIdforSQ}.
Instead we have to prove the following: Given B\"odeker's equation in the
form (\ref{eq:BoedekerWithEandB}) and a gauge invariant functional
$F[A^0,\myvec{A}]$, then any choice of a flow gauge leads to the same result.
Or put in different words, calculating
$\langle F[\,v[\myvec{A}],\myvec{A}\,] \rangle$
by means of the equation Eq.~(\ref{eq:Boedeker+v}) gives always the same
value, independent of $v[\myvec{A}]$.

We now proceed in a similar manner to \cite{ZinnJustinZwanziger:WardIdforSQ}.
Let us consider the left-hand side of Eq.~(\ref{eq:Boedeker+v}) where we add
a small variation of the $v^a[\myvec{A}]$ term. We evaluate this expression
for a gauge field that is subject to an arbitrary, infinitesimal gauge
transformation
\begin{math}
  \myvec{A}^{\!\prime a} = \myvec{A}^{\!a} + \myvec{D}^{ab} \omega^b
\end{math}
and find
\begin{eqnarray} \label{eq:TFofBoedeker+dv}
\lefteqn{
  \myvec{D}^{\prime ab}\!\times \myvec{B}^{\prime b}
  + \sigma (\dot{\myvec{A}}^{\!\prime a} 
            + \myvec{D}^{\prime ab} v^b[\myvec{A}^{\prime}]
            + \myvec{D}^{\prime ab} \delta v^b[\myvec{A}^{\prime}]            
           )} \\
  & = & 
  (\delta^{ab} + g f^{abc} \omega^c)
  \left[
    \myvec{D}^{bd}\!\times \myvec{B}^{d}
    + \sigma (\dot{\myvec{A}}^b 
              + \myvec{D}^{bd} v^d[\myvec{A}]
             )
  \right]
  \,+\, 
  \sigma \myvec{D}^{ab} \!
  \left[
    \frac{\partial\omega^b}{\partial t}
    + \left[ H[\myvec{A}]\omega\right]^b
    + \delta v^b[\myvec{A}]    
  \right] \nonumber
\end{eqnarray}
Here we have used
\begin{eqnarray}
  \myvec{D}^{\prime ab}\!\times \myvec{B}^{\prime b} & = &
  (\delta^{ab} + g f^{abc} \omega^c) \,
  \myvec{D}^{bd} \!\times \myvec{B}^d \\
  \dot{\myvec{A}}^{\!\prime a} & = &
  (\delta^{ab} + g f^{abc} \omega^c) \,
  \dot{\myvec{A}}^{\! b}
  \,+\, \myvec{D}^{ab}
  \,\frac{\partial\omega^b}{\partial t} \label{eq:TFofdotA}
\end{eqnarray}
i.e.~the product $\myvec{D}^{ab} \!\times \myvec{B}^b$ transforms
covariantly whereas the transformation of $\dot{\myvec{A}}^{\! a}$
has a covariant and non-covariant contribution. 
In the same way we have split the transformation of $v^a[\myvec{A}]$ 
into a covariant and non-covariant part: Starting from
\begin{equation} \label{eq:VarOfV}
  v^a[\myvec{A}^{\prime}](t,\myvec{x}) =   
  v^a[\myvec{A}](t,\myvec{x}) +
  \int\!\!d^{D-1}\!y\,
  \frac{\delta v^a[\myvec{A}](t,\myvec{x})}{\delta A^{bi}(t,\myvec{y})}
  \;\delta A^{bi}(t,\myvec{y})
\end{equation}
we have indeed
\begin{equation} \label{eq:GaugeTF:v[A]}
  v^a[\myvec{A}^{\prime}](t,\myvec{x}) =  
  (\delta^{ab} + g f^{abc} \omega^c) \,
  v^b[\myvec{A}](t,\myvec{x})
  \,+\,
  \left[ H[\myvec{A}]\omega\right]^a\!(t,\myvec{x})
\end{equation}
where $\delta A^{bi} = D^{bc}_i \omega^c$ has been used and
we have introduced the abbreviation
\begin{equation} \label{def:H[A]}
  \left[ H[\myvec{A}]\omega\right]^a\!(t,\myvec{x}) =
  \int\!\!d^{D-1}\!y\,
  \frac{\delta v^a[\myvec{A}](t,\myvec{x})}{\delta A^{bi}(t,\myvec{y})}
  \; (D^{bc}_i \omega^c)(t,\myvec{y})
  -g f^{abc} v^b[\myvec{A}](t,\myvec{x})\,\omega^c(t,\myvec{x})
\end{equation}
Note that the functional derivatives in Eqs.~(\ref{eq:VarOfV}) 
and (\ref{def:H[A]}) are only with respect to a spacial variation
because $v^a[\myvec{A}]$ does not contain any time derivatives 
(otherwise we would also have to integrate over time).
Let us give the explicit form of this somewhat frightening expression
for $H[A]\omega$ in the case of the choice
$v^a[\myvec{A}]= -\frac{1}{\kappa}\nabla\cdot\myvec{A}^a$. One
simply obtains
\begin{equation} \label{eq:H[A]forKappaGauge}
  \left[ H[\myvec{A}]\omega\right]^a\!(t,\myvec{x}) =
  -\frac{1}{\kappa}\,(\myvec{D}^{ab}\cdot\nabla\omega^b)(t,\myvec{x})
\end{equation}
Finally, Eq.~(\ref{eq:GaugeTF:v[A]}) leads to
\begin{equation} \label{eq:TFofDv}
  \myvec{D}^{\prime ab} v^b[\myvec{A}^{\prime}] = 
  (\delta^{ab} + g f^{abc} \omega^c) \,
  \myvec{D}^{bd} v^d[\myvec{A}]
  \, +\, 
  \myvec{D}^{ab} \left[ H[\myvec{A}]\omega\right]^b  
\end{equation}
where it was used that $\omega$ is infinitesimal and of course
\begin{equation} \label{eq:TFofDdv}
  \myvec{D}^{\prime ab} \delta v^b[\myvec{A}^{\prime}] =
  \myvec{D}^{ab} \delta v^b[\myvec{A}]
\end{equation}
because $\delta v$ is infinitesimal itself.

Let us now come back to Eq.~(\ref{eq:TFofBoedeker+dv}) and its 
meaning. Suppose the gauge field, before the gauge transformation
has been performed, was a solution of B\"odeker's equation with 
the $v^a[\myvec{A}]$ term present, but without the additional 
$\delta v^a[\myvec{A}]$ term.
In other words, the original gauge field was a solution of 
Eq.~(\ref{eq:Boedeker+v}). We can then replace the first square 
bracket on the right-hand side of Eq.~(\ref{eq:TFofBoedeker+dv}) by
the stochastic force and find
\begin{eqnarray} \label{eq:TFofBoedekerWithZeta}
\lefteqn{
  \myvec{D}^{\prime ab}\!\times \myvec{B}^{\prime b}
  + \sigma (\dot{\myvec{A}}^{\!\prime a} 
            + \myvec{D}^{\prime ab} v^b[\myvec{A}^{\prime}]
            + \myvec{D}^{\prime ab} \delta v^b[\myvec{A}^{\prime}]            
           )} \nonumber\\
  & = & 
  \gvec{\zeta}^{\prime a}
  \,+\, 
  \sigma \myvec{D}^{ab} \!
  \left[
    \frac{\partial\omega^b}{\partial t}
    + \left[ H[\myvec{A}]\omega\right]^b
    + \delta v^b[\myvec{A}]    
  \right]
\end{eqnarray}
This means, if we subject the original gauge field to an arbitrary,
infinitesimal gauge transformation with parameter $\omega$, then the
gauge transformed field will be a solution of 
Eq.~(\ref{eq:TFofBoedekerWithZeta}), i.e.~of the original equation
with $v$ replaced by $v + \delta v$ and the stochastic force 
transformed in the same way as the gauge field \dots but with an 
ugly additional term on the right-hand side.
However, one can play a dirty trick: What was said so far was true for
an \emph{arbitrary} gauge transformation. But if we demand $\omega$
to be a solution of
\begin{equation}\label{eq:OmegaConstraint}
  \frac{\partial\omega^b}{\partial t}
  + \left[ H[\myvec{A}]\omega\right]^b
  + \delta v^b[\myvec{A}] = 0  
\end{equation}
then the square bracket on the right 
of Eq.~(\ref{eq:TFofBoedekerWithZeta})
will vanish and we finally arrive at
\begin{equation}
  \myvec{D}^{\prime ab}\!\times \myvec{B}^{\prime b}
  + \sigma (\dot{\myvec{A}}^{\!\prime a} 
            + \myvec{D}^{\prime ab} v^b[\myvec{A}^{\prime}]
            + \myvec{D}^{\prime ab} \delta v^b[\myvec{A}^{\prime}]            
           )
   = 
  \gvec{\zeta}^{\prime a}
\end{equation}
However, there is a certain subtlety that we want to draw attention to.
To clarify this point, let us once again repeat the line of reasoning:
Starting with a gauge field being solution of
\begin{equation} \label{eq:Original}
  \myvec{D}^{ab} \!\times \myvec{B}^b 
  + \sigma (\dot{\myvec{A}}^{\!a} + \myvec{D}^{ab} v^b[\myvec{A}])
  = \gvec{\zeta}^a
\end{equation}
we search for a gauge transformation $\omega$ that obeys
\begin{equation} \label{eq:CondOnOmega}
  \frac{\partial\omega^a}{\partial t}
  + \left[ H[\myvec{A}]\omega\right]^a
  + \delta v^a[\myvec{A}] = 0  
\end{equation}
(and we can always find such an $\omega$ because (\ref{eq:CondOnOmega})
is a linear, inhomogeneous equation with given inhomogeneity 
$\delta v^a[\myvec{A}]$). Then the gauge field transformed with 
\emph{this} $\omega$, 
\begin{math}
  \myvec{A}^{\!\prime a} = \myvec{A}^{\!a} + \myvec{D}^{ab} \omega^b
\end{math},
is a solution of the original equation with $v$ replaced by $v + \delta v$
and the stochastic force also transformed by the same $\omega$
\begin{equation} \label{eq:TransformedEq}
  \myvec{D}^{\prime ab}\!\times \myvec{B}^{\prime b}
  + \sigma (\dot{\myvec{A}}^{\!\prime a} 
            + \myvec{D}^{\prime ab} v^b[\myvec{A}^{\prime}]
            + \myvec{D}^{\prime ab} \delta v^b[\myvec{A}^{\prime}]            
           )
   = 
  \gvec{\zeta}^{\prime a}
\end{equation}
The subtle point is the following: The original gauge field $\myvec{A}$ 
is a solution of Eq.~(\ref{eq:Original}) and thus depends on the stochastic
force $\gvec{\zeta}$, of course. But $\myvec{A}$ is an input of
Eq.~(\ref{eq:CondOnOmega}) that determines $\omega$. Therefore, $\omega$
via $\myvec{A}$ too depends on $\gvec{\zeta}$. As a consequence of this,
$\gvec{\zeta}^{\prime}$ inherits a non-trivial dependence on $\gvec{\zeta}$:
The stochastic force $\gvec{\zeta}^{\prime}$ not only depends on
$\gvec{\zeta}$ because it is the gauge transform of $\gvec{\zeta}$, but also 
because the gauge transformation itself depends on $\gvec{\zeta}$
\begin{equation} \label{eq:ZetaPrimeOfZeta}
  \gvec{\zeta}^{\prime a} =
  (\delta^{ab} + g f^{abc} \omega^c[\gvec{\zeta}]) \, \gvec{\zeta}^b
\end{equation}

We denote by 
$\myvec{A}^{\!s}[\gvec{\zeta},v,\myvec{A}_{\mathrm{ini}}]$
the solution of Eq.~(\ref{eq:Boedeker+v}) for the specific
realisation $\gvec{\zeta}$ of the stochastic force term and
initial conditions $\myvec{A}_{\mathrm{ini}}$.
Correspondingly, let
$\myvec{A}^{\!s}[\gvec{\zeta},v+\delta v,\myvec{A}_{\mathrm{ini}}]$
denote the solution of this equation with 
$v$ replaced by $v + \delta v$ and for the same stochastic force
and initial conditions. We can then express the contents of 
Eq.~(\ref{eq:TransformedEq}) in this new notation 
\begin{equation} \label{eq:TFnewNotation}
  \myvec{A}^{\!s}
  [{}^{\omega}\!\gvec{\zeta},v+\delta v,
   {}^{\omega}\!\myvec{A}_{\mathrm{ini}}]
  =
  {}^{\omega}\!\myvec{A}^{\!s}
  [\gvec{\zeta},v,\myvec{A}_{\mathrm{ini}}]
\end{equation}
where the superscript $\omega$ indicates gauge transformation with
the special parameter $\omega$ corresponding to the solution on the
right-hand side via Eq.~(\ref{eq:CondOnOmega}).

\noindent
After these preparations we can now show that gauge invariant
expectation values $\left\langle F[A^0,\myvec{A}] \right\rangle$ 
are independent of the choice of $v^a[\myvec{A}]$. To this end, let us
write the gauge invariant observable as functional of the non-abelian
electric and magnetic field
\begin{equation} \label{def:non-abelianE&B}
  \begin{array}{rcl}
    \myvec{E}^a & = & -\dot{\myvec{A}}^{\!a} \;
                      -\myvec{D}^{ab} A^{b0} \\[1.5ex]
    \myvec{B}^a & = & \nabla\times\myvec{A}^{\!a} 
    \;+\;\frac{1}{2}\,g f^{abc} \myvec{A}^{\!b} \times \myvec{A}^{\!c} 
  \end{array}
\end{equation}
We then have
\begin{equation}
  \left\langle F[\myvec{E},\myvec{B}]\right\rangle_{v + \delta v} = 
  \int\!\mathcal{D}\gvec{\zeta}^{\prime}\;
  \varrho[\gvec{\zeta}^{\prime}]\;
  F\bigl[\,\myvec{E}_{v + \delta v}[\myvec{A}],
      \myvec{B}_{v + \delta v}[\myvec{A}]\,\bigr]_{\myvec{A}=
        \myvec{A}^{\!s}[\zeta^{\prime},v+\delta v,
        \myvec{A}_{\mathrm{ini}}^{\prime}]}
\end{equation}
with
\begin{equation}
  \myvec{E}_{v + \delta v}^a[\myvec{A}] =
  -\dot{\myvec{A}}^{\!a} 
  -\myvec{D}^{ab} v^b[\myvec{A}]
  -\myvec{D}^{ab} \delta v^b[\myvec{A}]
\end{equation}
and 
\begin{math}
  \myvec{B}_{v + \delta v}[\myvec{A}] =
  \myvec{B}_v[\myvec{A}]
\end{math}
as in Eq.~(\ref{def:non-abelianE&B}). Changing variables 
according to Eq.~(\ref{eq:ZetaPrimeOfZeta}), one obtains
\begin{equation}\label{eq:F[A0,A]_ChangeofVariables}
  \left\langle F[\myvec{E},\myvec{B}]\right\rangle_{v + \delta v} = 
  \int\!\mathcal{D}\gvec{\zeta}\;
  \mathrm{Det}
  \left(
    \frac{\delta{}^{\omega}\!\gvec{\zeta}}{\delta\gvec{\zeta}}
  \right)
  \varrho[{}^{\omega}\!\gvec{\zeta}]\;
  F\bigl[\,\myvec{E}_{v + \delta v}[\myvec{A}],
      \myvec{B}_{v + \delta v}[\myvec{A}]\,\bigr]_{\myvec{A}=
        \myvec{A}^{\!s}[{}^{\omega}\!\zeta,v+\delta v,
        \myvec{A}_{\mathrm{ini}}^{\prime}]}
\end{equation}
We now use independence on the initial conditions, 
the transformation property (\ref{eq:TFnewNotation}), gauge
invariance of $\varrho[\gvec{\zeta}]$ and finally the fact 
that the determinant is unity (shown in Appendix \ref{Jacobian}).
This all together leads to
\begin{equation}
  \left\langle F[\myvec{E},\myvec{B}]\right\rangle_{v + \delta v} = 
  \int\!\mathcal{D}\gvec{\zeta}\;
  \varrho[\gvec{\zeta}]\;
  F\bigl[\,\myvec{E}_{v + \delta v}[{}^{\omega}\!\myvec{A}],
    \myvec{B}_{v + \delta v}[{}^{\omega}\!\myvec{A}]\,\bigr]_{\myvec{A}=
    \myvec{A}^{\!s}
    [\zeta,v,\myvec{A}_{\mathrm{ini}}]}
\end{equation}
Taking into account the transformation properties (\ref{eq:TFofdotA}),
(\ref{eq:TFofDv}) and (\ref{eq:TFofDdv}), we find
\begin{equation}
\begin{array}{rcl}
  \myvec{E}_{v + \delta v}^a[{}^{\omega}\!\myvec{A}] & = & 
  ({}^{\omega}\myvec{E}_v[\myvec{A}] )^a
  -
  \myvec{D}^{ab} \!
  \left[
    \frac{\partial\omega^b}{\partial t}
    + \left[ H[\myvec{A}]\omega\right]^b
    + \delta v^b[\myvec{A}]    
  \right]
  =
  ({}^{\omega}\myvec{E}_v[\myvec{A}] )^a \\[1.5ex]
  \myvec{B}_{v + \delta v}^a[{}^{\omega}\!\myvec{A}] & = &
  ({}^{\omega}\myvec{B}_v[\myvec{A}] )^a 
\end{array}
\end{equation}
and thus
\begin{equation}
  \left\langle F[\myvec{E},\myvec{B}]\right\rangle_{v + \delta v} = 
  \int\!\mathcal{D}\gvec{\zeta}\;
  \varrho[\gvec{\zeta}]\;
  F\bigl[\,{}^{\omega}\myvec{E}_v[\myvec{A}],
    {}^{\omega}\myvec{B}_v[\myvec{A}]\,\bigr]_{\myvec{A}=
    \myvec{A}^{\!s}
    [\zeta,v,\myvec{A}_{\mathrm{ini}}]}
  =
  \left\langle F[\myvec{E},\myvec{B}]\right\rangle_v
\end{equation}
because $F[\myvec{E},\myvec{B}]$ is a gauge invariant functional.

\noindent
Consequently, we have shown that B\"odeker's equation in $A_0 = 0$ gauge
\begin{equation}\label{eq:Bodeker2}
  \myvec{D}^{ab} \!\times \myvec{B}^b + \sigma\dot{\myvec{A}}^{\!a} 
  = \gvec{\zeta}^a
\end{equation}
can equivalently be formulated in any flow gauge
\begin{equation}
  \myvec{D}^{ab} \!\times \myvec{B}^b 
  + \sigma (\dot{\myvec{A}}^{\!a} + \myvec{D}^{ab} v^b[\myvec{A}])
  = \gvec{\zeta}^a
\end{equation}
without any time derivatives allowed inside the functional $v^a[\myvec{A}]$.
We will henceforth use the special choice
$A^{a0} = v^a[\myvec{A}]= -\frac{1}{\kappa}\nabla\cdot\myvec{A}^{\!a}$
and refer to it as $\kappa$ gauge. This is a natural choice for
$v^a[\myvec{A}]$, since it has the lowest order in $\Ab$, preserves colour
invariance, and with $\kappa>0$ the term $\myvec{D}^{ab} v^b[\myvec{A}]$
provides a globally restoring force along gauge orbits \cite{Zwanziger2003},
while at the same time having the correct dimensions.

\section{BRST Symmetric Action and Ward-Takahashi Identities}
\label{WardIDs}

We have argued that in order to derive any reliable statements from our
theory, it is essential to gain some control over the gauge dependence
possibly introduced by the truncation of the \textsc{Dyson--Schwinger} 
equations. This was our main motivation to generalise B\"odeker's
equation from $A_0\!=\!0$ gauge to a more general class of flow gauges.
In addition to this, the corresponding introduction of a gauge-fixing 
force has a welcome side-effect: It solves at the same time the
problem of undamped longitudinal components of the 
initial gauge field configuration.

However, the detection of an unphysical gauge dependence is not what 
we really want; in fact, we would rather like to avoid it. The ultimate 
goal is to construct a truncation scheme that is physically reasonable 
and does not (or, realistically speaking, only slightly) violate the 
gauge symmetry.

To this end, we need identities expressing the gauge symmetry on the 
level of n-point functions, i.e.~we need the \textsc{Ward--Takahashi} 
identities of the theory.\footnote{In the non-abelian context, these
identities are often referred to as \textsc{Slavnov--Taylor} identities.
However, following the terminology of 
Ref.~\cite{ZinnJustinZwanziger:WardIdforSQ}, we denote these identities
as gauge Ward identities in analogy to the stochastic Ward identities
also encountered in this work.}

Any physically reasonable truncation will have to respect these
identities. Besides this conceptual importance, we may also hope that 
some of the Ward identities to be derived in the following 
will be of some practical use in solving the 
\textsc{Dyson--Schwinger} equations: In ordinary QCD, for instance,
the full gluon propagator in covariant gauge is restricted to 
being purely transversal as a consequence of the Ward identities. 
This leads, of course, to a great simplification in the  
\textsc{Dyson--Schwinger} equations of QCD.

In this section, we study three different kinds of non-perturbative
identities: gauge Ward identities, i.e. \textsc{Slavnov--Taylor}
identities; stochastic Ward identities; and ghost number conservation. 

\subsection{Constructing a BRST Symmetric Action}
In Section \ref{se:UpgradeKappaGauge}, we saw that Eq.~(\ref{eq:Boedeker+v}) 
transforms covariantly only under a restricted class of gauge transformations.
Obtaining the gauge Ward identities with this restriction turns out to be
rather cumbersome. Instead, we will raise to life the gauge parameter
$\omega$ by introducing into the theory an additional (Grassmann valued)
field that realises the constraint on the gauge transformations. The
resulting action will be endowed with a BRST symmetry, and we will be able
to obtain the gauge Ward identities in a straight-forward manner. 

Setting $\delta v^a[\myvec{A}]$ to zero in Eq.~(\ref{eq:OmegaConstraint}) 
we see that Eq.~(\ref{eq:Boedeker+v}) transforms covariantly under gauge
transformations which obey
\begin{equation}\label{eq:OmegaConstraint2}
  \frac{\partial\omega^b}{\partial t}
  + \left[ H[\myvec{A}]\omega\right]^b
  = 0  
\end{equation}
Note that the introduction of $v^a[\myvec{A}]$ does not restrict the gauge
group any further than it already would be. Even without the extra term,
the gauge transformations would have to be restricted in order for 
Eq.~(\ref{eq:Boedeker+v}) to be gauge covariant.

The restriction in Eq.~(\ref{eq:OmegaConstraint2}) can be taken into account
in the path integral in the following way.
Define a term $\gamma^a[\omega,\myvec{A}]$ from the left-hand side of
Eq.~(\ref{eq:OmegaConstraint2}), which for our choice of the functional
$v^a[\myvec{A}]$ takes the form
\begin{equation}  \label{def:gamma[omega]}
  \gamma^a[\omega,\myvec{A}] =
  \frac{\partial\omega^a}{\partial t}
  -\frac{1}{\kappa}\,\myvec{D}^{ab}\!\cdot\nabla\omega^b
\end{equation}
Perform a change of variables from $\gamma$ to $\omega$ in the following
Grassmann integral representation of unity
\begin{equation} \label{eq:UnityByOmegas}
  1 = \int\!\mathcal{D}\gamma\; \delta(\gamma)
    = \int\!\mathcal{D}\omega\;
    \frac{1}{\mathrm{Det}\!\left(\frac{\delta \gamma[\omega,\myvec{A}]}
                                      {\delta\omega}\right)} \;
    \delta\Bigl(\,\frac{\partial\omega^a}{\partial t}
             -\frac{1}{\kappa}\,\myvec{D}^{ab}\!\cdot\nabla\omega^b
          \Bigr)
\end{equation}
Since the determinant is Grassmann even, it no longer depends on $\omega$
and it can be pulled out of the integral. The determinant is a constant,
and can be calculated in a similar manner to the determinant in
Eq.~(\ref{eq:PathIntegral3}); see Appendix~\ref{Jacobian} for an explicit
calculation.

\noindent
Inserting the integral representation of the Grassmann delta function
\begin{equation} \label{eq:GrassmannDelta}
  \delta(\gamma) = \int\!\mathcal{D}\bar{\omega}\;
  \exp\left\{
    \int\!\!dx\;\bar{\omega}^a(x)\,\gamma^a(x)
  \right\}
\end{equation}
and absorbing the constant determinant into the measure, we find the identity
\begin{equation} \label{eq:UnityWithOmegaOmegaBar}
  1 = \int\!\mathcal{D}\omega\mathcal{D}\bar{\omega}\;
      \exp\left\{
        \int\!\!dx\;\bar{\omega}^a(x)
        \Bigl(\,\delta^{ab}\frac{\partial}{\partial t}
               -\frac{1}{\kappa}\,\myvec{D}^{ab}\!\cdot\nabla
        \Bigr) \omega^b(x)
      \right\}
\end{equation}
which holds independently of the gauge field $\myvec{A}$. Therefore, it can
be inserted into the analogous of the path integral representation of the
generating functional, Eq.~(\ref{eq:PathIntegral4}), based on the generalised
version of B\"odeker's equation (\ref{eq:Boedeker+v}). This leads to
\begin{equation} \label{eq:GenFuncWithOmega}
  Z[\myvec{J}] =
  \int\!\mathcal{D}\!\myvec{A}\mathcal{D}\!\gvec{\lambda}
        \mathcal{D}\omega\mathcal{D}\bar{\omega}\;
  \exp\left\{-S[\myvec{A},\gvec{\lambda},\omega,\bar{\omega}]
             + \int\!\!dx\;\myvec{J}^a(x) \myvec{A}^{\!a}(x)
      \right\}
\end{equation}
with the action now given by
\begin{equation} \label{eq:S[A,Lambda,Omega,OmegaBar]}
  S[\myvec{A},\gvec{\lambda},\omega,\bar{\omega}] =
  S^{\scriptscriptstyle\mathrm{(D)}}[\myvec{A},\gvec{\lambda}]
  \,+\, S^{\scriptscriptstyle\mathrm{(GG)}}
  [\myvec{A},\omega,\bar{\omega}]
\end{equation}
where $S^{\scriptscriptstyle\mathrm{(D)}}[\myvec{A},\gvec{\lambda}]$ is
the generalised contribution of the dynamical fields
\begin{equation}
  S^{\scriptscriptstyle\mathrm{(D)}}[\myvec{A},\gvec{\lambda}] =
  \int\!\!dx\,
  \Bigl[
    \sigma T \,\gvec{\lambda}^a\!\cdot\!\gvec{\lambda}^a
    -i \gvec{\lambda}^a \!\cdot\!
    \left(
      \myvec{D}^{ab} \!\times \myvec{B}^b
      + \sigma (\dot{\myvec{A}}^{\!a}
                -{\textstyle\frac{1}{\kappa}\,}
                 \myvec{D}^{ab}\,\nabla\!\cdot\!\myvec{A}^{\!b}
               )
    \right)
  \Bigr]
\end{equation}
and
\begin{equation} \label{eq:S(GG)}
  S^{\scriptscriptstyle\mathrm{(GG)}}[\myvec{A},\omega,\bar{\omega}]
  =
  \int\!\!dx\,
  \Bigl[
    -\,\bar{\omega}^a \dot{\omega}^a
    +\frac{1}{\kappa}\,
      \bar{\omega}^a \myvec{D}^{ab}\!\cdot\nabla \omega^b
  \Bigr]
\end{equation}
is the new contribution containing the gauge ghosts $\omega$
and $\bar{\omega}$.

\subsection{Gauge Ward Identities}

\noindent
The \textsc{Slavnov--Taylor} identities can be derived by noting that the
action (\ref{eq:S[A,Lambda,Omega,OmegaBar]}) is invariant under the
following BRST transformation
\begin{equation} \label{eq:GaugeBRST}
  \begin{array}{rcl@{\qquad}rcl}
    \delta_{\varepsilon}\myvec{A}^{\!a}(x)
    \!\! & = & \!\!
    \myvec{D}^{ab}(x) \,\varepsilon\omega^b(x)
    &
    \delta_{\varepsilon} \omega^a(x)
    \!\! & = & \!\!
    \frac{1}{2} g f^{abc} \varepsilon\omega^c(x) \omega^b(x) \\[1.0ex]
    \delta_{\varepsilon}\gvec{\lambda}^{\!a}(x)
    \!\! & = & \!\!
    g f^{abc} \varepsilon\omega^c(x) \gvec{\lambda}^{\!b}(x)
    &
    \delta_{\varepsilon} \bar{\omega}^a(x)
    \!\! & = & \!\!
    g f^{abc} \varepsilon\omega^c(x) \bar{\omega}^b(x)
    + i\varepsilon\sigma\myvec{D}^{ab}(x)\!\cdot\!\gvec{\lambda}^{\!b}(x)
    \end{array}
\end{equation}
where $\varepsilon$ is a constant Grassmann parameter. It is convenient to
introduce the finite BRST operator $s$ such that the result of $s$ acting
on a functional of the fields $\myvec{A}$, $\gvec{\lambda}$, $\omega$ and
$\bar{\omega}$ is defined as (left) derivative with respect to the parameter
$\varepsilon$ of the variations in Eq.~(\ref{eq:GaugeBRST}). We thus have
\begin{equation} \label{def:finiteBRST}
  s F[\myvec{A},\gvec{\lambda},\omega,\bar{\omega}]
  \;=\;
  \frac{\partial}{\partial\varepsilon}\,
  \delta_{\varepsilon}
  F[\myvec{A},\gvec{\lambda},\omega,\bar{\omega}]
\end{equation}
or conversely
\begin{equation}
  \delta_{\varepsilon}
  F[\myvec{A},\gvec{\lambda},\omega,\bar{\omega}]
  \;=\;
  \varepsilon\, s F[\myvec{A},\gvec{\lambda},\omega,\bar{\omega}]
\end{equation}
From Eq.~(\ref{def:finiteBRST}) one finds the following representation
\begin{equation} \label{eq:FuncDiffRepOfS}
  s = \int\!\!dx\left[
         (sA^{ai}) \,\frac{\delta}{\delta\! A^{ai}}
      \,+\,  (s\lambda^{ai}) \,\frac{\delta}{\delta\lambda^{ai}}
      \,+\,  (s\omega^a) \,\frac{\delta}{\delta\omega^a}
      \,+\,  (s\bar{\omega}^a) \,\frac{\delta}{\delta\bar{\omega}^a}
    \right]
\end{equation}
with the finite BRST transforms of the fundamental fields given by
Eq.~(\ref{eq:GaugeBRST})
\begin{equation} \label{eq:finiteBRST}
  \begin{array}{rcl@{\qquad}rcl}
    s\myvec{A}^{\!a}(x)
    \!\! & = & \!\!
    \myvec{D}^{ab}(x) \,\omega^b(x)
    &
    s\omega^a(x)
    \!\! & = & \!\!
    \frac{1}{2} g f^{abc} \omega^c(x) \omega^b(x) \\[1.0ex]
    s\gvec{\lambda}^{\!a}(x)
    \!\! & = & \!\!
    g f^{abc} \omega^c(x) \gvec{\lambda}^{\!b}(x)
    &
    s\bar{\omega}^a(x)
    \!\! & = & \!\!
    g f^{abc} \omega^c(x) \bar{\omega}^b(x)
    + i\sigma\myvec{D}^{ab}(x)\!\cdot\!\gvec{\lambda}^{\!b}(x)
    \end{array}
\end{equation}
The BRST operator $s$ has two essential properties: it annihilates the
complete action (\ref{eq:S[A,Lambda,Omega,OmegaBar]})
\begin{equation}
  sS[\myvec{A},\gvec{\lambda},\omega,\bar{\omega}]=0
\end{equation}
expressing the invariance of
$S[\myvec{A},\gvec{\lambda},\omega,\bar{\omega}]$ under the BRST
transformation (\ref{eq:GaugeBRST}), and it's nilpotency
\begin{equation}
  s^2 = 0
\end{equation}
Using the operator $s$, we now define the generating functional in the
following way
\begin{eqnarray} \label{eq:Z[J,I]}
  Z[J,I] & = &
  \int\!\mathcal{D}\!\myvec{A}\mathcal{D}\!\gvec{\lambda}
        \mathcal{D}\omega\mathcal{D}\bar{\omega}\;
  \exp\biggl\{-S[\myvec{A},\gvec{\lambda},\omega,\bar{\omega}]
             + \int\!\!dx\,
   \Bigl[
       \myvec{A}^{\!a} \!\cdot \myvec{J}_{\!A}^a
    \;+\;  \gvec{\lambda}^{\!a} \!\cdot \myvec{J}_{\!\lambda}^a
    \;+\;  \omega^a J_{\omega}^a
    \;+\;  \bar{\omega}^a J_{\bar{\omega}}^a
   \Bigr.\biggr. \nonumber\\ 
   & & \hspace{4.0cm}
   \biggl.\Bigl.
    + \ \myvec{I}_{s\!A}^a \!\cdot s\myvec{A}^{\!a}
    +   \myvec{I}_{s\!\lambda}^a \!\cdot s\gvec{\lambda}^{\!a}
    +   I_{s\omega}^a s\omega^a
    +   I_{s\bar{\omega}}^a s\bar{\omega}^a
   \Bigr]
   \biggr\}
\end{eqnarray}
Note that $\omega$, $\bar{\omega}$, $s\myvec{A}$ and
$s\gvec{\lambda}$ together with their sources
$J_{\omega}$, $J_{\bar{\omega}}$, $\myvec{I}_{s\!A}$,
$\myvec{I}_{s\!\lambda}$ are Grassmann odd, the remaining
quantities Grassmann even.

We proceed to vary the fields in Eq.~(\ref{eq:Z[J,I]}) according to
Eq.~(\ref{eq:GaugeBRST}). The Jacobian of such a transformation is unity
due to Eq.~(\ref{eq:finiteBRST}) as can be seen from the explicit calculation
in Appendix~\ref{Jacobian}.
We also know that the action is invariant under this change of variables
\begin{math}
  S[\myvec{A},\gvec{\lambda},\omega,\bar{\omega}] =
  S[\myvec{A}^{\!\prime},\gvec{\lambda}^{\!\prime},
    \omega^{\prime},\bar{\omega}^{\prime}]
\end{math}.
In addition, the source terms of the BRST transformed fields are
invariant due to the nilpotency of $s$ and the fact that the variations are
$s$-transforms themselves, e.g.~$\delta_{\varepsilon}\myvec{A}^{\!\prime}
= \varepsilon s\!\myvec{A}^{\!\prime}$. Only the source terms of the
fundamental fields are not invariant and transform according to
\begin{equation} \label{eq:TFofFundamentalSourceTerms}
  \myvec{A}^{\!a} \!\cdot \myvec{J}_{\!A}^a \;=\;
  \myvec{A}^{\!\prime a} \!\cdot \myvec{J}_{\!A}^a
  \,+\, \delta_{\varepsilon} \myvec{A}^{\!\prime a}\!\cdot \myvec{J}_{\!A}^a
  \;=\;
  \myvec{A}^{\!\prime a} \!\cdot \myvec{J}_{\!A}^a
  \,+\, \varepsilon \,s\myvec{A}^{\!\prime a}\!\cdot \myvec{J}_{\!A}^a
\end{equation}
and likewise for the other fields. Thus, under the change of variables
(\ref{eq:GaugeBRST}), the integrand in Eq.~(\ref{eq:Z[J,I]}) 
is simply reproduced with all fields replaced by their
primed counterparts and an additional factor
\begin{equation}
  \exp\biggl\{ \varepsilon
    \int\!\!dx\,
    \Bigl[
       s\myvec{A}^{\!\prime a} \!\cdot \myvec{J}_{\!A}^a
    \;+\;  s\gvec{\lambda}^{\!\prime a} \!\cdot \myvec{J}_{\!\lambda}^a
    \;+\;  s\omega^{\prime a} J_{\omega}^a
    \;+\;  s\bar{\omega}^{\prime a} J_{\bar{\omega}}^a
    \Bigr]
   \biggr\}
\end{equation}
generated by the transformation of the fundamental source terms,
Eq.~(\ref{eq:TFofFundamentalSourceTerms}).
Because $\varepsilon$ is Grassmann odd we have
\begin{displaymath}
    \exp\biggl\{ \varepsilon\!
    \int\!\!dx\,
    \Bigl[
       s\myvec{A}^{\!a} \!\cdot \myvec{J}_{\!A}^a
    +  s\gvec{\lambda}^{\!a} \!\cdot \myvec{J}_{\!\lambda}^a
    +  s\omega^a J_{\omega}^a
    +  s\bar{\omega}^a J_{\bar{\omega}}^a
    \Bigr]
   \biggr\} =
   1 +
   \varepsilon\!\!
    \int\!\!dx\,
    \Bigl[
       s\myvec{A}^{\!\prime a} \!\cdot \myvec{J}_{\!A}^a
    \;+\;  s\gvec{\lambda}^{\!\prime a} \!\cdot \myvec{J}_{\!\lambda}^a
    \;+\;  s\omega^{\prime a} J_{\omega}^a
    \;+\;  s\bar{\omega}^{\prime a} J_{\bar{\omega}}^a
    \Bigr]
\end{displaymath}
Inserted back into the path integral Eq.~(\ref{eq:Z[J,I]}),
the one just gives $Z[J,I]$, which cancels the
left-hand side of the equation. Hence, we obtain
\begin{equation} \label{eq:someInt=0}
  0 =
  \int\!\mathcal{D}\!\myvec{A}\mathcal{D}\!\gvec{\lambda}
        \mathcal{D}\omega\mathcal{D}\bar{\omega}\;
   \varepsilon\!\!
    \int\!\!dx\,
    \Bigl[
       s\myvec{A}^{\!a} \!\cdot \myvec{J}_{\!A}^a
    +  s\gvec{\lambda}^{\!a} \!\cdot \myvec{J}_{\!\lambda}^a
    +  s\omega^a J_{\omega}^a
    +  s\bar{\omega}^a J_{\bar{\omega}}^a
    \Bigr]
   \exp\biggl\{ (\dots)
   \biggr\}
\end{equation}
where the dots represent the exponential in Eq.~(\ref{eq:Z[J,I]}).
This has to be true for any $\varepsilon$ and thus the expression
without $\varepsilon$ has to vanish itself. Changing the BRST transformed
fields for functional derivatives with respect to their sources, we find
the following identity
\begin{equation} \label{eq:ST-Z}
  \int\!\!dx\,
  \Bigl[
    J_{\!A}^{ai}(x) \,\frac{\delta}{\delta I_{s\!A}^{ai}(x)}
  + J_{\!\lambda}^{ai}(x) \,\frac{\delta}{\delta I_{s\!\lambda}^{ai}(x)}
  + J_{\omega}^a(x) \,\frac{\delta}{\delta I_{s\omega}^a(x)}
  + J_{\bar{\omega}}^a(x) \,\frac{\delta}{\delta I_{s\bar{\omega}}^a(x)}
  \Bigr]
  Z[J,I] = 0
\end{equation}
Finally, let us transcribe this relation in an identity
for the generating functional of one-particle irreducible (1PI)
correlation functions. To this end, we first express it by
the generating functional of connected correlation
functions $W[J,I] = \ln Z[J,I]$.
In terms of $W[J,I]$ the relation (\ref{eq:ST-Z}) reads
\begin{equation} \label{eq:ST-W}
  \int\!\!dx\,
  \Bigl[
    J_{\!A}^{ai}(x) \,\frac{\delta W[J,I]}
                           {\delta I_{s\!A}^{ai}(x)}
  + J_{\!\lambda}^{ai}(x) \,\frac{\delta W[J,I]}
                                 {\delta I_{s\!\lambda}^{ai}(x)}
  + J_{\omega}^a(x) \,\frac{\delta W[J,I]}
                           {\delta I_{s\omega}^a(x)}
  + J_{\bar{\omega}}^a(x) \,\frac{\delta W[J,I]}
                                 {\delta I_{s\bar{\omega}}^a(x)}
  \Bigr]
   = 0
\end{equation}
To define the generating functional of one-particle irreducible
correlation functions, we introduce the usual expectation
values for the fields in the presence of the external sources
\begin{equation} \label{def:classicalFields}
  \begin{array}{rcl@{\qquad\quad}rcl}
    \displaystyle
    A^{ai}(x)
    \!\! & = & \!\!
    \displaystyle
    \frac{\delta W[J,I]}{\delta J_A^{ai}(x)}
    &
    \displaystyle
    \omega^a(x)
    \!\! & = & \!\!
    \displaystyle
    -\frac{\delta W[J,I]}{\delta J_{\omega}^a(x)}
    \\[3.5ex]
    \displaystyle
    \lambda^{ai}(x)
    \!\! & = & \!\!
    \displaystyle
    \frac{\delta W[J,I]}{\delta J_{\lambda}^{ai}(x)}
    &
    \displaystyle
    \bar{\omega}^a(x)
    \!\! & = & \!\!
    \displaystyle
    -\frac{\delta W[J,I]}{\delta J_{\bar{\omega}}^a(x)}
    \end{array}
\end{equation}
The minus signs in the case of the ghost fields are a consequence
of our definition of the generating functional, Eq.~(\ref{eq:Z[J,I]}),
where we ordered the sources to the right of the fundamental fields.

Assuming that the relations (\ref{def:classicalFields}) can be solved
for the sources $J$, we can define the 1PI generating functional $\Gamma$
as the \textsc{Legendre} transform of $W[J,I]$ with respect
to the sources $J$. The sources of the BRST transformed fields
are not \textsc{Legendre} transformed and play the
role of spectators only. With the definition
\begin{equation}
  \Gamma[\myvec{A},\gvec{\lambda},\omega,\bar{\omega};I] =
  \int\!\!dx\left[
    \myvec{A}^{\!a} \!\cdot \myvec{J}_{\!A}^a
    \;+\;  \gvec{\lambda}^{\!a} \!\cdot \myvec{J}_{\!\lambda}^a
    \;+\;  \omega^a J_{\omega}^a
    \;+\;  \bar{\omega}^a J_{\bar{\omega}}^a
   \right] - W[J,I]
\end{equation}
one finds
\begin{equation} \label{eq:DerivativesOfGamma}
  \begin{array}{rcl@{\qquad\quad}rcl}
    \displaystyle
    \frac{\delta \Gamma}{\delta A^{ai}(x)}
    \!\! & = & \!\!
    \displaystyle
    J^{ai}_A(x)
    &
    \displaystyle
    \frac{\delta \Gamma}{\delta \omega^a(x)}
    \!\! & = & \!\!
    \displaystyle
    J^a_{\omega}(x)
    \\[3ex]
    \displaystyle
    \frac{\delta \Gamma}{\delta \lambda^{ai}(x)}
    \!\! & = & \!\!
    \displaystyle
    J^{ai}_{\lambda}(x)
    &
    \displaystyle
    \frac{\delta \Gamma}{\delta \bar{\omega}^a(x)}
    \!\! & = & \!\!
    \displaystyle
    J^a_{\bar{\omega}}(x)
    \end{array}
\end{equation}
and also
\begin{equation} \label{eq:MoreDerivativesOfGamma}
  \begin{array}{rcl@{\qquad\quad}rcl}
    \displaystyle
    \frac{\delta \Gamma}{\delta I^{ai}_{s\!A}(x)}
    \!\! & = & \!\!
    \displaystyle
    -\frac{\delta W}{\delta I^{ai}_{s\!A}(x)}
    &
    \displaystyle
    \frac{\delta \Gamma}{\delta I_{s\omega}^a(x)}
    \! & = & \!\!
    \displaystyle
    -\frac{\delta W}{\delta I_{s\omega}^a(x)}
    \\[3ex]
    \displaystyle
    \frac{\delta \Gamma}{\delta I_{s\lambda}^{ai}(x)}
    \!\! & = & \!\!
    \displaystyle
    -\frac{\delta W}{\delta I_{s\lambda}^{ai}(x)}
    &
    \displaystyle
    \frac{\delta \Gamma}{\delta I_{s\bar{\omega}}^a(x)}
    \!\! & = & \!\!
    \displaystyle
    -\frac{\delta W}{\delta I_{s\bar{\omega}}^a(x)}
    \end{array}
\end{equation}\\[1ex]
which may be used to reexpress the gauge Ward identity (\ref{eq:ST-W})
in terms of $\Gamma$
\begin{equation} \label{eq:ST-Gamma}
  \int\!\!dx\,
  \Bigl[
    \frac{\delta \Gamma}{\delta \!A^{ai}(x)}
    \,\frac{\delta \Gamma}{\delta I_{s\!A}^{ai}(x)}
  + \frac{\delta \Gamma}{\delta \lambda^{ai}(x)}
    \,\frac{\delta \Gamma}{\delta I_{s\!\lambda}^{ai}(x)}
  +     \frac{\delta \Gamma}{\delta \omega^a(x)}
    \,\frac{\delta \Gamma}{\delta I_{s\omega}^a(x)}
  + \frac{\delta \Gamma}{\delta \bar{\omega}^a(x)}
    \,\frac{\delta \Gamma}{\delta I_{s\bar{\omega}}^a(x)}
  \Bigr]
   = 0
\end{equation}

\subsection{Stochastic Ward Identities}

We have included in Eq.~(\ref{eq:Z[J,I]}) the auxiliary field $\Lb$ and
the ghost fields $\omega$ and $\bar{\omega}$, all of which were not
strictly necessary, but rather were included so as to facilitate our work.
They could, in principle, be integrated out and we would be left
with Eq.~(\ref{eq:PathIntegral4}), except that we have now also introduced
sources for the extra fields, as well as for the BRST transformed ones.
This would suggest that there could be some sort of relations for the
generating functional in Eq.~(\ref{eq:Z[J,I]}) resulting from our choice
to include the extra fields and sources.

To derive these relations for B\"odeker's effective theory, one starts
from the generating functional (\ref{eq:Z[J,I]}), including sources
of the fundamental as well as the (gauge) BRST transformed fields.
Inserting the action and the BRST transforms according to
Eqs.~(\ref{eq:S[A,Lambda,Omega,OmegaBar]}) -- (\ref{eq:S(GG)})
and Eq.~(\ref{eq:finiteBRST}) with the definitions
(\ref{eq:E}) and (\ref{def:gamma[omega]})
in use, the generating functional $Z[J,I]$ may be written
\begin{eqnarray}
  Z[J,I] \!\!\! & = & \!\!\!\!\!
  \int\!\mathcal{D}\!\myvec{A}\mathcal{D}\!\gvec{\lambda}
        \mathcal{D}\omega\mathcal{D}\bar{\omega}\;
  \exp\biggl\{ \int\!\!dx
   \Bigl[
     -\sigma T \gvec{\lambda}^{\!a}\!\cdot\gvec{\lambda}^{\!a}
     + i \gvec{\lambda}^{\!a} \!\cdot\!
       \bigl(
         \myvec{E}^{a}[\myvec{A}]
         - i\myvec{J}_{\!\lambda}^a
         + igf^{abc} \omega^b \myvec{I}_{s\!\lambda}^c
         - \sigma \myvec{D}^{ab} I_{s\bar{\omega}}^b
       \bigr)
       \nonumber\\
   & & \hspace{3.95cm}
     + \;\bar{\omega}^a
       \bigl(
         \gamma^a[\omega,\myvec{A}]
         + J_{\bar{\omega}}^a
         - gf^{abc} \omega^b I_{s\bar{\omega}}^c
       \bigr)
     \;+\, \myvec{A}^{\!a} \!\cdot \myvec{J}_{\!A}^a
     +  \omega^a J_{\omega}^a
  \Bigr.\biggr. \nonumber\\[1.0ex]
   & & \hspace{3.95cm}
   \biggl.\Bigl.
     + \;\myvec{I}_{s\!A}^a \!\cdot \myvec{D}^{ab} \omega^b
     +  I_{s\omega}^a {\textstyle\frac{1}{2}} g f^{abc} \omega^c \omega^b
     +
   \Bigr]
   \biggr\}
\end{eqnarray}
where terms multiplying $\gvec{\lambda}$ and $\bar{\omega}$ have been
collected. Because the exponent is quadratic in the former and linear
in the latter, both of these fields can be integrated. One obtains
\begin{eqnarray} \label{eq:Z-forStochWIafterInt}
  Z[J,I] \!\!\! & = & \!\!\!\!\!
  \int\!\mathcal{D}\!\myvec{A}\mathcal{D}\omega\;
  \delta(\gamma^{\prime})
  \exp\biggl\{ \int\!\!dx
   \Bigl[
     - {\textstyle\frac{1}{4\sigma T}}\,
       \myvec{E}^{\prime a} \!\cdot \myvec{E}^{\prime a}
     + \myvec{A}^{\!a} \!\cdot \myvec{J}_{\!A}^a
     + \omega^a J_{\omega}^a
       \nonumber\\
   & & \hspace{4.25cm}
     + \;\myvec{I}_{s\!A}^a \!\cdot \myvec{D}^{ab} \omega^b
     + I_{s\omega}^a {\textstyle\frac{1}{2}} g f^{abc} \omega^c \omega^b
   \Bigr]
   \biggr\}
\end{eqnarray}
with the new functionals $\myvec{E}^{\prime}$ and $\gamma^{\prime}$
defined as
\begin{eqnarray} \label{def:E_prime}
  \myvec{E}^{\prime a}[\omega,\myvec{A};\,
                       \myvec{J}_{\!\lambda},
                       \myvec{I}_{s\!\lambda},
                       I_{s\bar{\omega}}]
  &=&
  \myvec{E}^{a}[\myvec{A}]
  - i\myvec{J}_{\!\lambda}^a
  + igf^{abc} \omega^b \myvec{I}_{s\!\lambda}^c
  - \sigma \myvec{D}^{ab} I_{s\bar{\omega}}^b
  \\[0.5ex] \label{def:gamma_prime}
  \gamma^{\prime a}[\omega,\myvec{A};\,
                    J_{\bar{\omega}}, I_{s\bar{\omega}}]
  &=&
  \gamma^a[\omega,\myvec{A}]
  + J_{\bar{\omega}}^a
  - gf^{abc} \omega^b I_{s\bar{\omega}}^c
\end{eqnarray}
Hence, when restricting to vanishing sources
\begin{math}
  \myvec{J}_{\!A} \!= \myvec{I}_{s\!A} \!= 0
\end{math}
and
\begin{math}
  J_{\omega} \!= I_{s\omega} \!= 0
\end{math}
the exponent 
becomes purely quadratic in $\myvec{E}^{\prime}$. Defining for brevity
\begin{equation}
  Z_1[\myvec{J}_{\!\lambda}, J_{\bar{\omega}},
      \myvec{I}_{s\!\lambda},I_{s\bar{\omega}}] =
    Z[\myvec{J}_{\!A}\!=0,\,
      \myvec{J}_{\!\lambda},\,
      J_{\omega}\!=0,\,
      J_{\bar{\omega}},\,
      \myvec{I}_{s\!A}\!=0,\,
      \myvec{I}_{s\!\lambda},\,
      I_{s\omega}\!= 0,\, I_{s\bar{\omega}}]
\end{equation}
we have
\begin{equation} \label{eq:Z1_int:A,omega}
  Z_1[\myvec{J}_{\!\lambda}, J_{\bar{\omega}},
      \myvec{I}_{s\!\lambda},I_{s\bar{\omega}}] =
  \int\!\mathcal{D}\!\myvec{A}\mathcal{D}\omega\;
  \delta(\gamma^{\prime})
  \exp\biggl\{
   - {\frac{1}{4\sigma T}}
   \int\!\!dx\;\myvec{E}^{\prime a} \!\cdot \myvec{E}^{\prime a}
   \biggr\}
\end{equation}
where $\myvec{E}^{\prime}$ and $\gamma^{\prime}$ both depend
on $\myvec{A}$ and $\omega$ as indicated in Eqs.~(\ref{def:E_prime})
and (\ref{def:gamma_prime}). Thus, it is quite natural to attempt
a change of variables from $\myvec{A}$ and $\omega$ to $\myvec{E}^{\prime}$
and $\gamma^{\prime}$. The Jacobian can be calculated in a similar manner
as the Jacobian of Eq.~(\ref{eq:UnityByOmegas}), and again can be shown
to be a constant. The resulting integral is gaussian and evaluates to a
constant functional $Z_1$ leading to
\begin{equation}
  Z_1[\myvec{J}_{\!\lambda}, J_{\bar{\omega}},
      \myvec{I}_{s\!\lambda},I_{s\bar{\omega}}] = \mathrm{const.}
\end{equation}
or likewise for $W_1 = \ln Z_1$
\begin{equation} \label{eq:W1=const}
  W_1[\myvec{J}_{\!\lambda}, J_{\bar{\omega}},
      \myvec{I}_{s\!\lambda},I_{s\bar{\omega}}] = \mathrm{const.}
\end{equation}
As a consequence, any combination of functional derivatives with
respect to sources chosen from the class
\begin{math}
  \{\myvec{J}_{\!\lambda}, J_{\bar{\omega}},
  \myvec{I}_{s\!\lambda},I_{s\bar{\omega}}\}
\end{math}
yields zero when acting on the full generating functionals
and evaluated for vanishing sources:
\begin{equation} \label{eq:stochasticWI}
  \underbrace{
  \frac{\delta}{\delta\dots}\,
  \frac{\delta}{\delta\dots}\,
  \cdots
  \frac{\delta}{\delta\dots}
  }_{\parbox{2.7cm}{\footnotesize any combination\\
                  of
            \begin{math} J_{\lambda}       \end{math},
                    \begin{math} J_{\bar{\omega}}  \end{math},
                    \begin{math} I_{s\!\lambda}    \end{math},
                    \begin{math} I_{s\bar{\omega}} \end{math}
                   }}
  W[J,I]\Bigr|_{J=I=0} = 0
\end{equation}
with the same relation holding for derivatives of $Z[J,I]$.
To obtain a corresponding identity for the 1PI generating functional
$\Gamma$, note that due to Eq.~(\ref{eq:W1=const}) one has on the
submanifold defined by the vanishing of the four sources
\begin{math}
  \myvec{J}_{\!A}, \myvec{I}_{s\!A},
  J_{\omega}
\end{math} and $I_{s\omega}$
\begin{equation}
  \lambda^{ai}(x)
  \Bigr|_{\scriptstyle\myvec{J}_{\!A} = \myvec{I}_{s\!A} = 0 \atop
          \scriptstyle J_{\omega} = I_{s\omega} =0}
 =
 \frac{\delta W_1}{\delta J_{\lambda}^{ai}(x)}
 = 0
 \qquad\mathrm{and}\qquad
  \bar{\omega}^{a}(x)
  \Bigr|_{\scriptstyle\myvec{J}_{\!A} = \myvec{I}_{s\!A} = 0 \atop
          \scriptstyle J_{\omega} = I_{s\omega} =0}
 =
 -\frac{\delta W_1}{\delta J_{\bar{\omega}}^{a}(x)}
 = 0
\end{equation}
So $\Gamma$ could at most depend on $\Ab$, $\omega$,
$\myvec{I}_{s\!\lambda}$ and $I_{s\bar{\omega}}$. However,
from Eq.~(\ref{eq:DerivativesOfGamma}) we have
\begin{equation}
  \frac{\delta \Gamma}{\delta A^{ai}(x)} = J^{ai}_A(x)
  \qquad\qquad
  \frac{\delta \Gamma}{\delta \omega^a(x)} = J^a_{\omega}(x)
\end{equation}
and therefore $\Gamma$ may not depend on $\myvec{A}$ or $\omega$ anymore.
The same conclusion can be reached for the $\myvec{I}_{s\!\lambda}$ and
$I_{s\bar{\omega}}$ by looking at Eq.~(\ref{eq:MoreDerivativesOfGamma})
and Eq.~(\ref{eq:stochasticWI}). Therefore, $\Gamma$ must be a constant,
this leads to
\begin{equation} \label{eq:GammastochasticWI}
  \underbrace{
  \frac{\delta}{\delta\dots}\,
  \frac{\delta}{\delta\dots}\,
  \cdots
  \frac{\delta}{\delta\dots}
  }_{\parbox{2.7cm}{\footnotesize any combination\\
                    of \begin{math} A              \end{math},
                    \begin{math} \omega            \end{math},
                    \begin{math} I_{s\!\lambda}    \end{math},
                    \begin{math} I_{s\bar{\omega}} \end{math}
                   }}
  \Gamma[\myvec{A},\gvec{\lambda},\omega,\bar{\omega};I]\Bigr|_{J=I=0} = 0
\end{equation}
which is the equivalent of the stochastic Ward identity
(\ref{eq:stochasticWI}) in terms of the 1PI generating functional $\Gamma$.

\subsection{Ghost Number Conservation}

We will discuss one last symmetry of the
action (\ref{eq:S[A,Lambda,Omega,OmegaBar]}). The action is invariant
under the global transformation
\begin{equation} \label{eq:GhostNumberTF}
  \begin{array}{rcr}
    \omega^a(x)       &=& e^{ i\alpha} \omega^{\prime a}(x) \\[1.0ex]
    \bar{\omega}^a(x) &=& e^{-i\alpha} \bar{\omega}^{\prime a}(x)
  \end{array}
\end{equation}
of the ghost and anti-ghost fields. In addition to this, subjecting the
measure $\mathcal{D}\omega\mathcal{D}\bar{\omega}$ to the transformation
(\ref{eq:GhostNumberTF}), i.e.~to
\begin{eqnarray}
\lefteqn{
  (\omega^a(x_1),\,\bar{\omega}^a(x_1),\,
   \omega^a(x_2),\,\bar{\omega}^a(x_2),\,\dots)} \\[1ex]
  &=&
  (e^{ i\alpha}\omega^{\prime a}(x_1),\,
   e^{-i\alpha}\bar{\omega}^{\prime a}(x_1),\,
   e^{ i\alpha}\omega^{\prime a}(x_2),\,
   e^{-i\alpha}\bar{\omega}^{\prime a}(x_2),\,\dots)
   \nonumber
\end{eqnarray}
one finds
\begin{equation}
  \mathcal{D}\omega\mathcal{D}\bar{\omega} =
  \prod_{a,n} [d\omega^a(x_n)\,d\bar{\omega}^a(x_n)] =
  \prod_{a,n} [d\omega^{\prime a}(x_n)\,d\bar{\omega}^{\prime a}(x_n)]
  \;J(\omega^{\prime},\bar{\omega}^{\prime})
\end{equation}
with the Jacobian
\begin{equation}
  J^{-1}(\omega^{\prime},\bar{\omega}^{\prime}) = \det
  \left(
    \begin{array}{ccccc}
      e^{+i\alpha} & 0 & 0 & 0 & \cdots \\
      0 & e^{-i\alpha} & 0 & 0 & \cdots \\
      0 & 0 & e^{+i\alpha} & 0 & \cdots \\
      0 & 0 & 0 & e^{-i\alpha} & \cdots \\
      \vdots & \vdots & \vdots & \vdots & \ddots
    \end{array}
  \right)
  = 1
\end{equation}
Hence, the measure is also invariant under the transformation
(\ref{eq:GhostNumberTF})
\begin{equation}
  \mathcal{D}\omega\mathcal{D}\bar{\omega} =
  \mathcal{D}\omega^{\prime}\mathcal{D}\bar{\omega}^{\prime}
\end{equation}
Together with the invariance of the action, this symmetry leads to
ghost number conservation, which poses another restriction on the form of
the generating functionals and their derivatives.
Indeed, taking the parameter $\alpha$ in Eq.~(\ref{eq:GhostNumberTF}) to
be infinitesimal and performing the corresponding change of variables
\begin{equation} \label{eq:GhostNumberTFinf}
  \begin{array}{rcr}
    \omega^a(x)       &=& \omega^{\prime a}(x)
                          + i\alpha\,\omega^{\prime a}(x) \\[1.0ex]
    \bar{\omega}^a(x) &=& \bar{\omega}^{\prime a}(x)
                          -i\alpha\,\bar{\omega}^{\prime a}(x)
  \end{array}
\end{equation}
in the defining path integral (\ref{eq:Z[J,I]}) of the generating functional
$Z[J,I]$ yields
\begin{displaymath}
  Z[J,I] = \!\!
  \int\!\mathcal{D}\!\myvec{A}\mathcal{D}\!\gvec{\lambda}
        \mathcal{D}\omega\mathcal{D}\bar{\omega}\;
  \exp\biggl\{ i\alpha\!
    \int\!\!dx\,
    \Bigl[
       \omega^a J_{\omega}^a
    -  \bar{\omega}^a J_{\bar{\omega}}^a
    +  \myvec{I}_{s\!A}^a \!\cdot s\myvec{A}^{\!a}
    +  \myvec{I}_{s\!\lambda}^a \!\cdot s\gvec{\lambda}^{\!a}
    +  2 I_{s\omega}^a s\omega^a
    \Bigr]
   \biggr\}
   \exp\biggl\{ (\dots)
   \biggr\}
\end{displaymath}
Here we have already renamed the primed symbols again to unprimed ones
after the change of variables has been completed. As before, the dots
represent the original exponent as it occurs in Eq.~(\ref{eq:Z[J,I]}).
Using the fact that $\alpha$ is assumed to be infinitesimal, we can expand
the first exponential and replace any fields that appear by functional
derivatives acting on the exponential (after interchanging the order of
the ghost and anti-ghost field and their corresponding sources leading
to a minus sign in either case). The derivatives can finally be pulled
out of the functional integral and we obtain
\begin{equation} \label{eq:GhostNumber-Z}
  \int\!\!dx\,
  \Bigl[
    J_{\omega}^a(x) \,\frac{\delta}{\delta J_{\omega}^{a}(x)}
  - J_{\bar{\omega}}^a(x) \,\frac{\delta}{\delta J_{\bar{\omega}}^{a}(x)}
  + I_{s\!A}^{ai}(x) \,\frac{\delta}{\delta I_{s\!A}^{ai}(x)}
  + I_{s\!\lambda}^{ai}(x) \,\frac{\delta}{\delta I_{s\!\lambda}^{ai}(x)}
  + 2I_{s\omega}^a(x) \,\frac{\delta}{\delta I_{s\omega}^a(x)}
  \Bigr]
  Z[J,I] = 0
\end{equation}
Again, the definition $W[J,I] = \ln Z[J,I]$ implies that the same
identity holds for the generating functional $W[J,I]$ of connected
correlation functions
\begin{equation} \label{eq:GhostNumber-W}
  \int\!\!dx\,
  \Bigl[
    J_{\omega}^a \,
    \frac{\delta W}{\delta J_{\omega}^{a}}
  - J_{\bar{\omega}}^a \,
    \frac{\delta W}{\delta J_{\bar{\omega}}^{a}}
  + I_{s\!A}^{ai} \,
    \frac{\delta W}{\delta I_{s\!A}^{ai}}
  + I_{s\!\lambda}^{ai} \,
    \frac{\delta W}{\delta I_{s\!\lambda}^{ai}}
  + 2I_{s\omega}^a
    \frac{\delta W}{\delta I_{s\omega}^a}
  \Bigr] = 0
\end{equation}
where we have suppressed the space-time argument $x$ and the dependence
of $W$ on the sources $J$ and $I$.
This identity in turn can easily be translated to the corresponding
restriction on the 1PI generating functional
$\Gamma[\myvec{A},\gvec{\lambda},\omega,\bar{\omega};I]$. By means
of Eqs.~(\ref{def:classicalFields}), (\ref{eq:DerivativesOfGamma})
and (\ref{eq:MoreDerivativesOfGamma}) one finds
\begin{equation} \label{eq:GhostNumber-Gamma}
  \int\!\!dx\,
  \Bigl[\,
    \frac{\delta\Gamma}{\delta\omega^{a}} \,
    \omega^a
  - \frac{\delta\Gamma}{\delta\bar{\omega}^{a}} \,
    \bar{\omega}^a
  + I_{s\!A}^{ai} \,
    \frac{\delta\Gamma}{\delta I_{s\!A}^{ai}}
  + I_{s\!\lambda}^{ai} \,
    \frac{\delta\Gamma}{\delta I_{s\!\lambda}^{ai}}
  + 2I_{s\omega}^a
    \frac{\delta\Gamma}{\delta I_{s\omega}^a}\,
  \Bigr] = 0
\end{equation}
This concludes our derivation of non-perturbative identities for the
generating functional (\ref{eq:Z[J,I]}). Explicit forms of these
identities for lower N-point functions are shown in
Appendix~\ref{NPointFunctions}.


\section{Dyson--Schwinger Equations}
\label{DSE}

To derive the \textsc{Dyson--Schwinger} equations, we observe that the
path integral of a functional derivative vanishes, i.e.
\begin{equation}
  \int\!\mathcal{D}\phi\;
  \frac{\delta}{\delta\phi(x)}\,F[\phi] = 0
\end{equation}
for any functional $F[\phi]$. Hence, in the case of B\"odeker's theory,
we obtain four different equations by inserting a functional derivative
with respect to each of the fields $\myvec{A}$, $\gvec{\lambda}$, $\omega$
or $\bar{\omega}$ into the generating functional
\begin{eqnarray} \label{eq:Z-for-DSE}
  Z[J,I] & = & 
  \int\!\mathcal{D}\!\myvec{A}\mathcal{D}\!\gvec{\lambda}
        \mathcal{D}\omega\mathcal{D}\bar{\omega}\;
  \exp\biggl\{-S[\myvec{A},\gvec{\lambda},\omega,\bar{\omega}]
             + \int\!\!dx\,
   \Bigl[
       \myvec{A}^{\!a} \!\cdot \myvec{J}_{\!A}^a
    \;+\;  \gvec{\lambda}^{\!a} \!\cdot \myvec{J}_{\!\lambda}^a
    \;+\;  \omega^a J_{\omega}^a
    \;+\;  \bar{\omega}^a J_{\bar{\omega}}^a
   \Bigr.\biggr. \nonumber\\
   & & \hspace{4.0cm}
   \biggl.\Bigl.
    + \ \myvec{I}_{s\!A}^a \!\cdot s\myvec{A}^{\!a} 
    +   \myvec{I}_{s\!\lambda}^a \!\cdot s\gvec{\lambda}^{\!a}
    +   I_{s\omega}^a s\omega^a 
    +   I_{s\bar{\omega}}^a s\bar{\omega}^a 
   \Bigr]
   \biggr\}
\end{eqnarray}

\subsection{General Dyson--Schwinger Equations}
\subsubsection{Ghost ($\omega$) and Anti-ghost ($\bar{\omega}$) equations}

\noindent
Starting from the identity
\begin{equation} \label{eq:DSE-ghost-start}
  0 =
  \int\!\mathcal{D}\!\myvec{A}\mathcal{D}\!\gvec{\lambda}
        \mathcal{D}\omega\mathcal{D}\bar{\omega}\;
  \frac{\delta}{\delta\omega^a(x)}\,
  \exp\bigl\{(\dots)\bigr\}
\end{equation}
where the dots represent the exponent of
Eq.~(\ref{eq:Z-for-DSE}), gives
\begin{eqnarray} \label{eq:DSE-ghost:withDerivative}
  0 \!\!\! &=& \!\!\!\!
  \int\!\mathcal{D}\!\myvec{A}\mathcal{D}\!\gvec{\lambda}
        \mathcal{D}\omega\mathcal{D}\bar{\omega}\;
  \Bigl[
    \dot{\bar{\omega}}^a(x)
    +{\textstyle\frac{1}{\kappa}}\, 
     \myvec{D}^{ab}(x) \cdot\nabla\bar{\omega}^b(x) 
    -{\textstyle\frac{g}{\kappa}}\, 
    f^{abc} \,\bar{\omega}^b(x) \,\nabla\!\cdot\!\myvec{A}^{\!c}(x)
    + J_{\omega}^a(x)
    + \nabla\!\cdot\myvec{I}^a_{s\!A}(x)
\nonumber\\
  & & \!\!\!\!\!\!\!
    + \, g f^{abc} 
    \bigl(
      -\myvec{I}^b_{s\!A}(x) \!\cdot\! \myvec{A}^{\!c}(x)
      -\myvec{I}^b_{s\lambda}(x) \!\cdot\! \gvec{\lambda}^{\!c}(x)
      +I_{s\omega}^b(x) \,{\omega}^c(x) 
      +I_{s\bar{\omega}}^b(x) \,\bar{\omega}^c(x) 
    \bigr)
  \Bigr]
  \exp\bigl\{(\dots)\bigr\}
\end{eqnarray}
Expressing the fields by derivatives acting on the exponential, one obtains
\begin{eqnarray}
  \hspace{-7ex}
  Z J_{\omega}^a(x) \!\!\! &=& \!\!\!
  \left(
    \partial_t + {\textstyle\frac{1}{\kappa}} \Delta
  \right)
  \frac{\delta Z}{\delta J_{\bar{\omega}}^a(x)}
  \,-\,
  \frac{g}{\kappa} f^{abc} \,\partial_j
  \frac{\delta^2 Z}{\delta J_{\bar{\omega}}^b(x)\,\delta J_A^{cj}(x)}
  - Z \,\nabla\!\cdot\myvec{I}^a_{s\!A}(x)
  \nonumber\\
  & & \!\!\!\!\!\!\!\!\!\!\!\!\!\!\!\!\!\!\!\!\! +\;
  g f^{abc} 
  \biggl[
    I_{s\!A}^{bj}(x) \,
    \frac{\delta Z}{\delta J_A^{cj}(x)}
    +
    I_{s\lambda}^{bj}(x) \,
    \frac{\delta Z}{\delta J_{\lambda}^{cj}(x)}
    +
    I_{s\omega}^b(x) \,
    \frac{\delta Z}{\delta J_{\omega}^c(x)}
    +
    I_{s\bar{\omega}}^b(x) \,
    \frac{\delta Z}{\delta J_{\bar{\omega}}^c(x)}
  \biggr]
\end{eqnarray}
or in terms of $W$
\begin{eqnarray}
  J_{\omega}^a(x) \!\!\! &=& \!\!\!
  \left(
    \partial_t + {\textstyle\frac{1}{\kappa}} \Delta
  \right)
  \frac{\delta W}{\delta J_{\bar{\omega}}^a(x)}
  \,-\,
  \frac{g}{\kappa} f^{abc} \,\partial_j
  \biggl[
    \frac{\delta^2 W}{\delta J_{\bar{\omega}}^b(x)\,\delta J_A^{cj}(x)}
    +
    \frac{\delta W}{\delta J_{\bar{\omega}}^b(x)} \,
    \frac{\delta W}{\delta J_A^{cj}(x)}
  \biggr]
  - \nabla\!\cdot\myvec{I}^a_{s\!A}(x)
  \nonumber \\
  & & \!\!\!\!\!\!\!\!\!\!\!\!\!\! +\;
  g f^{abc} 
  \biggl[
    I_{s\!A}^{bj}(x) \,
    \frac{\delta W}{\delta J_A^{cj}(x)}
    +
    I_{s\lambda}^{bj}(x) \,
    \frac{\delta W}{\delta J_{\lambda}^{cj}(x)}
    +
    I_{s\omega}^b(x) \,
    \frac{\delta W}{\delta J_{\omega}^c(x)}
    +
    I_{s\bar{\omega}}^b(x) \,
    \frac{\delta W}{\delta J_{\bar{\omega}}^c(x)}
  \biggr]
\end{eqnarray}
Transcription to the 1PI generating functional $\Gamma$ yields
\begin{eqnarray}
  \frac{\delta\Gamma}{\delta \omega^a(x)}
  \!\!\! &=& \!\!\!
  -\left(
    \partial_t + {\textstyle\frac{1}{\kappa}} \Delta
  \right)
  \bar{\omega}^a(x)
  \,-\,
  \frac{g}{\kappa} f^{abc} \,\partial_j
  \biggl[
    \frac{\delta^2 W}{\delta J_{\bar{\omega}}^b(x)\,\delta J_A^{cj}(x)}
    -
    \bar{\omega}^b(x) \,
    A^{cj}(x)
  \biggr]
  - \nabla\!\cdot\myvec{I}^a_{s\!A}(x)
  \nonumber \\ \label{eq:DSE-ghost:Gamma}
  & & \!\! +\;
  g f^{abc} 
  \biggl[
    I_{s\!A}^{bj}(x) \,
    A^{cj}(x)
    +
    I_{s\lambda}^{bj}(x) \,
    {\lambda}^{cj}(x)
    -
    I_{s\omega}^b(x) \,
    {\omega}^c(x)
    -
    I_{s\bar{\omega}}^b(x) \,
    \bar{\omega}^c(x)
  \biggr]
\end{eqnarray}

\noindent
The antighost equation is obtained in a similar manner and reads
\begin{equation} \label{eq:DSE-anti-ghost:Gamma}
  \frac{\delta\Gamma}{\delta \bar{\omega}^a(x)}
  = 
  \left(
    -\partial_t + {\textstyle\frac{1}{\kappa}} \Delta
  \right)
  \omega^a(x)
  \,+\,
  \frac{g}{\kappa} f^{abc} \,\partial_j
  \biggl[
      \frac{\delta^2 W}{\delta J_{\omega}^b(x)\,\delta J_A^{cj}(x^{\prime})}
      -
      \omega^b(x)
      A^{cj}(x^{\prime})
  \biggr]_{x^{\prime}=x}
  \!\!\!\!\!\! -
  g f^{abc} I_{s\bar{\omega}}^b(x) \, \omega^c(x)
\end{equation}
where $x^{\prime}$ is set to $x$ after the space-time derivative is
carried out, i.e.~the derivative acts on the argument of $J_{\omega}^b(x)$
only.

\subsubsection{Auxiliary field ($\gvec{\lambda}$) equation}

\noindent
To deduce the auxiliary field equation from
\begin{equation} \label{eq:DSE-lambda-start}
  0 =
  \int\!\mathcal{D}\!\myvec{A}\mathcal{D}\!\gvec{\lambda}
        \mathcal{D}\omega\mathcal{D}\bar{\omega}\;
  \frac{\delta}{\delta\lambda^{ai}(x)}\,
  \exp\bigl\{(\dots)\bigr\}
\end{equation}
we need, among other things, the functional derivative of the action
\begin{math}
  S[\myvec{A},\gvec{\lambda},\omega,\bar{\omega}] =
  S^{\scriptscriptstyle\mathrm{(D)}}[\myvec{A},\gvec{\lambda}] +
  S^{\scriptscriptstyle\mathrm{(GG)}}[\myvec{A},\omega,\bar{\omega}]
\end{math}.
However, in the present case the corresponding expression becomes rather
cumbersome. 

As for the \textsc{Feynman} rules in Appendix \ref{FeynmanRules}, we want
to use a 
symmetrised $\gvec{\lambda}\myvec{A}^2$ and $\gvec{\lambda}\myvec{A}^3$ 
vertex. The $\gvec{\lambda}$ dependence of the action spreads out 
over the three contributions to the dynamical action
\begin{math}
  S^{\scriptscriptstyle\mathrm{(D)}}[\myvec{A},\gvec{\lambda}] =  
  S^{\scriptscriptstyle\mathrm{(D)}}_{0}
  [\myvec{A},\gvec{\lambda}] +
  S^{\scriptscriptstyle\mathrm{(D)}}_{\,\mathrm{int},3}
  [\myvec{A},\gvec{\lambda}] +
  S^{\scriptscriptstyle\mathrm{(D)}}_{\,\mathrm{int},4}
  [\myvec{A},\gvec{\lambda}]
\end{math}.
The corresponding derivatives can be written in the form
\begin{eqnarray}
  \frac{\delta S^{\scriptscriptstyle\mathrm{(D)}}_{0}
                  [\myvec{A},\gvec{\lambda}]}
       {\delta\lambda^{ai}(x)}
  &=&
  2\sigma T \lambda^{ai}(x)
  -i\left[
    \delta^{ij}\!\left(\sigma\partial_t - \Delta \right)
    +
    \left(1-{\textstyle\frac{\sigma}{\kappa}}\right)
    \partial_i \partial_j
  \right] \!
  A^{aj}(x) \\ 
  \frac{\delta S^{\scriptscriptstyle\mathrm{(D)}}_{\,\mathrm{int},3}
                  [\myvec{A},\gvec{\lambda}]}
       {\delta\lambda^{ai}(x)}
  &=&  
  \frac{1}{2!}\,(-ig) f^{abc}
  \Bigl[
    \left(1-{\textstyle\frac{\sigma}{\kappa}}\right) \!
    \bigl[
      \delta^{ij} \partial_k^{\,\prime} \!
     -\delta^{ik} \partial_j
    \bigr]
    + \,2\,
    \bigl[
      \delta^{ij} \partial_k \!
     -\delta^{ik} \partial_j^{\,\prime}
    \bigr] \nonumber\\[1ex]
  & &  \hspace{14.6ex}
    + \,
    \bigl[
      \delta^{jk} \partial_i^{\,\prime} \!
     -\delta^{kj} \partial_i
    \bigr]\,
  \Bigr]
  A^{bj}(x) A^{ck}(x^{\prime})
  \Bigr|_{x^{\prime}=x} \\ 
  \frac{\delta S^{\scriptscriptstyle\mathrm{(D)}}_{\,\mathrm{int},4}
                  [\myvec{A},\gvec{\lambda}]}
       {\delta\lambda^{ai}(x)}
  &=&  
  \frac{1}{3!}\,(-ig^2) \, V^{abcd}_{\hspace{0.6ex}ijkl} \,
  A^{bj}(x) A^{ck}(x) A^{dl}(x)
\end{eqnarray}
where $V^{abcd}_{\hspace{0.6ex}ijkl}$ is defined in Eq.~(\ref{def:V4}). One then obtains in terms of the 1PI generating functional
\begin{eqnarray}
  \frac{\delta\Gamma}{\delta\lambda^{ai}(x)}
  &=&
  2\sigma T \lambda^{ai}(x)
  -i\left[
    \delta^{ij}\!\left(\sigma\partial_t - \Delta \right)
    +
    \left(1-{\textstyle\frac{\sigma}{\kappa}}\right)
    \partial_i \partial_j
  \right] A^{aj}(x)
\nonumber\\
  &-&\!\!
  \frac{ig}{2!}\, f^{abc}
  \Bigl[
    \left(1-{\textstyle\frac{\sigma}{\kappa}}\right) \!
    \bigl[
      \delta^{ij} \partial_k^{\,\prime} \!
     -\delta^{ik} \partial_j
    \bigr]
    + \,2\,
    \bigl[
      \delta^{ij} \partial_k \!
     -\delta^{ik} \partial_j^{\,\prime}
    \bigr]
  + 
  \bigl[
    \delta^{jk} \partial_i^{\,\prime} \!
    -\delta^{kj} \partial_i
  \bigr]\,\Bigr]
   \nonumber\\
& & \hspace{36ex}
  \times\biggl[\,
    \frac{\delta^2 W}{\delta J_A^{bj}(x)\,\delta J_A^{ck}(x^{\prime})}
    +
    A^{bj}(x)\,
    A^{ck}(x^{\prime})
  \,\biggr]_{x^{\prime}=x}
  \nonumber\\
  &-& \!\!
  \frac{ig^2}{3!} V^{abcd}_{\hspace{0.6ex}ijkl} 
  \biggl[\,
  \frac{\delta^3 W}{\delta J_A^{bj}(x)\,\delta J_A^{ck}(x)\,
                    \delta J_A^{dl}(x)}
  + 3\,
  \frac{\delta^2 W}{\delta J_A^{bj}(x)\,\delta J_A^{ck}(x)}\,
  A^{dl}(x)
  +
  A^{bj}(x) \,
  A^{ck}(x) \,
  A^{dl}(x)
  \,\biggr]
  \nonumber\\ \label{eq:DSE-lambda:Gamma}
  &-&\!\!
    g f^{abc} 
    \biggl[
      -I_{s\lambda}^{bi}(x) \,
      {\omega}^c(x)
      +
      i\sigma
      I_{s\bar{\omega}}^b(x) \,
      A^{ci}(x)
    \biggr] 
    + i\sigma \,\partial_i I_{s\bar{\omega}}^a(x)
\end{eqnarray}

\subsubsection{Gauge field ($\myvec{A}$) equation}

\noindent
Finally, coming to the gauge field equation
\begin{equation} \label{eq:DSE-gauge-start}
  0 =
  \int\!\mathcal{D}\!\myvec{A}\mathcal{D}\!\gvec{\lambda}
        \mathcal{D}\omega\mathcal{D}\bar{\omega}\;
  \frac{\delta}{\delta A^{ai}(x)}\,
  \exp\bigl\{(\dots)\bigr\}
\end{equation}
and using the derivatives
\begin{eqnarray}
  \frac{\delta S^{\scriptscriptstyle\mathrm{(D)}}_{0}
                  [\myvec{A},\gvec{\lambda}]}
       {\delta A^{ai}(x)}
  &=&
  -i\left[
    \delta^{ij}\!\left(-\sigma\partial_t - \Delta \right)
    +
    \left(1-{\textstyle\frac{\sigma}{\kappa}}\right)
    \partial_i \partial_j
  \right] \!
  \lambda^{aj}(x) \\ 
  \frac{\delta S^{\scriptscriptstyle\mathrm{(D)}}_{\,\mathrm{int},3}
                  [\myvec{A},\gvec{\lambda}]}
       {\delta A^{ai}(x)}
  &=&  
  -ig f^{abc}
  \Bigl[
    -\!\left(1-{\textstyle\frac{\sigma}{\kappa}}\right) \!
    \bigl[
      \delta^{ij} \partial_k^{\,\prime} \!
     +\delta^{jk} (\partial_i +  \partial_i^{\,\prime})
    \bigr]
    + \,2\,
    \bigl[
      \delta^{jk} \partial_i^{\,\prime} \!
     +\delta^{ij} (\partial_k + \partial_k^{\,\prime})
    \bigr] \nonumber\\[1ex]
  & &  \hspace{10.6ex}
    - \,
    \bigl[
      \delta^{ik} \partial_j^{\,\prime} \!
     +\delta^{ik} (\partial_j + \partial_j^{\,\prime})
    \bigr]\,
  \Bigr]
  \lambda^{bj}(x) A^{ck}(x^{\prime})
  \Bigr|_{x^{\prime}=x} \\ 
  \frac{\delta S^{\scriptscriptstyle\mathrm{(D)}}_{\,\mathrm{int},4}
                  [\myvec{A},\gvec{\lambda}]}
       {\delta A^{ai}(x)}
  &=&  
  \frac{1}{2!}\,(-ig^2) \, V^{dabc}_{\hspace{0.6ex}lijk} \,
  \lambda^{dl}(x) A^{bj}(x) A^{ck}(x)
\end{eqnarray}
where the symmetry of $V^{abcd}_{\hspace{0.6ex}ijkl}$ has been
exploited, together with
\begin{equation}
  \frac{\delta S^{\scriptscriptstyle\mathrm{(GG)}}
                  [\myvec{A},\omega,\bar{\omega}]}
       {\delta A^{ai}(x)}
  =
  -\frac{g}{\kappa}\,f^{abc} \bar{\omega}^b(x)\,\partial_i\,\omega^c(x)
\end{equation}
one arrives at
\begin{eqnarray}
\lefteqn{
  \frac{\delta \Gamma}{\delta A^{ai}(x)} =
  -i\left[
    \delta^{ij}\!\left(-\sigma\partial_t - \Delta \right)
    +
    \left(1-{\textstyle\frac{\sigma}{\kappa}}\right)
    \partial_i \partial_j
  \right] \lambda^{aj}(x)
  \;-\;
  \frac{ig^2}{2!}\,V^{dabc}_{\hspace{0.6ex}lijk}
  \biggl[
  \frac{\delta^3 W}{\delta J_{\lambda}^{dl}(x)\,\delta J_A^{bj}(x)\,
                    \delta J_A^{ck}(x)}
} \nonumber\\
  & &
  + 2 \,
  \frac{\delta^2 W}{\delta J_{\lambda}^{dl}(x)\,\delta J_A^{bj}(x)}\,
  A^{ck}(x)
  \,+\,
  \lambda^{dl}(x)\,
  \frac{\delta^2 W}{\delta J_A^{bj}(x)\,\delta J_A^{ck}(x)}
  \,+\,
  \lambda^{dl}(x)\,
  A^{bj}(x)\,
  A^{ck}(x)  
  \biggr]
  \nonumber\\
  & &
  -ig f^{abc}
  \Bigl[
    -\!\left(1-{\textstyle\frac{\sigma}{\kappa}}\right) \!
    \bigl[
      \delta^{ij} \partial_k^{\,\prime} \!
     +\delta^{jk} (\partial_i +  \partial_i^{\,\prime})
    \bigr]
    + \,2\,
    \bigl[
      \delta^{jk} \partial_i^{\,\prime} \!
     +\delta^{ij} (\partial_k + \partial_k^{\,\prime})
    \bigr]
  \nonumber\\
  & &  \hspace{23.5ex}
    - \,
    \bigl[
      \delta^{ik} \partial_j^{\,\prime} \!
     +\delta^{ik} (\partial_j + \partial_j^{\,\prime})
    \bigr]\,
  \Bigr]
  \biggl[
  \frac{\delta^2 W}{\delta J_\lambda^{bj}(x)\,\delta J_A^{ck}(x^{\prime})}
  \,+\,
  \lambda^{bj}(x)\,
  A^{ck}(x^{\prime})
  \bigg]_{x^{\prime}=x}
  \nonumber\\
  & &
  - \frac{g}{\kappa} f^{abc} \,\partial_i
  \biggl[
    \frac{\delta^2 W}
         {\delta J_{\bar{\omega}}^b(x^{\prime})\,
          \delta J_{\omega}^c(x)}
    \,+\,
    \bar{\omega}^b(x^{\prime})\,
    \omega^c(x)
  \biggr]_{x^{\prime}=x}
  \!\!\!\!
  + g f^{abc} 
    \biggl[
      I_{s\!A}^{bi}(x) \,
      \omega^c(x)
      +
      i\sigma
      I_{s\bar{\omega}}^b(x) \,
      \lambda^{ci}(x)
    \biggr]
  \nonumber\\ \label{eq:DSE-A:Gamma}
\end{eqnarray}

\subsection{Explicit Equations for Lower N-Point Functions}
\begin{fmffile}{fmDSE}
%
%
\unitlength=1mm


\newsavebox{\DSEO}
\savebox{\DSEO}(60,32){
\begin{fmfgraph*}(60,25)
  \fmfleft{x}
  \fmfright{y}
  \fmftop{t}
  \fmfbottom{b}
  \fmf{dots_arrow,tension=3}{y,vy}
  \fmf{dots_arrow,tension=3}{vx,x}
  \fmf{phantom}{vy,vx}
  \fmffreeze
  \fmf{dots_arrow,right=0.30}{vy,t}
  \fmf{dots_arrow,right=0.30}{t,vx}
  \fmf{gluon,right=0.30}{vx,b}
  \fmf{plain,right=0.30}{b,vy}
  \fmfv{decor.shape=circle,decor.filled=30,decor.size=0.113w}{t}
  \fmfv{decor.shape=circle,decor.filled=30,decor.size=0.113w}{b}
  \fmfv{decor.shape=circle,decor.filled=empty,decor.size=0.167w}{vy}
  \fmfdot{vx}
  \put(47.0,11.3){\large $\Gamma$}
\end{fmfgraph*}
}

\newsavebox{\DSEOG}
\savebox{\DSEOG}(60,32){
\begin{fmfgraph*}(60,25)
  \fmfleft{x}
  \fmfright{y}
  \fmftop{t}
  \fmfbottom{b}
  \fmf{dots_arrow,tension=3}{y,vy}
  \fmf{dots_arrow,tension=3}{vx,x}
  \fmf{phantom}{vy,vx}
  \fmffreeze
  \fmf{dots_arrow,right=0.30}{vy,t}
  \fmf{dots_arrow,right=0.30}{t,vx}
  \fmf{gluon,right=0.30}{vx,b}
  \fmf{gluon,right=0.30}{b,vy}
  \fmfv{decor.shape=circle,decor.filled=30,decor.size=0.113w}{t}
  \fmfv{decor.shape=circle,decor.filled=30,decor.size=0.113w}{b}
  \fmfv{decor.shape=circle,decor.filled=empty,decor.size=0.167w}{vy}
  \fmfdot{vx}
  \put(47.0,11.3){\large $\Gamma$}
\end{fmfgraph*}
}

\newsavebox{\DSEOL}
\savebox{\DSEOL}(60,32){
\begin{fmfgraph*}(60,25)
  \fmfleft{x}
  \fmfright{y}
  \fmftop{t}
  \fmfbottom{b}
  \fmf{dots_arrow,tension=3}{y,vy}
  \fmf{dots_arrow,tension=3}{vx,x}
  \fmf{phantom}{vy,vx}
  \fmffreeze
  \fmf{dots_arrow,right=0.30}{vy,t}
  \fmf{dots_arrow,right=0.30}{t,vx}
  \fmf{gluon,right=0.30}{vx,b}
  \fmf{dbl_curly,right=0.30}{b,vy}
  \fmfv{decor.shape=circle,decor.filled=30,decor.size=0.113w}{t}
  \fmfv{decor.shape=circle,decor.filled=30,decor.size=0.113w}{b}
  \fmfv{decor.shape=circle,decor.filled=empty,decor.size=0.167w}{vy}
  \fmfdot{vx}
  \put(47.0,11.3){\large $\Gamma$}
\end{fmfgraph*}
}

\newsavebox{\DSEOALT}
\savebox{\DSEOALT}(60,32){
\begin{fmfgraph*}(60,25)
  \fmfleft{x}
  \fmfright{y}
  \fmftop{t}
  \fmfbottom{b}
  \fmf{dots_arrow,tension=3}{y,vy}
  \fmf{dots_arrow,tension=3}{vx,x}
  \fmf{phantom}{vy,vx}
  \fmffreeze
  \fmf{dots_arrow,right=0.30}{vy,t}
  \fmf{dots_arrow,right=0.30}{t,vx}
  \fmf{plain,right=0.30}{vx,b}
  \fmf{gluon,right=0.30}{b,vy}
  \fmfv{decor.shape=circle,decor.filled=30,decor.size=0.113w}{t}
  \fmfv{decor.shape=circle,decor.filled=30,decor.size=0.113w}{b}
  \fmfv{decor.shape=circle,decor.filled=empty,decor.size=0.167w}{vx}
  \fmfdot{vy}
  \put(10.75,11.3){\large $\Gamma$}
\end{fmfgraph*}
}

\newsavebox{\DSELLLP}
\savebox{\DSELLLP}(60,32){
\begin{fmfgraph*}(60,25)
  \fmfleft{x}
  \fmfright{y}
  \fmftop{t}
  \fmfbottom{b}
  \fmf{dbl_curly,tension=3}{vy,y}
  \fmf{dbl_curly,tension=3}{x,vx}
  \fmf{phantom}{vy,vx}
  \fmffreeze
  \fmf{plain,right=0.30}{vy,t}
  \fmf{gluon,right=0.30}{t,vx}
  \fmf{gluon,right=0.30}{vx,b}
  \fmf{plain,right=0.30}{b,vy}
  \fmfv{decor.shape=circle,decor.filled=30,decor.size=0.113w}{t}
  \fmfv{decor.shape=circle,decor.filled=30,decor.size=0.113w}{b}
  \fmfv{decor.shape=circle,decor.filled=empty,decor.size=0.167w}{vy}
  \fmfdot{vx}
  \put(47.0,11.3){\large $\Gamma$}
\end{fmfgraph*}
}

\newsavebox{\DSELALP}
\savebox{\DSELALP}(60,32){
\begin{fmfgraph*}(60,25)
  \fmfleft{x}
  \fmfright{y}
  \fmftop{t}
  \fmfbottom{b}
  \fmf{gluon,tension=3}{vy,y}
  \fmf{dbl_curly,tension=3}{x,vx}
  \fmf{phantom}{vy,vx}
  \fmffreeze
  \fmf{plain,right=0.30}{vy,t}
  \fmf{gluon,right=0.30}{t,vx}
  \fmf{gluon,right=0.30}{vx,b}
  \fmf{plain,right=0.30}{b,vy}
  \fmfv{decor.shape=circle,decor.filled=30,decor.size=0.113w}{t}
  \fmfv{decor.shape=circle,decor.filled=30,decor.size=0.113w}{b}
  \fmfv{decor.shape=circle,decor.filled=empty,decor.size=0.167w}{vy}
  \fmfdot{vx}
  \put(47.0,11.3){\large $\Gamma$}
\end{fmfgraph*}
}

\newsavebox{\DSEALLP}
\savebox{\DSEALLP}(60,32){
\begin{fmfgraph*}(60,25)
  \fmfleft{x}
  \fmfright{y}
  \fmftop{t}
  \fmfbottom{b}
  \fmf{dbl_curly,tension=3}{vy,y}
  \fmf{gluon,tension=3}{x,vx}
  \fmf{phantom}{vy,vx}
  \fmffreeze
  \fmf{plain,right=0.30}{vy,t}
  \fmf{gluon,right=0.30}{t,vx}
  \fmf{dbl_curly,right=0.30}{vx,b}
  \fmf{gluon,right=0.30}{b,vy}
  \fmfv{decor.shape=circle,decor.filled=30,decor.size=0.113w}{t}
  \fmfv{decor.shape=circle,decor.filled=30,decor.size=0.113w}{b}
  \fmfv{decor.shape=circle,decor.filled=empty,decor.size=0.167w}{vy}
  \fmfdot{vx}
  \put(47.0,11.3){\large $\Gamma$}
\end{fmfgraph*}
}

\newsavebox{\DSEAALP}
\savebox{\DSEAALP}(60,32){
\begin{fmfgraph*}(60,25)
  \fmfleft{x}
  \fmfright{y}
  \fmftop{t}
  \fmfbottom{b}
  \fmf{gluon,tension=3}{vy,y}
  \fmf{gluon,tension=3}{x,vx}
  \fmf{phantom}{vy,vx}
  \fmffreeze
  \fmf{dbl_curly,right=0.30}{vy,t}
  \fmf{gluon,right=0.30}{t,vx}
  \fmf{dbl_curly,right=0.30}{vx,b}
  \fmf{gluon,right=0.30}{b,vy}
  \fmfv{decor.shape=circle,decor.filled=30,decor.size=0.113w}{t}
  \fmfv{decor.shape=circle,decor.filled=30,decor.size=0.113w}{b}
  \fmfv{decor.shape=circle,decor.filled=empty,decor.size=0.167w}{vy}
  \fmfdot{vx}
  \put(47.0,11.3){\large $\Gamma$}
\end{fmfgraph*}
}

\newsavebox{\DSEALLG}
\savebox{\DSEALLG}(60,32){
\begin{fmfgraph*}(60,25)
  \fmfleft{x}
  \fmfright{y}
  \fmftop{t}
  \fmfbottom{b}
  \fmf{dbl_curly,tension=3}{vy,y}
  \fmf{gluon,tension=3}{x,vx}
  \fmf{phantom}{vy,vx}
  \fmffreeze
  \fmf{dots_arrow,right=0.30}{vy,t}
  \fmf{dots_arrow,right=0.30}{t,vx}
  \fmf{dots_arrow,right=0.30}{vx,b}
  \fmf{dots_arrow,right=0.30}{b,vy}
  \fmfv{decor.shape=circle,decor.filled=30,decor.size=0.113w}{t}
  \fmfv{decor.shape=circle,decor.filled=30,decor.size=0.113w}{b}
  \fmfv{decor.shape=circle,decor.filled=empty,decor.size=0.167w}{vy}
  \fmfdot{vx}
  \put(47.0,11.3){\large $\Gamma$}
\end{fmfgraph*}
}

\newsavebox{\DSEAALG}
\savebox{\DSEAALG}(60,32){
\begin{fmfgraph*}(60,25)
  \fmfleft{x}
  \fmfright{y}
  \fmftop{t}
  \fmfbottom{b}
  \fmf{gluon,tension=3}{vy,y}
  \fmf{gluon,tension=3}{x,vx}
  \fmf{phantom}{vy,vx}
  \fmffreeze
  \fmf{dots_arrow,right=0.30}{vy,t}
  \fmf{dots_arrow,right=0.30}{t,vx}
  \fmf{dots_arrow,right=0.30}{vx,b}
  \fmf{dots_arrow,right=0.30}{b,vy}
  \fmfv{decor.shape=circle,decor.filled=30,decor.size=0.113w}{t}
  \fmfv{decor.shape=circle,decor.filled=30,decor.size=0.113w}{b}
  \fmfv{decor.shape=circle,decor.filled=empty,decor.size=0.167w}{vy}
  \fmfdot{vx}
  \put(47.0,11.3){\large $\Gamma$}
\end{fmfgraph*}
}

\newsavebox{\DSELLSS}
\savebox{\DSELLSS}(60,32){
\begin{fmfgraph*}(60,25)
  \fmfleft{x}
  \fmfright{y}
  \fmftop{t}
  \fmfbottom{b}
  \fmf{dbl_curly,tension=3}{vy,y}
  \fmf{dbl_curly,tension=3}{x,vx}
  \fmf{phantom}{vy,vx}
  \fmffreeze
  \fmf{plain,right=0.30}{vy,t}
  \fmf{gluon,right=0.30}{t,vx}
  \fmf{gluon,right=0.30}{vx,b}
  \fmf{plain,right=0.30}{b,vy}
  \fmf{plain}{m,vy}
  \fmf{gluon}{vx,m}
  \fmfv{decor.shape=circle,decor.filled=30,decor.size=0.113w}{t}
  \fmfv{decor.shape=circle,decor.filled=30,decor.size=0.113w}{b}
  \fmfv{decor.shape=circle,decor.filled=30,decor.size=0.113w}{m}
  \fmfv{decor.shape=circle,decor.filled=empty,decor.size=0.167w}{vy}
  \fmfdot{vx}
  \put(47.0,11.3){\large $\Gamma$}
\end{fmfgraph*}
}

\newsavebox{\DSELASS}
\savebox{\DSELASS}(60,32){
\begin{fmfgraph*}(60,25)
  \fmfleft{x}
  \fmfright{y}
  \fmftop{t}
  \fmfbottom{b}
  \fmf{gluon,tension=3}{vy,y}
  \fmf{dbl_curly,tension=3}{x,vx}
  \fmf{phantom}{vy,vx}
  \fmffreeze
  \fmf{plain,right=0.30}{vy,t}
  \fmf{gluon,right=0.30}{t,vx}
  \fmf{gluon,right=0.30}{vx,b}
  \fmf{plain,right=0.30}{b,vy}
  \fmf{plain}{m,vy}
  \fmf{gluon}{vx,m}
  \fmfv{decor.shape=circle,decor.filled=30,decor.size=0.113w}{t}
  \fmfv{decor.shape=circle,decor.filled=30,decor.size=0.113w}{b}
  \fmfv{decor.shape=circle,decor.filled=30,decor.size=0.113w}{m}
  \fmfv{decor.shape=circle,decor.filled=empty,decor.size=0.167w}{vy}
  \fmfdot{vx}
  \put(47.0,11.3){\large $\Gamma$}
\end{fmfgraph*}
}

\newsavebox{\DSEALSS}
\savebox{\DSEALSS}(60,32){
\begin{fmfgraph*}(60,25)
  \fmfleft{x}
  \fmfright{y}
  \fmftop{t}
  \fmfbottom{b}
  \fmf{dbl_curly,tension=3}{vy,y}
  \fmf{gluon,tension=3}{x,vx}
  \fmf{phantom}{vy,vx}
  \fmffreeze
  \fmf{plain,right=0.30}{vy,t}
  \fmf{gluon,right=0.30}{t,vx}
  \fmf{dbl_curly,right=0.30}{vx,b}
  \fmf{gluon,right=0.30}{b,vy}
  \fmf{plain}{m,vy}
  \fmf{gluon}{vx,m}
  \fmfv{decor.shape=circle,decor.filled=30,decor.size=0.113w}{t}
  \fmfv{decor.shape=circle,decor.filled=30,decor.size=0.113w}{b}
  \fmfv{decor.shape=circle,decor.filled=30,decor.size=0.113w}{m}
  \fmfv{decor.shape=circle,decor.filled=empty,decor.size=0.167w}{vy}
  \fmfdot{vx}
  \put(47.0,11.3){\large $\Gamma$}
\end{fmfgraph*}
}

\newsavebox{\DSEAASS}
\savebox{\DSEAASS}(60,32){
\begin{fmfgraph*}(60,25)
  \fmfleft{x}
  \fmfright{y}
  \fmftop{t}
  \fmfbottom{b}
  \fmf{gluon,tension=3}{vy,y}
  \fmf{gluon,tension=3}{x,vx}
  \fmf{phantom}{vy,vx}
  \fmffreeze
  \fmf{plain,right=0.30}{vy,t}
  \fmf{gluon,right=0.30}{t,vx}
  \fmf{dbl_curly,right=0.30}{vx,b}
  \fmf{gluon,right=0.30}{b,vy}
  \fmf{plain}{m,vy}
  \fmf{gluon}{vx,m}
  \fmfv{decor.shape=circle,decor.filled=30,decor.size=0.113w}{t}
  \fmfv{decor.shape=circle,decor.filled=30,decor.size=0.113w}{b}
  \fmfv{decor.shape=circle,decor.filled=30,decor.size=0.113w}{m}
  \fmfv{decor.shape=circle,decor.filled=empty,decor.size=0.167w}{vy}
  \fmfdot{vx}
  \put(47.0,11.3){\large $\Gamma$}
\end{fmfgraph*}
}

\newsavebox{\DSELLG}
\savebox{\DSELLG}(80,33){
\fmfframe(0,7)(0,-7){
\begin{fmfgraph*}(80,30)
  \fmfleft{x}
  \fmfright{y}
  \fmf{dbl_curly,tension=2}{x,vx}
  \fmf{dbl_curly,tension=2}{vy,y}
  \fmf{phantom}{vx,m}
  \fmf{phantom}{m,vy}
  \fmffreeze
  \fmfv{decor.shape=circle,decor.filled=empty,decor.size=0.125w}{m}
  \fmfv{decor.shape=circle,decor.filled=empty,decor.size=0.125w}{vy}
  \fmfdot{vx}
  \fmf{plain}{m,vr}
  \fmf{plain}{vr,vy}
  \fmfv{decor.shape=circle,decor.filled=30,decor.size=0.085w}{vr}
  \fmf{plain,right=0.35}{vy,t}
  \fmf{gluon,right=0.35}{t,vx}
  \fmfv{decor.shape=circle,decor.filled=30,decor.size=0.085w}{t}
  \fmffreeze
  \fmfshift{(0, 0.25w)}{t}
  \fmf{plain,right=0.2}{m,tx}
  \fmf{gluon,right=0.2}{tx,vx}
  \fmf{plain,left=0.2}{m,bx}
  \fmf{gluon,right=0.2}{vx,bx}
  \fmffreeze
  \fmfshift{(0, 0.065w)}{tx}
  \fmfshift{(0,-0.065w)}{bx}
  \fmfv{decor.shape=circle,decor.filled=30,decor.size=0.085w}{tx}
  \fmfv{decor.shape=circle,decor.filled=30,decor.size=0.085w}{bx}
  \put(65.5,13.75){\large $\Gamma$}
  \put(38.75,13.75){\large $\Gamma$}
\end{fmfgraph*}
}}

\newsavebox{\DSELAG}
\savebox{\DSELAG}(80,33){
\fmfframe(0,7)(0,-7){
\begin{fmfgraph*}(80,30)
  \fmfleft{x}
  \fmfright{y}
  \fmf{dbl_curly,tension=2}{x,vx}
  \fmf{gluon,tension=2}{vy,y}
  \fmf{phantom}{vx,m}
  \fmf{phantom}{m,vy}
  \fmffreeze
  \fmfv{decor.shape=circle,decor.filled=empty,decor.size=0.125w}{m}
  \fmfv{decor.shape=circle,decor.filled=empty,decor.size=0.125w}{vy}
  \fmfdot{vx}
  \fmf{plain}{m,vr}
  \fmf{plain}{vr,vy}
  \fmfv{decor.shape=circle,decor.filled=30,decor.size=0.085w}{vr}
  \fmf{plain,right=0.35}{vy,t}
  \fmf{gluon,right=0.35}{t,vx}
  \fmfv{decor.shape=circle,decor.filled=30,decor.size=0.085w}{t}
  \fmffreeze
  \fmfshift{(0, 0.25w)}{t}
  \fmf{plain,right=0.2}{m,tx}
  \fmf{gluon,right=0.2}{tx,vx}
  \fmf{plain,left=0.2}{m,bx}
  \fmf{gluon,right=0.2}{vx,bx}
  \fmffreeze
  \fmfshift{(0, 0.065w)}{tx}
  \fmfshift{(0,-0.065w)}{bx}
  \fmfv{decor.shape=circle,decor.filled=30,decor.size=0.085w}{tx}
  \fmfv{decor.shape=circle,decor.filled=30,decor.size=0.085w}{bx}
  \put(65.5,13.75){\large $\Gamma$}
  \put(38.75,13.75){\large $\Gamma$}
\end{fmfgraph*}
}}

\newsavebox{\DSEALG}
\savebox{\DSEALG}(80,33){
\fmfframe(0,7)(0,-7){
\begin{fmfgraph*}(80,30)
  \fmfleft{x}
  \fmfright{y}
  \fmf{gluon,tension=2}{x,vx}
  \fmf{dbl_curly,tension=2}{vy,y}
  \fmf{phantom}{vx,m}
  \fmf{phantom}{m,vy}
  \fmffreeze
  \fmfv{decor.shape=circle,decor.filled=empty,decor.size=0.125w}{m}
  \fmfv{decor.shape=circle,decor.filled=empty,decor.size=0.125w}{vy}
  \fmfdot{vx}
  \fmf{plain}{m,vr}
  \fmf{plain}{vr,vy}
  \fmfv{decor.shape=circle,decor.filled=30,decor.size=0.085w}{vr}
  \fmf{plain,right=0.35}{vy,t}
  \fmf{gluon,right=0.35}{t,vx}
  \fmfv{decor.shape=circle,decor.filled=30,decor.size=0.085w}{t}
  \fmffreeze
  \fmfshift{(0, 0.25w)}{t}
  \fmf{plain,right=0.2}{m,tx}
  \fmf{gluon,right=0.2}{tx,vx}
  \fmf{gluon,right=0.2}{bx,m}
  \fmf{dbl_curly,right=0.2}{vx,bx}
  \fmffreeze
  \fmfshift{(0, 0.065w)}{tx}
  \fmfshift{(0,-0.065w)}{bx}
  \fmfv{decor.shape=circle,decor.filled=30,decor.size=0.085w}{tx}
  \fmfv{decor.shape=circle,decor.filled=30,decor.size=0.085w}{bx}
  \put(65.5,13.75){\large $\Gamma$}
  \put(38.75,13.75){\large $\Gamma$}
\end{fmfgraph*}
}}

\newsavebox{\DSEAAG}
\savebox{\DSEAAG}(80,33){
\fmfframe(0,7)(0,-7){
\begin{fmfgraph*}(80,30)
  \fmfleft{x}
  \fmfright{y}
  \fmf{gluon,tension=2}{x,vx}
  \fmf{gluon,tension=2}{vy,y}
  \fmf{phantom}{vx,m}
  \fmf{phantom}{m,vy}
  \fmffreeze
  \fmfv{decor.shape=circle,decor.filled=empty,decor.size=0.125w}{m}
  \fmfv{decor.shape=circle,decor.filled=empty,decor.size=0.125w}{vy}
  \fmfdot{vx}
  \fmf{plain}{m,vr}
  \fmf{plain}{vr,vy}
  \fmfv{decor.shape=circle,decor.filled=30,decor.size=0.085w}{vr}
  \fmf{plain,right=0.35}{vy,t}
  \fmf{gluon,right=0.35}{t,vx}
  \fmfv{decor.shape=circle,decor.filled=30,decor.size=0.085w}{t}
  \fmffreeze
  \fmfshift{(0, 0.25w)}{t}
  \fmf{plain,right=0.2}{m,tx}
  \fmf{gluon,right=0.2}{tx,vx}
  \fmf{gluon,right=0.2}{bx,m}
  \fmf{dbl_curly,right=0.2}{vx,bx}
  \fmffreeze
  \fmfshift{(0, 0.065w)}{tx}
  \fmfshift{(0,-0.065w)}{bx}
  \fmfv{decor.shape=circle,decor.filled=30,decor.size=0.085w}{tx}
  \fmfv{decor.shape=circle,decor.filled=30,decor.size=0.085w}{bx}
  \put(65.5,13.75){\large $\Gamma$}
  \put(38.75,13.75){\large $\Gamma$}
\end{fmfgraph*}
}}

\newsavebox{\DSEALH}
\savebox{\DSEALH}(80,33){
\fmfframe(0,7)(0,-7){
\begin{fmfgraph*}(80,30)
  \fmfleft{x}
  \fmfright{y}
  \fmf{gluon,tension=2}{x,vx}
  \fmf{dbl_curly,tension=2}{vy,y}
  \fmf{phantom}{vx,m}
  \fmf{phantom}{m,vy}
  \fmffreeze
  \fmfv{decor.shape=circle,decor.filled=empty,decor.size=0.125w}{m}
  \fmfv{decor.shape=circle,decor.filled=empty,decor.size=0.125w}{vy}
  \fmfdot{vx}
  \fmf{plain}{m,vr}
  \fmf{plain}{vr,vy}
  \fmfv{decor.shape=circle,decor.filled=30,decor.size=0.085w}{vr}
  \fmf{gluon,right=0.35}{vy,t}
  \fmf{dbl_curly,right=0.35}{t,vx}
  \fmfv{decor.shape=circle,decor.filled=30,decor.size=0.085w}{t}
  \fmffreeze
  \fmfshift{(0, 0.25w)}{t}
  \fmf{plain,right=0.2}{m,tx}
  \fmf{gluon,right=0.2}{tx,vx}
  \fmf{plain,left=0.2}{m,bx}
  \fmf{gluon,right=0.2}{vx,bx}
  \fmffreeze
  \fmfshift{(0, 0.065w)}{tx}
  \fmfshift{(0,-0.065w)}{bx}
  \fmfv{decor.shape=circle,decor.filled=30,decor.size=0.085w}{tx}
  \fmfv{decor.shape=circle,decor.filled=30,decor.size=0.085w}{bx}
  \put(65.5,13.75){\large $\Gamma$}
  \put(38.75,13.75){\large $\Gamma$}
\end{fmfgraph*}
}}

\newsavebox{\DSEAAH}
\savebox{\DSEAAH}(80,33){
\fmfframe(0,7)(0,-7){
\begin{fmfgraph*}(80,30)
  \fmfleft{x}
  \fmfright{y}
  \fmf{gluon,tension=2}{x,vx}
  \fmf{gluon,tension=2}{vy,y}
  \fmf{phantom}{vx,m}
  \fmf{phantom}{m,vy}
  \fmffreeze
  \fmfv{decor.shape=circle,decor.filled=empty,decor.size=0.125w}{m}
  \fmfv{decor.shape=circle,decor.filled=empty,decor.size=0.125w}{vy}
  \fmfdot{vx}
  \fmf{gluon}{m,vr}
  \fmf{dbl_curly}{vr,vy}
  \fmfv{decor.shape=circle,decor.filled=30,decor.size=0.085w}{vr}
  \fmf{gluon,right=0.35}{vy,t}
  \fmf{dbl_curly,right=0.35}{t,vx}
  \fmfv{decor.shape=circle,decor.filled=30,decor.size=0.085w}{t}
  \fmffreeze
  \fmfshift{(0, 0.25w)}{t}
  \fmf{plain,right=0.2}{m,tx}
  \fmf{gluon,right=0.2}{tx,vx}
  \fmf{plain,left=0.2}{m,bx}
  \fmf{gluon,right=0.2}{vx,bx}
  \fmffreeze
  \fmfshift{(0, 0.065w)}{tx}
  \fmfshift{(0,-0.065w)}{bx}
  \fmfv{decor.shape=circle,decor.filled=30,decor.size=0.085w}{tx}
  \fmfv{decor.shape=circle,decor.filled=30,decor.size=0.085w}{bx}
  \put(65.5,13.75){\large $\Gamma$}
  \put(38.75,13.75){\large $\Gamma$}
\end{fmfgraph*}
}}

\newsavebox{\DSELATP}
\savebox{\DSELATP}(30,32){
\begin{fmfgraph*}(30,25)
  \fmfstraight
  \fmftop{t}
  \fmfbottom{x,y}
  \fmf{gluon,tension=3}{v,y}
  \fmf{dbl_curly,tension=3}{x,v}
  \fmffreeze
  \fmf{gluon,right=0.80}{v,t}
  \fmf{gluon,right=0.80}{t,v}
  \fmfv{decor.shape=circle,decor.filled=30,decor.size=0.226w}{t}
  \fmfdot{v}
\end{fmfgraph*}
}

\newsavebox{\DSEALTP}
\savebox{\DSEALTP}(30,32){
\begin{fmfgraph*}(30,25)
  \fmfstraight
  \fmftop{t}
  \fmfbottom{x,y}
  \fmf{dbl_curly,tension=3}{v,y}
  \fmf{gluon,tension=3}{x,v}
  \fmffreeze
  \fmf{gluon,right=0.80}{v,t}
  \fmf{gluon,right=0.80}{t,v}
  \fmfv{decor.shape=circle,decor.filled=30,decor.size=0.226w}{t}
  \fmfdot{v}
\end{fmfgraph*}
}

\newsavebox{\DSEAATP}
\savebox{\DSEAATP}(30,32){
\begin{fmfgraph*}(30,25)
  \fmfstraight
  \fmftop{t}
  \fmfbottom{x,y}
  \fmf{gluon,tension=3}{v,y}
  \fmf{gluon,tension=3}{x,v}
  \fmffreeze
  \fmf{gluon,right=0.80}{v,t}
  \fmf{dbl_curly,right=0.80}{t,v}
  \fmfv{decor.shape=circle,decor.filled=30,decor.size=0.226w}{t}
  \fmfdot{v}
\end{fmfgraph*}
}


\newsavebox{\PiOmega}
\savebox{\PiOmega}(30,7){
\begin{fmfgraph*}(30,7)
  \fmfleft{x}
  \fmfright{y}
  \fmf{dots_arrow}{y,v}
  \fmf{dots_arrow}{v,x}
  \fmfv{decor.shape=circle,decor.filled=empty,decor.size=0.227w}{v}
  \put(13.8,2.1){\large $\Pi$}
  \put(12.5,3.4){\line(1,0){1.35}}
\end{fmfgraph*}
}

\newsavebox{\PiLL}
\savebox{\PiLL}(30,7){
\begin{fmfgraph*}(30,7)
  \fmfleft{x}
  \fmfright{y}
  \fmf{dbl_curly}{v,y}
  \fmf{dbl_curly}{x,v}
  \fmfv{decor.shape=circle,decor.filled=empty,decor.size=0.227w}{v}
  \put(13.8,2.1){\large $\Pi$}
  \put(12.5,3.4){\line(1,0){1.35}}
\end{fmfgraph*}
}

\newsavebox{\PiLA}
\savebox{\PiLA}(30,7){
\begin{fmfgraph*}(30,7)
  \fmfleft{x}
  \fmfright{y}
  \fmf{gluon}{v,y}
  \fmf{dbl_curly}{x,v}
  \fmfv{decor.shape=circle,decor.filled=empty,decor.size=0.227w}{v}
  \put(13.8,2.1){\large $\Pi$}
  \put(12.5,3.4){\line(1,0){1.35}}
\end{fmfgraph*}
}

\newsavebox{\PiAL}
\savebox{\PiAL}(30,7){
\begin{fmfgraph*}(30,7)
  \fmfleft{x}
  \fmfright{y}
  \fmf{dbl_curly}{v,y}
  \fmf{gluon}{x,v}
  \fmfv{decor.shape=circle,decor.filled=empty,decor.size=0.227w}{v}
  \put(13.8,2.1){\large $\Pi$}
  \put(12.5,3.4){\line(1,0){1.35}}
\end{fmfgraph*}
}

\newsavebox{\PiAA}
\savebox{\PiAA}(30,7){
\begin{fmfgraph*}(30,7)
  \fmfleft{x}
  \fmfright{y}
  \fmf{gluon}{v,y}
  \fmf{gluon}{x,v}
  \fmfv{decor.shape=circle,decor.filled=empty,decor.size=0.227w}{v}
  \put(13.8,2.1){\large $\Pi$}
  \put(12.5,3.4){\line(1,0){1.35}}
\end{fmfgraph*}
}


\newsavebox{\GLL}
\savebox{\GLL}(30,7){
\begin{fmfgraph*}(30,7)
  \fmfleft{x}
  \fmfright{y}
  \fmf{dbl_curly}{v,y}
  \fmf{dbl_curly}{x,v}
  \fmfv{decor.shape=circle,decor.filled=30,decor.size=0.227w}{v}
\end{fmfgraph*}
}

\newsavebox{\GOmega}
\savebox{\GOmega}(30,7){
\begin{fmfgraph*}(30,7)
  \fmfleft{x}
  \fmfright{y}
  \fmf{dots_arrow}{y,v}
  \fmf{dots_arrow}{v,x}
  \fmfv{decor.shape=circle,decor.filled=30,decor.size=0.227w}{v}
\end{fmfgraph*}
} 


\newsavebox{\DOmega}
\savebox{\DOmega}(20,7){
\begin{fmfgraph*}(20,7)
  \fmfleft{x}
  \fmfright{y}
  \fmf{dots_arrow}{y,x}
\end{fmfgraph*}
}


\newsavebox{\DPGO}
\savebox{\DPGO}(60,7){
\begin{fmfgraph*}(60,7)
  \fmfleft{x}
  \fmfright{y}
  \fmf{phantom}{y,v}
  \fmf{dots_arrow, tension=2}{v,x} 
  \fmffreeze
  \fmf{dots_arrow,tension=1.5}{y,b1}
  \fmf{dots_arrow}{b1,b2}
  \fmf{dots_arrow,tension=1.5}{b2,v}
  \fmfv{decor.shape=circle,decor.filled=30,decor.size=0.113w}{b1}
  \fmfv{decor.shape=circle,decor.filled=empty,decor.size=0.113w}{b2}
  \put(30.3,2.1){\large $\Pi$}
  \put(29.0,3.4){\line(1,0){1.35}}
\end{fmfgraph*}
}


\subsubsection{Definitions and General Relations}
Concerning the propagators, mixing will occur
between the gauge field $\myvec{A}$ and the auxiliary field 
$\gvec{\lambda}$, resulting in four possible propagators 
from the gauge/auxiliary field sector that can be combined into 
one matrix propagator. These are completed by the propagator 
of the gauge ghosts. Altogether, we define the full (connected)
propagators as\,
\begin{eqnarray} \label{def:ConProp(AA)}
  G^{(AA)}{}^{ab}_{ij}(x,y) &=&
  \makebox[16.7ex]{\begin{math}
  \left\langle A^{ai}(x) \, A^{bj}(y) \right\rangle_c 
  \end{math}}
  \ =\ 
  \frac{\delta^2 W[J,I]}{\delta J_A^{ai}(x) \, \delta J_A^{bj}(y)}
  \biggr|_{J=I=0}\\[2ex] \label{def:ConProp(LambdaA)}
  G^{(\lambda A)}{}^{ab}_{ij}(x,y) &=&
  \makebox[16.7ex]{\begin{math}
  \left\langle \lambda^{ai}(x) \, A^{bj}(y) \right\rangle_c 
  \end{math}}
  \ =\ 
  \frac{\delta^2 W[J,I]}{\delta J_{\lambda}^{ai}(x) \, \delta J_A^{bj}(y)}
  \biggr|_{J=I=0}\\[2ex] \label{def:ConProp(LambdaLambda)}
  G^{(\lambda\lambda)}{}^{ab}_{ij}(x,y) &=&
  \makebox[16.7ex]{\begin{math}
  \left\langle \lambda^{ai}(x) \, \lambda^{bj}(y) \right\rangle_c
  \end{math}}
  \ =\ 
  \frac{\delta^2 W[J,I]}{\delta J_{\lambda}^{ai}(x) \, 
    \delta J_{\lambda}^{bj}(y)}
  \biggr|_{J=I=0}\\[2ex] \label{def:ConProp(Omega)}
  G^{(\omega) \,ab}(x,y) &=&
  \makebox[16.7ex]{\begin{math}
  \left\langle \omega^a(x) \, \bar{\omega}^b(y) \right\rangle_c     
  \end{math}}
  \ =\ 
  \hspace{1.3ex}
  \frac{\delta^2 W[J,I]}{\delta J_{\omega}^a(x) \, 
    \delta J_{\bar{\omega}}^b(y)}
  \biggr|_{J=I=0}
\end{eqnarray}
and
\begin{math}
  G^{(A \lambda)}{}^{ab}_{ij}(x,y) = 
  G^{(\lambda A)}{}^{ba}_{ji}(y,x)
\end{math}
of course. In graphical representations we denote the gauge field
by curly lines, the auxiliary field by double curly lines and the gauge 
ghosts by dotted lines. Thus, the full propagators are
represented by

\noindent
\unitlength=1mm
\begin{center}
\begin{picture}(77,44) 
  \put(0,36){ 
  \begin{picture}(75,7) 
  \put(35,0){
  \begin{fmfgraph*}(30,7)
    \fmfleft{x}
    \fmfright{y}
    \fmf{gluon}{x,v}
    \fmf{gluon}{v,y}
    \fmfdot{x,y}
    \fmfv{decor.shape=circle,decor.filled=30,decor.size=0.227w}{v}
    \put(-1,6){$x$}
    \put(29,6){$y$}
    \put(-7,2.5){$a,i$}
    \put(32.5,2.5){$b,j$}
    \put(-35,2.5){$G^{(AA)}{}^{ab}_{ij}(x,y)$}
    \put(-12.5,2.5){\Large =}
  \end{fmfgraph*}}
  \end{picture} 
  } 
  \put(0,24){ 
  \begin{picture}(75,7) 
  \put(35,0){
  \begin{fmfgraph*}(30,7)
    \fmfleft{x}
    \fmfright{y}
    \fmf{gluon}{x,v}
    \fmf{dbl_curly}{v,y}
    \fmfdot{x,y}
    \fmfv{decor.shape=circle,decor.filled=30,decor.size=0.227w}{v}
    \put(-1,6){$x$}
    \put(29,6){$y$}
    \put(-7,2.5){$a,i$}
    \put(32.5,2.5){$b,j$}
    \put(-35,2.5){$G^{(A\lambda)}{}^{ab}_{ij}(x,y)$}
    \put(-12.5,2.5){\Large =}
  \end{fmfgraph*}}
  \end{picture} 
  } 
  \put(0,12){ 
  \begin{picture}(75,7) 
  \put(35,0){
  \begin{fmfgraph*}(30,7)
    \fmfleft{x}
    \fmfright{y}
    \fmf{dbl_curly}{x,v}
    \fmf{gluon}{v,y}
    \fmfdot{x,y}
    \fmfv{decor.shape=circle,decor.filled=30,decor.size=0.227w}{v}
    \put(-1,6){$x$}
    \put(29,6){$y$}
    \put(-7,2.5){$a,i$}
    \put(32.5,2.5){$b,j$}
    \put(-35,2.5){$G^{(\lambda A)}{}^{ab}_{ij}(x,y)$}
    \put(-12.5,2.5){\Large =}
  \end{fmfgraph*}}
  \end{picture} 
  } 
  \put(0,0){ 
  \begin{picture}(75,7) 
  \put(35,0){
  \begin{fmfgraph*}(30,7)
    \fmfleft{x}
    \fmfright{y}
    \fmf{dbl_curly}{x,v}
    \fmf{dbl_curly}{v,y}
    \fmfdot{x,y}
    \fmfv{decor.shape=circle,decor.filled=30,decor.size=0.227w}{v}
    \put(-1,6){$x$}
    \put(29,6){$y$}
    \put(-7,2.5){$a,i$}
    \put(32.5,2.5){$b,j$}
    \put(-35,2.5){$G^{(\lambda\lambda)}{}^{ab}_{ij}(x,y)$}
    \put(-12.5,2.5){\Large =}
  \end{fmfgraph*}}
  \end{picture} 
  } 
\end{picture} 
\end{center}
and finally
\unitlength=1mm
\begin{center}
\begin{picture}(77,8) 
  \put(0,0){ 
  \begin{picture}(75,7) 
  \put(35,0){
  \begin{fmfgraph*}(30,7)
    \fmfleft{x}
    \fmfright{y}
    \fmf{dots_arrow}{y,v}
    \fmf{dots_arrow}{v,x}
    \fmfdot{x,y}
    \fmfv{decor.shape=circle,decor.filled=30,decor.size=0.227w}{v}
    \put(-1,6){$x$}
    \put(29,6){$y$}
    \put(-4,2.5){$a$}
    \put(32.5,2.5){$b$}
    \put(-33,2.5){$G^{(\omega)\,ab}(x,y)$}
    \put(-12.5,2.5){\Large =}
  \end{fmfgraph*}}
  \end{picture} 
  } 
\end{picture} 
\end{center}
\noindent
Besides the propagators, we have to set out our definition for the 
self-energies. To this end, let us summarise the two left-hand 
equations of (\ref{eq:DerivativesOfGamma}) in the form
\begin{equation}\label{eq:JF}
  J_F^{ai}(x) = 
  \frac{\delta \Gamma[\myvec{A},\gvec{\lambda},\omega,\bar{\omega};I]}
       {\delta F^{ai}(x)}
\end{equation}
where the index $F$ stands for any of the fields $A$ or $\lambda$.
Taking the functional derivative of this equation with respect 
to $J_G^{bj}(y)$, where again $G\in\{A,\lambda\}$, then yields 
(observing that $\myvec{A}$, 
$\gvec{\lambda}$, $\omega$ and $\bar{\omega}$ are functionals of the
sources $J$ and $I$)
\begin{eqnarray}
  \delta^{ab} \,\delta^{ij} \,\delta_{FG} \,\delta(x-y) \!\!\!&=&\!\!\!
  \int\!\!dz
  \biggl[ \hspace{1ex}
    \frac{\delta A^{ck}(z)}{\delta J_G^{bj}(y)} \,
    \frac{\delta^2 \Gamma}{\delta A^{ck}(z)\,\delta F^{ai}(x)}
    \;+\;
    \frac{\delta \lambda^{ck}(z)}{\delta J_G^{bj}(y)} \,
    \frac{\delta^2 \Gamma}{\delta \lambda^{ck}(z)\,\delta F^{ai}(x)}
    \hspace{5ex}\nonumber \\
    & & \hspace{4.5ex}
    + \;
    \frac{\delta \omega^c(z)}{\delta J_G^{bj}(y)} \,
    \frac{\delta^2 \Gamma}{\delta \omega^c(z)\,\delta F^{ai}(x)}
    \;+\;
    \frac{\delta \bar{\omega}^c(z)}{\delta J_G^{bj}(y)} \,
    \frac{\delta^2 \Gamma}{\delta \bar{\omega}^c(z)\,\delta F^{ai}(x)}\;
  \biggr]
\end{eqnarray}
Thus, using the Eqs.~(\ref{def:classicalFields}) to express the 
first factor in each term as a second derivative of $W$ and finally 
setting the sources to zero leads to
\begin{eqnarray} \label{eq:2GammaInvTo2-W}
\lefteqn{
  \delta^{ab} \,\delta^{ij} \,\delta_{FG} \,\delta(x-y) =}\\[2ex]
  & &
  \int\!\!dz\left[
  G^{(GA)}{}^{bc}_{jk}(y,z) \,
  \frac{\delta^2 \Gamma}{\delta A^{ck}(z)\,\delta F^{ai}(x)}\biggr|_{J=I=0}
  \!\!\!\!\!+\;
  G^{(G\lambda)}{}^{bc}_{jk}(y,z) \,
  \frac{\delta^2 \Gamma}{\delta \lambda^{ck}(z)\,\delta F^{ai}(x)}
  \biggr|_{J=I=0}
  \right] \nonumber
\end{eqnarray}
Here, the definitions (\ref{def:ConProp(AA)}) -- 
(\ref{def:ConProp(LambdaLambda)}) have been used and the terms 
involving ghost and anti-ghost fields have vanished due to ghost 
number conservation.

In the following we will often encounter multiple derivatives
of the generating functionals $W$ and $\Gamma$ evaluated for 
vanishing sources. Let us therefore introduce a shorthand notation
where we indicate the fields with respect to which the derivatives
are taken as superscripts. Possible \textsc{Lorentz} or colour 
indices as well as space-time arguments appear in the order of the 
fields they belong to. For instance, we abbreviate
\begin{equation} \label{def:shorthand}
  \Gamma^{(\lambda A \bar{\omega})}{}^{abc}_{ij}(x,y,z) =
  \frac{\delta^3 \Gamma}{\delta \lambda^{ai}(x) \,\delta A^{bj}(y)
        \,\delta\bar{\omega}^c(z)}\biggr|_{J=I=0}
\end{equation}
In the case of $W$, we also use the \emph{fields} as superscripts though
the derivatives are taken with respect to the corresponding \emph{sources}, 
of course. 

\noindent
In this new notation, Eq.~(\ref{eq:2GammaInvTo2-W}) reads
\begin{equation}\label{eq:W2XGamma2=1}
  \delta^{ab} \,\delta^{ij} \,\delta_{FG} \,\delta(x-y) =
  \int\!\!dz\;
    G^{(GH)}{}^{bc}_{jk}(y,z)\,
    \Gamma^{(HF)}{}^{ca}_{ki}(z,x)
\end{equation}
where $H$ is a summation index running over the fields $A$ and $\lambda$.
This equation expresses the fact that the matrix propagator of the
gauge/auxiliary field sector
\begin{equation} \label{def:MatrixProp(AL)}
  \hat{G}^{ab}_{ij}\,(x,y) =
  \left(
  \begin{array}{cc}
  \!\! G^{(\lambda\lambda)}{}^{ab}_{ij}(x,y)   & 
       G^{(\lambda A)}{}^{ab}_{ij}(x,y) \!\! 
  \\[2ex] 
  \!\! G^{(A\lambda)}{}^{ab}_{ij}(x,y) & 
       G^{(AA)}{}^{ab}_{ij}(x,y) \!\!
  \end{array}
  \right)
\end{equation}
is inverse to the matrix 
\begin{equation}
  \hat{\Gamma}^{ab}_{ij}\,(x,y) =
  \left(
  \begin{array}{cc}
  \!\! \Gamma^{(\lambda\lambda)}{}^{ab}_{ij}(x,y)   & 
       \Gamma^{(\lambda A)}{}^{ab}_{ij}(x,y) \!\! 
  \\[2ex] 
  \!\! \Gamma^{(A\lambda)}{}^{ab}_{ij}(x,y) & 
       \Gamma^{(AA)}{}^{ab}_{ij}(x,y) \!\!
  \end{array}
  \right)
\end{equation}
constructed of the second derivatives of $\Gamma$. Consequently, 
the self-energy 
\begin{math}
  \hat{\Pi}^{ab}_{ij}(x,y)
\end{math}
is determined via the relation
\begin{equation} \label{def:self-energy(AL)}
  \Gamma^{(FG)}{}^{ab}_{ij}(x,y) \;=\;
  (\Delta^{-1})^{(FG)}{}^{ab}_{ij}(x,y) \;+\;
  \Pi^{(FG)}{}^{ab}_{ij}(x,y)
\end{equation}
where
\begin{math}
  (\Delta^{-1})^{(FG)}{}^{ab}_{ij}(x,y)
\end{math}
are the components of the inverse free propagator of perturbation
theory (see Appendix \ref{FeynmanRules},
Eqs.~(\ref{eq:Delta^-1:LL}) -- (\ref{eq:Delta^-1:AA})), 
and where $F,G \in\{\lambda, A\}$ as before.

\noindent
Analogously, taking the derivative with respect 
to $J_{\omega}^b(y)$ of
\begin{equation}
  J_{\omega}^a(x) = 
  \frac{\delta\Gamma[\myvec{A},\gvec{\lambda},\omega,\bar{\omega};I]}
       {\delta\omega^a(x)}
\end{equation}
and performing the same manipulations as described above leads to
\begin{equation} \label{eq:G(omega)inversGamma(2)}
  \delta^{ab} \,\delta(x-y) =
  -\int\!\!dz\;
  G^{(\omega) \,bc}(y,z) \,
  \frac{\delta^2 \Gamma}{\delta \bar{\omega}^c(z)\,
                         \delta \omega^a(x)}
  \biggr|_{J=I=0}
\end{equation}
Hence, we define the self-energy of the gauge ghosts via
\begin{equation} \label{def:GG-self-energy}
  \Gamma^{(\bar{\omega}\omega)\,ab}(x,y) \;=\;
  -\left[
    (\Delta^{-1})^{(\omega)\,ab}(x,y) \;+\;
    \Pi^{(\omega)\,ab}(x,y)
  \right]
\end{equation}
with the free inverse propagator $(\Delta^{-1})^{(\omega)\,ab}(x,y)$
given in Eq.~(\ref{eq:Delta^-1:omega}).
In our graphical representations we denote self-energies and other
one-particle irreducible quantities by open circles. 

Though generally we are using three-vectors, 
in the \textsc{Fourier} transformation we use four-vector
notation
\begin{equation}
  f(x) = \intDk e^{-ikx} f(k)
\end{equation}
with
\begin{math}
  -ikx = -ik_0t + i\,\myvec{k}\cdot\myvec{x}
\end{math}.
The proper vertex functions in momentum space are basically given by the
\textsc{Fourier} transforms of the various functional derivatives of the 
1PI generating functional $\Gamma$. However, due to translational 
invariance of the theory, all these \textsc{Fourier} transforms contain
a delta function expressing momentum conservation at the vertex.
It is therefore convenient to pull these delta functions out of the
definitions of the vertex functions. In this way, the latter become 
functions of one momentum variable less than indicated by the 
number of external legs. For instance, we define
\begin{equation}\label{DSE:defGamma_xyz}
  (2\pi)^D \delta^D(k_1 \!+ k_2 \!+ k_3)\,
  \Gamma^{(\bar{\omega}\omega G)}{}^{ab}{}^{c}_{j}(k_1,k_2)
  =
  \int\!\!dx\,dy\,dz\;
  e^{-ik_1x -ik_2y -ik_3z}\;
  \Gamma^{(\bar{\omega}\omega G)}{}^{ab}{}^{c}_{j}(x,y,z)
\end{equation}
or equivalently
\begin{equation}
  \Gamma^{(\bar{\omega}\omega G)}{}^{ab}{}^{c}_{j}(x,y,z) =
  \int\!\frac{d^D \! k_1}{(2\pi)^D}\frac{d^D \! k_2}{(2\pi)^D}
  \;e^{-ik_1(z-x) -ik_2(z-y)}\;
  \Gamma^{(\bar{\omega}\omega G)}{}^{ab}{}^{c}_{j}(k_1,k_2)
\end{equation}
Here, the two arguments of the proper vertex function
\begin{math}
  \Gamma^{(\bar{\omega}\omega G)}{}^{ab}{}^{c}_{j}(k_1,k_2)
\end{math}
refer to the (incoming) momenta along the ghost lines leaving and
entering the vertex in this order.

The choice of the $N-1$ momenta that are used as arguments of a 
vertex with $N$ external legs is, of course, arbitrary and thereby 
a source of possible confusion. We therefore explicitly list 
the definitions of the other relevant vertex functions used 
in this work
\begin{equation} \label{def:1PI-3-momentum}
  \Gamma^{(FGH)}{}^{abc}_{ijk}(x,y,z) =
  -\int\!\frac{d^D \! k_2}{(2\pi)^D}\frac{d^D \! k_3}{(2\pi)^D}
  \;e^{-ik_2(x-y) -ik_3(x-z)}\;
  \Gamma^{(FGH)}{}^{abc}_{ijk}(k_2,k_3)
\end{equation}
with $k_2$ and $k_3$ denoting the incoming momenta along the $G$
and $H$ line respectively, and
\begin{eqnarray}
  \Gamma^{(FGHK)}{}^{abcd}_{ijkl}(x,y,z,w) \!\!\!&=&\!\!\!
  -\int\!\frac{d^D \! k_2}{(2\pi)^D}\frac{d^D \! k_3}{(2\pi)^D}
         \frac{d^D \! k_4}{(2\pi)^D}
  \;e^{-ik_2(x-y) -ik_3(x-z)} 
  \nonumber\\[1.5ex] \label{def:1PI-4-momentum}
  & & \hspace{14ex}
  e^{-ik_4(x-w)}\;
  \Gamma^{(FGHK)}{}^{abcd}_{ijkl}(k_2,k_3,k_4)
\end{eqnarray}
with incoming momenta $k_2$, $k_3$, $k_4$ along the $G$, $H$ and $K$
line. Note the minus signs in the last two equations. The definitions 
above are chosen in such a way that they reduce at leading order to
the corresponding vertices of the \textsc{Feynman} rules, i.e.
\begin{eqnarray}
  \Gamma^{(\bar{\omega}\omega A)}{}^{ab}{}^{c}_{j}(k_1,k_2)
  &=&
  \frac{ig}{\kappa} \, f^{abc} k_2^j
  \ +\ \dots
  \\
  \Gamma^{(\lambda AA)}{}^{abc}_{ijk}(k_2,k_3)
  &=& \!\!
  -g\,V^{abc}_{ijk}(\myvec{k}_2,\myvec{k}_3)
  \ +\ \dots
  \\[0.5ex]
  \Gamma^{(\lambda AAA)}{}^{abcd}_{ijkl}(k_2,k_3,k_4)
  &=&
  ig^2\,V^{abcd}_{ijkl}
  \ +\ \dots
\end{eqnarray}
\subsubsection[DSE for $\Pi^{(\omega)}$]{DSE for $\Pi^{(\omega)}(k)$}

Let us again start with the ghost equations, being much simpler than
the equations for the gauge/auxiliary field sector. By taking the 
derivative of Eq.~(\ref{eq:DSE-anti-ghost:Gamma}) with respect 
to $\omega^b(y)$, one finds evaluated for vanishing sources
\begin{equation}
  \frac{\delta^2 \Gamma}{\delta\omega^b(y) \,\delta\bar{\omega}^a(x)}  
  \biggr|_{J=I=0}
  = \ 
  \underbrace{
  \delta^{ab} \! 
  \left(
    -\partial_t + {\textstyle\frac{1}{\kappa}} \Delta
  \right)
  \delta(x-y)
  }_{\textstyle (\Delta^{-1})^{(\omega)\,ab}(x,y)}
  \;+\;
  \frac{g}{\kappa} f^{ade} \,\partial_j
  \frac{\delta}{\delta\omega^b(y)} \,
  \frac{\delta^2 W[J,I]}{\delta J_{\omega}^d(x)\,
                    \delta J_A^{ej}(x^{\prime})}
  \biggr|_{\!\!\!\begin{array}{lcl}
    \scriptstyle x^{\prime} = \,x \\[-1ex]
    \scriptstyle J \,=\, I \,=\, 0
  \end{array}}
\end{equation}
Comparing to the definition of the self-energy of the gauge ghosts
in Eq.~(\ref{def:GG-self-energy}) then leads to the relation
\begin{equation} \label{eq:startExplicitBarOmega}
  \Pi^{(\omega)\,ab}(x,y) =
  \frac{g}{\kappa} f^{ade} \,\partial_j
  \frac{\delta}{\delta\omega^b(y)} \,
  \frac{\delta^2 W[J,I]}{\delta J_{\omega}^d(x)\,
                    \delta J_A^{ej}(x^{\prime})}
  \biggr|_{\!\!\!\begin{array}{lcl}
    \scriptstyle x^{\prime} = \,x \\[-1ex]
    \scriptstyle J \,=\, I \,=\, 0
  \end{array}}
\end{equation}
for the gauge ghost self-energy. If we carry out the functional
derivative with respect to $\omega^b(y)$, four terms arise because
any of the sources $\myvec{J}_A$, $\myvec{J}_{\lambda}$, $J_{\omega}$
and $J_{\bar{\omega}}$ depends on $\omega$. 
However, due to ghost number conservation three of these 
terms vanish when the sources are set to zero and one is left 
with\footnote{It should be clear that $x^{\prime}$ is set to $x$ only
after the space-time derivative is carried out. In order to avoid
an extensive use of brackets we decided to assume in this and similar 
cases some thoughtfulness on the part of the reader.}
\begin{equation} \label{eq:DSE:omega:step1}
  \Pi^{(\omega)\,ab}(x,y) =
  \frac{g}{\kappa} f^{ade} \,\partial_j
  \!\int\!\!dv
  \biggl[
  \frac{\delta^2 \Gamma}{\delta\omega^b(y)\,\delta\bar{\omega}^c(v)}\,
  \frac{\delta^3 W}{\delta J_{\bar{\omega}}^c(v)\,
                         \delta J_{\omega}^d(x)\,
                         \delta J_A^{ej}(x^{\prime})}
  \biggr]_{\!\!\!\begin{array}{lcl}
    \scriptstyle x^{\prime} = \,x \\[-1ex]
    \scriptstyle J \,=\, I \,=\, 0
  \end{array}}
\end{equation}
Finally, we express the connected three-point function by its 1PI 
counterpart
\begin{equation} \label{eq:W3-by-1PI}
  W^{(\bar{\omega}\omega F)}{}^{ab}{}^{c}_{j}(x,y,z) =
  \int\!\!du\,du^{\prime} du^{\prime\prime}\;
  G^{(\omega)\,a^{\prime}\!a}(u,x) \,
  G^{(\omega)\,b b^{\prime}}\!(y,u^{\prime}) \,
  G^{(FG)}{}^{cc^{\prime}}_{jj^{\prime}}(z,u^{\prime\prime})\,
  \Gamma^{(\omega\bar{\omega} G)}{}^{a^{\prime}
  b^{\prime}}{}^{c^{\prime}}_{j^{\prime}}
  (u,u^{\prime}\!,u^{\prime\prime})
\end{equation}
where $F$ represents one of the fields $\lambda$ or $A$ and $G$ is a 
summation index taking these two values. The shorthand notation used
here 
was introduced in Eq.~(\ref{def:shorthand}).
Note that the order of the ghost and anti-ghost fields in
Eq.~(\ref{eq:W3-by-1PI}) is changed from
$W^{(\bar{\omega}\omega F)}$ to $\Gamma^{(\omega\bar{\omega} G)}$
and that the (full) gauge ghost propagator is
$G^{(\omega)\,ab}(x,y) = W^{(\omega\bar{\omega})\,ab}(x,y)$, 
as defined in Eq.~(\ref{def:ConProp(Omega)}).

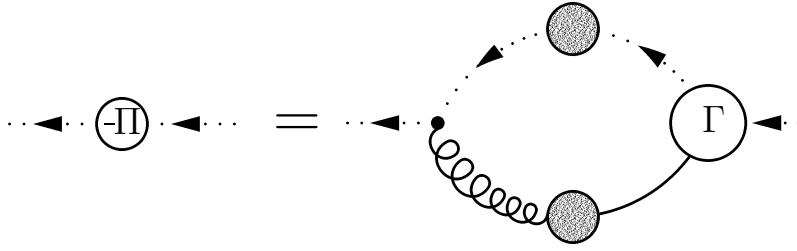
\begin{figure}[t]
\begin{center}
  \begin{picture}(110,40) 
  \put(0,12.5){\usebox{\PiOmega}}
  \put(35,14.1){\huge $=$}
  \put(45,0){\usebox{\DSEO}}
  \end{picture}
  \caption{\label{fig:DSE(Omega)} {
    DSE of the gauge ghost self-energy, Eqs.~(\ref{eq:DSE:Pi(omega)(k)}). 
    Filled circles denote full propagators. Empty circles are used for 
    one-particle irreducible quantities, i.e.~self-energies and proper 
    vertices. The solid line represents a summation of one graph with 
    the line replaced by a gauge field and a second diagram with an 
    auxiliary field instead.}}
\end{center} 
\end{figure}

Now, inserting relation (\ref{eq:W3-by-1PI}) into 
Eq.~(\ref{eq:DSE:omega:step1}), using the property 
(\ref{eq:G(omega)inversGamma(2)}) of the two-point functions
and
\begin{equation}
  \Gamma^{(\omega\bar{\omega} G)}{}^{a^{\prime}
  b^{\prime}}{}^{c^{\prime}}_{j^{\prime}}
  (u,u^{\prime}\!,u^{\prime\prime}) =
  -\Gamma^{(\bar{\omega}\omega G)}{}^{b^{\prime}
  a^{\prime}}{}^{c^{\prime}}_{j^{\prime}}
  (u^{\prime}\!,u,u^{\prime\prime})
\end{equation}
yields the \textsc{Dyson--Schwinger} equation
\begin{equation} \label{eq:DSE:Pi(omega)(x,y)}
  \Pi^{(\omega)\,ab}(x,y) =
  -\!\int\!\!du^{\prime} du^{\prime\prime}\;
  G^{(AG)}{}^{ee^{\prime}}_{jj^{\prime}}(x,u^{\prime\prime})\,
  \frac{g}{\kappa} f^{ade} \,\partial_j
  G^{(\omega)\,d d^{\prime}}\!(x,u^{\prime})\,
  \Gamma^{(\bar{\omega}\omega G)}{}^{d^{\prime}b}{}^{e^{\prime}}_{j^{\prime}}
  (u^{\prime}\!,y,u^{\prime\prime})
\end{equation}
Using the definition for the momentum space proper vertex
Eq.(\ref{DSE:defGamma_xyz}), we transform to momentum space
\begin{equation} \label{eq:DSE:Pi(omega)(k)}
  \Pi^{(\omega)\,ab}(k) =
  -\!\intDkp \frac{ig}{\kappa} f^{ade} k^{\prime j}\,
  G^{(AG)}{}^{ee^{\prime}}_{jj^{\prime}}(k-k^{\prime})\,
  G^{(\omega)\,d d^{\prime}}\!(k^{\prime})\,
  \Gamma^{(\bar{\omega}\omega G)}{}^{d^{\prime}b}{}^{e^{\prime}}_{j^{\prime}}
  (-k^{\prime}\!,k)
\end{equation}
\noindent
The structure of the \textsc{Dyson--Schwinger}
equation (\ref{eq:DSE:Pi(omega)(k)}) 
is illustrated in Fig.~\ref{fig:DSE(Omega)}.
In Eq.~(\ref{eq:DSE:Pi(omega)(k)}) the 
field index $G$ has a summation index taking the values $G=\lambda$
and $G=A$. In the graphical representation of
Eq.~(\ref{eq:DSE:Pi(omega)(k)}) such a summation is symbolised by a
solid line. This short-hand notation will become even more important in
the other \textsc{Dyson--Schwinger} equations to follow. 
Thus, the right-hand side of Fig.~\ref{fig:DSE(Omega)} stands for
two individual diagrams. 
%
%

\noindent
Above we have deduced the \textsc{Dyson--Schwinger} equation of
the gauge ghost self-energy from the general \emph{anti-ghost} equation
(\ref{eq:DSE-anti-ghost:Gamma}). A complementary relation can be
obtained from the \emph{ghost} equation (\ref{eq:DSE-ghost:Gamma}).
By taking the derivative with respect to $\bar{\omega}^b(y)$ of
Eq.~(\ref{eq:DSE-ghost:Gamma}), one obtains
\begin{equation} \label{eq:DSE:Pi(omega)(k)II}
  \Pi^{(\omega)\,ab}(k) =
  -\intDkp \frac{ig}{\kappa} f^{dbe} k^j\,
  G^{(GA)}{}^{e^{\prime}e}_{j^{\prime}j}(k-k^{\prime})\,
  G^{(\omega)\,d^{\prime}\!d}(k^{\prime})\,
  \Gamma^{(\bar{\omega}\omega G)}{}^{ad^{\prime}}{}^{e^{\prime}}_{j^{\prime}}
  (-k,k^{\prime})
\end{equation}
\subsubsection[DSE for $\Pi^{(\lambda\lambda)}$]
           {DSE for $\Pi^{(\lambda\lambda)}(k)$}

We come now to the \textsc{Dyson--Schwinger} equations of the 
gauge/auxiliary field sector. Taking the derivative with respect 
to $\lambda^{bj}(y)$ of the auxiliary field equation 
(\ref{eq:DSE-lambda:Gamma}) yields after setting the sources to zero 
\begin{eqnarray}
  & & \hspace{-5ex} 
  \frac{\delta^2\Gamma}{\delta\lambda^{ai}(x)\,\delta\lambda^{bj}(y)}
  \biggr|_{J=I=0}
  \!\! = \ 
  \overbrace{2\sigma T \,\delta^{ab}\,\delta^{ij}\,\delta(x-y)}^{
           \textstyle    
           (\Delta^{-1})^{(\lambda\lambda)}{}^{ab}_{ij}(x,y)}
  \;-\;
  \frac{ig^2}{3!} V^{acde}_{iklm} 
  \frac{\delta}{\delta\lambda^{bj}(y)}\,
  \frac{\delta^3 W}{\delta J_A^{ck}(x)\,\delta J_A^{dl}(x)\,
                    \delta J_A^{em}(x)}
  \biggr|_{J=I=0}
  \nonumber\\
 & & \hspace{20ex}
  -\frac{ig}{2!}\, f^{acd}
  \Bigl[
    \left(1-{\textstyle\frac{\sigma}{\kappa}}\right) \!
    \bigl[
      \delta^{ik} \partial_l^{\,\prime} \!
     -\delta^{il} \partial_k
    \bigr]
    + \,2\,
    \bigl[
      \delta^{ik} \partial_l \!
     -\delta^{il} \partial_k^{\,\prime}
    \bigr]
  \nonumber\\
 & & \hspace{32ex}
    + \,
    \bigl[
      \delta^{kl} \partial_i^{\,\prime} \!
      -\delta^{lk} \partial_i
    \bigr]\,
  \Bigr]
  \frac{\delta}{\delta\lambda^{bj}(y)}\,
  \frac{\delta^2 W}{\delta J_A^{ck}(x)\,\delta J_A^{dl}(x^{\prime})}
  \biggr|_{\!\!\!\begin{array}{lcl}
    \scriptstyle x^{\prime} = \,x \\[-1ex]
    \scriptstyle J \,=\, I \,=\, 0
  \end{array}}
\end{eqnarray}
Thus, comparing to Eq.~(\ref{def:self-energy(AL)}) one reads off the 
self-energy component
\begin{eqnarray}
  \Pi^{(\lambda\lambda)}{}^{ab}_{ij}(x,y) 
  \!\!\! &=& \!\!\!
  -\frac{ig}{2!}\, f^{acd}
  \Bigl[
    \left(1-{\textstyle\frac{\sigma}{\kappa}}\right) \!
    \bigl[
      \delta^{ik} \partial_l^{\,\prime} \!
     -\delta^{il} \partial_k
    \bigr]
    + \,2\,
    \bigl[
      \delta^{ik} \partial_l \!
     -\delta^{il} \partial_k^{\,\prime}
    \bigr]
  \nonumber\\
  & & \hspace{18ex}
    + \,
    \bigl[
      \delta^{kl} \partial_i^{\,\prime} \!
      -\delta^{lk} \partial_i
    \bigr]\,
  \Bigr]
  \frac{\delta}{\delta\lambda^{bj}(y)}\,
  \frac{\delta^2 W}{\delta J_A^{ck}(x)\,\delta J_A^{dl}(x^{\prime})}
  \biggr|_{\!\!\!\begin{array}{lcl}
    \scriptstyle x^{\prime} = \,x \\[-1ex]
    \scriptstyle J \,=\, I \,=\, 0
  \end{array}}
  \nonumber\\ \label{eq:Pi(LL):x-space}
  \!\!\! & & \!\!\!
  -\frac{ig^2}{3!} V^{acde}_{iklm} 
  \frac{\delta}{\delta\lambda^{bj}(y)}\,
  \frac{\delta^3 W}{\delta J_A^{ck}(x)\,\delta J_A^{dl}(x)\,
                    \delta J_A^{em}(x)}
  \biggr|_{J=I=0}
\end{eqnarray}

\noindent
To evaluate Eq.~(\ref{eq:Pi(LL):x-space}), we have to calculate the
remaining functional derivatives and finally transform into momentum
space. Let us start with the $\lambda$ derivative of the connected
two-point function. Because we will encounter similar expressions
also in the \textsc{Dyson--Schwinger} equations of the other
self-energy components, it is useful to generalise a bit and
do the work once and for all. 
Thus, with $F$, $G$ and $H$ chosen from the set $\{\lambda,A\}$, we 
find by means of the chain rule and using ghost number conservation, 
together with the identities (\ref{eq:DerivativesOfGamma})
\begin{displaymath}
  \frac{\delta}{\delta F^{bj}(y)}\,
  \frac{\delta^2 W}{\delta J_G^{ck}(x)\,\delta J_H^{dl}(x^{\prime})}
  \biggr|_{J=I=0} \!\!\! = 
  \int\!\!dv\,
  \biggl[
  \frac{\delta^2 \Gamma}{\delta F^{bj}(y)\,\delta K^{em}(v)}\,
  \frac{\delta^3 W}{\delta J_K^{em}(v)\,\delta J_G^{ck}(x)\,
                    \delta J_H^{dl}(x^{\prime})}
  \biggr]_{J=I=0}
\end{displaymath}
The field index $K$ in this equation is summed over the two 
values $\lambda$ and $A$. Expressing the connected three-point 
function by its one-particle irreducible counterpart
\begin{eqnarray} 
  W^{(FGH)}{}^{abc}_{ijk}(x,y,z) &=&
  -\int\!\!du\,du^{\prime} du^{\prime\prime}\;
  G^{(FF^{\prime})}{}^{aa^{\prime}}_{i\,i^{\prime}}(x,u)\,
  G^{(GG^{\prime})}{}^{b b^{\prime}}_{jj^{\prime}}(y,u^{\prime})\,
  \hspace{28.3ex}\nonumber\\ \label{eq:W3byGamma3}
  & & \hspace{15ex}
  G^{(HH^{\prime})}{}^{cc^{\prime}}_{kk^{\prime}}
     (z,u^{\prime\prime})\,
  \Gamma^{(F^{\prime}G^{\prime}H^{\prime})}{}^{a^{\prime}
           b^{\prime}c^{\prime}}_{
           i^{\prime}j^{\prime}k^{\prime}}
     (u,u^{\prime}\!,u^{\prime\prime})
\end{eqnarray}
and exploiting the relation
(\ref{eq:W2XGamma2=1}) then leads to the identity
\begin{eqnarray}
\lefteqn{
  \frac{\delta}{\delta F^{bj}(y)}\,
  \frac{\delta^2 W}{\delta J_G^{ck}(x)\,\delta J_H^{dl}(x^{\prime})}
  \biggr|_{J=I=0}
  =} \nonumber\\[1.5ex]
  & &
  -\int\!\!du^{\prime} du^{\prime\prime}\;
  G^{(GG^{\prime})}{}^{cc^{\prime}}_{kk^{\prime}}(x,u^{\prime})\,
  G^{(HH^{\prime})}{}^{dd^{\prime}}_{l\,l^{\prime}}
     (x^{\prime},u^{\prime\prime})\,
  \Gamma^{(FG^{\prime}H^{\prime})}{}^{bc^{\prime}d^{\prime}}_{jk^{\prime}
     l^{\prime}}
     (y,u^{\prime}\!,u^{\prime\prime})
\end{eqnarray}
Again, doubled field indices are summed over $\lambda$ and $A$
(which we will assume from now on in all relevant cases). Finally, 
transforming into momentum space and inserting the 
definition of the three-point vertex function 
(\ref{def:1PI-3-momentum}) yields 
\begin{eqnarray}
  \frac{\delta}{\delta F^{bj}(y)}\,
  \frac{\delta^2 W}{\delta J_G^{ck}(x)\,\delta J_H^{dl}(x^{\prime})}
  \biggr|_{J=I=0}
  \!\!\!\!\!\!\!\! &=& \!\!\!
  \int\!\frac{d^D\!k}{(2\pi)^D} \frac{d^D\!k^{\prime}}{(2\pi)^D}\,
  e^{-ik(x-y)} e^{ik^{\prime}(x-x^{\prime})}
  G^{(GG^{\prime})}{}^{cc^{\prime}}_{kk^{\prime}}
    (k-k^{\prime})\,
  \nonumber\\[1.5ex] \label{eq:3th-derivative-W} 
  & & \hspace{7ex}
  G^{(HH^{\prime})}{}^{dd^{\prime}}_{l\,l^{\prime}}
    (k^{\prime})\,
  \Gamma^{(FG^{\prime}H^{\prime})}{}^{bc^{\prime}d^{\prime}}_{jk^{\prime}
      l^{\prime}}
     (k^{\prime}\!-k,-k^{\prime})
\end{eqnarray}

\noindent
Analogously, one can derive a general expression for the fourth
functional derivative in Eq.~(\ref{eq:Pi(LL):x-space}). Using the chain rule 
as above, exploiting ghost number conservation and the identity
(\ref{eq:W2XGamma2=1}), translating connected into one-particle irreducible
quantities as in Eq.~(\ref{eq:W3byGamma3}) and finally 
introducing the momentum space vertex functions (\ref{def:1PI-3-momentum}) 
and (\ref{def:1PI-4-momentum}) leads to
\begin{eqnarray} 
\lefteqn{
  \frac{\delta}{\delta E^{bj}(y)}\,
  \frac{\delta^3 W}{\delta J_F^{ck}(x)\,\delta J_G^{dl}(x)\,
                    \delta J_H^{em}(x)} \biggr|_{J=I=0}
  =
  \intDk
  e^{-ik(x-y)}
  \int\!\frac{d^D\!k^{\prime}}{(2\pi)^D} 
        \frac{d^D\!k^{\prime\prime}}{(2\pi)^D} \biggl[}
  \nonumber\\[1.5ex]
  & & +\;
  G^{(FF^{\prime})}{}^{cc^{\prime}}_{kk^{\prime}}
     (k-k^{\prime})\,
  G^{(GG^{\prime})}{}^{dd^{\prime}}_{l\,l^{\prime}}
     (k^{\prime}\!-k^{\prime\prime})\,
  G^{(H\!H^{\prime})}{}^{ee^{\prime}}_{mm^{\prime}}
     (k^{\prime\prime})\,
  \Gamma^{(L^{\prime}G^{\prime}H^{\prime})}{}^{h^{\prime}
           d^{\prime}e^{\prime}}_{
           s^{\prime}l^{\prime}m^{\prime}}
     (k^{\prime\prime}\!-k^{\prime}\!, -k^{\prime\prime})
  \nonumber\\
  & & \hspace{9ex}
  G^{(L^{\prime}\!K^{\prime})}{}^{h^{\prime}g^{\prime}}_{
     s^{\prime}r^{\prime}}
     (k^{\prime})\,
  \Gamma^{(EF^{\prime}K^{\prime})}{}^{bc^{\prime}g^{\prime}}_{
           jk^{\prime}r^{\prime}}
     (k^{\prime}\!-k, -k^{\prime})
  \hspace{10ex}
  \nonumber\\ 
  & & +\;
  G^{(GF^{\prime})}{}^{dc^{\prime}}_{lk^{\prime}}
     (k-k^{\prime})\,
  G^{(HG^{\prime})}{}^{ed^{\prime}}_{ml^{\prime}}
     (k^{\prime}\!-k^{\prime\prime})\,
  G^{(F\!H^{\prime})}{}^{ce^{\prime}}_{km^{\prime}}
     (k^{\prime\prime})\,
  \Gamma^{(L^{\prime}G^{\prime}H^{\prime})}{}^{h^{\prime}d^{\prime}
           e^{\prime}}_{
           s^{\prime}l^{\prime}m^{\prime}}
     (k^{\prime\prime}\!-k^{\prime}\!, -k^{\prime\prime})
  \nonumber\\
  & & \hspace{9ex}
  G^{(L^{\prime}\!K^{\prime})}{}^{h^{\prime}g^{\prime}}_{
     s^{\prime}r^{\prime}}
     (k^{\prime})\,
  \Gamma^{(EF^{\prime}K^{\prime})}{}^{
           bc^{\prime}g^{\prime}}_{
           jk^{\prime}r^{\prime}}
     (k^{\prime}\!-k, -k^{\prime})
  \hspace{10ex}
  \nonumber\\ 
  & & +\;
  G^{(H\!F^{\prime})}{}^{ec^{\prime}}_{mk^{\prime}}
     (k-k^{\prime})\,
  G^{(FG^{\prime})}{}^{cd^{\prime}}_{kl^{\prime}}
     (k^{\prime}\!-k^{\prime\prime})\,
  G^{(G\!H^{\prime})}{}^{de^{\prime}}_{lm^{\prime}}
     (k^{\prime\prime})\,
  \Gamma^{(L^{\prime}G^{\prime}H^{\prime})}{}^{h^{\prime}d^{\prime}
           e^{\prime}}_{
           s^{\prime}l^{\prime}m^{\prime}}
     (k^{\prime\prime}\!-k^{\prime}\!, -k^{\prime\prime})
  \nonumber\\
  & & \hspace{9ex}
  G^{(L^{\prime}\!K^{\prime})}{}^{h^{\prime}g^{\prime}}_{
     s^{\prime}r^{\prime}}
     (k^{\prime})\,
  \Gamma^{(EF^{\prime}K^{\prime})}{}^{
           bc^{\prime}g^{\prime}}_{
           jk^{\prime}r^{\prime}}
     (k^{\prime}\!-k, -k^{\prime})
  \hspace{10ex}
  \nonumber\\ 
  & & + \;
  G^{(FF^{\prime})}{}^{cc^{\prime}}_{kk^{\prime}}
     (k-k^{\prime}\!-k^{\prime\prime})\,
  G^{(GG^{\prime})}{}^{dd^{\prime}}_{l\,l^{\prime}}
     (k^{\prime})\,
  \nonumber\\ \label{eq:4th-derivative-W} 
  & & \hspace{9ex}
  G^{(HH^{\prime})}{}^{ee^{\prime}}_{mm^{\prime}}
     (k^{\prime\prime})\,
  \Gamma^{(EF^{\prime}G^{\prime}H^{\prime})}{}^{
           bc^{\prime}d^{\prime}e^{\prime}}_{
           jk^{\prime}l^{\prime}m^{\prime}}
  (k^{\prime}\! + k^{\prime\prime}\!-k, -k^{\prime}, -k^{\prime\prime})\,
  \biggr]
\end{eqnarray} 

\begin{figure}[t]
\begin{center}
  \begin{picture}(180,70) 
  \put(-5.5,52.5){\usebox{\PiLL}}
  \put(29.5,54.1){\huge $=$}
  \put(39.5,54.1){\huge $\frac{1}{2}$}
  \put(46,40){\usebox{\DSELLLP}}
  \put(-10.25,14.1){\huge $+\,\frac{1}{2}$}
  \put(1.75,0){\usebox{\DSELLG}}
  \put(85.75,14.1){\huge $+\,\frac{1}{6}$}
  \put(98.25,0){\usebox{\DSELLSS}}
  \end{picture}
  \caption{\label{fig:DSE(LL)}
    {
    \textsc{Dyson--Schwinger} equation of the $\Pi^{(\lambda\lambda)}$ 
    self-energy component, Eq.~(\ref{eq:Pi(LL)Result}).
    }}
\end{center} 
\end{figure}
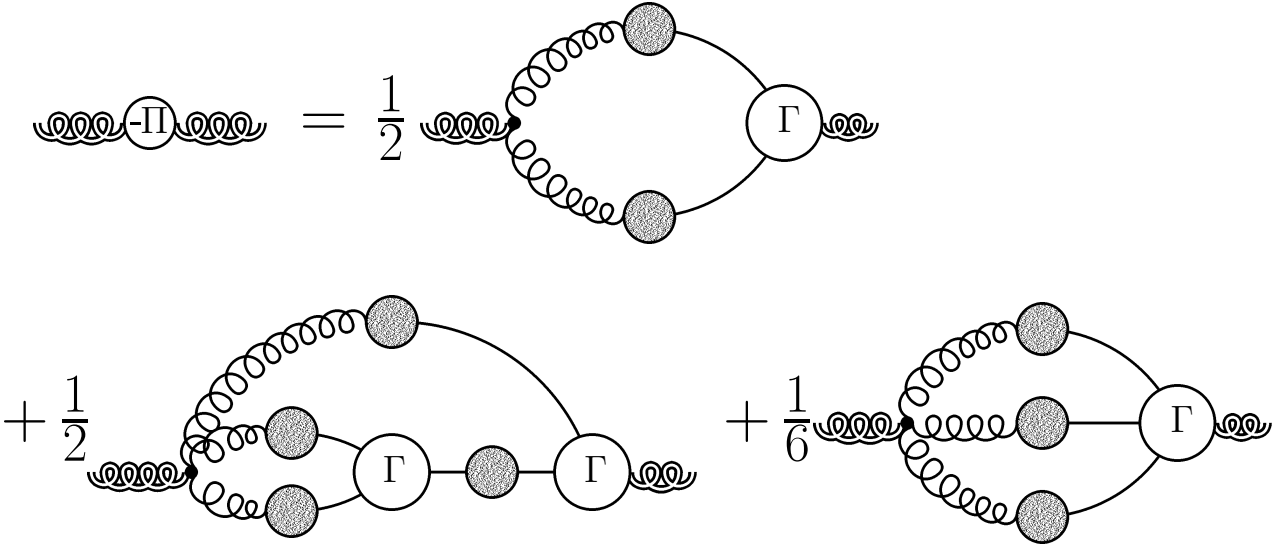

Exploiting the identities (\ref{eq:3th-derivative-W}) and 
(\ref{eq:4th-derivative-W}) one can now readily obtain
the \textsc{Dyson--Schwinger} equation of the $\Pi^{(\lambda\lambda)}$
self-energy component from Eq.~(\ref{eq:Pi(LL):x-space}). One finds
\begin{eqnarray} 
%
%
\lefteqn{
  \Pi^{(\lambda\lambda)}{}^{ab}_{ij}(k) 
  =
  -\frac{1}{2}
  \intDkp (-g) V^{acd}_{i\,kl}
                 (\myvec{k}-\myvec{k}^{\prime}, \myvec{k}^{\prime})\,
  G^{(AG^{\prime})}{}^{cc^{\prime}}_{kk^{\prime}}
     (k-k^{\prime})\,
  G^{(AH^{\prime})}{}^{dd^{\prime}}_{l\,l^{\prime}}
     (k^{\prime})\,
  }
  \nonumber\\
  & & \hspace{45.0ex}
  \Gamma^{(\lambda G^{\prime}H^{\prime})}{}^{
           bc^{\prime}d^{\prime}}_{jk^{\prime}l^{\prime}}
     (k^{\prime}\!-k, -k^{\prime})
  \nonumber\\ 
  & &
  -\frac{1}{2}
  \int\!\frac{d^D\!k^{\prime}}{(2\pi)^D} 
        \frac{d^D\!k^{\prime\prime}}{(2\pi)^D}\;
  ig^2 V^{acde}_{iklm}\;
  G^{(AF^{\prime})}{}^{cc^{\prime}}_{kk^{\prime}}
     (k-k^{\prime})\,
  G^{(AG^{\prime})}{}^{dd^{\prime}}_{l\,l^{\prime}}
     (k^{\prime}\!-k^{\prime\prime})\,
  G^{(AH^{\prime})}{}^{ee^{\prime}}_{mm^{\prime}}
     (k^{\prime\prime})
  \nonumber\\
  & & \hspace{17ex}
  \Gamma^{(L^{\prime}G^{\prime}H^{\prime})}{}^{
           h^{\prime}d^{\prime}e^{\prime}}_{s^{\prime}l^{\prime}m^{\prime}}
     (k^{\prime\prime}\!-k^{\prime}\!, -k^{\prime\prime})\,
  G^{(L^{\prime}\!K^{\prime})}{}^{h^{\prime}
      g^{\prime}}_{s^{\prime}\hspace{0.0pt}r^{\prime}}
     (k^{\prime})\,
  \Gamma^{(\lambda F^{\prime}K^{\prime})}{}^{
           bc^{\prime}g^{\prime}}_{jk^{\prime}r^{\prime}}
     (k^{\prime}\!-k, -k^{\prime})
  \nonumber\\ 
  & &
  -\frac{1}{6}
  \int\!\frac{d^D\!k^{\prime}}{(2\pi)^D} 
        \frac{d^D\!k^{\prime\prime}}{(2\pi)^D}\;
  ig^2 V^{acde}_{iklm}\;
  G^{(AF^{\prime})}{}^{cc^{\prime}}_{kk^{\prime}}
     (k-k^{\prime}\!-k^{\prime\prime})\,
  G^{(AG^{\prime})}{}^{dd^{\prime}}_{l\,l^{\prime}}
     (k^{\prime})\,
  G^{(AH^{\prime})}{}^{ee^{\prime}}_{mm^{\prime}}
     (k^{\prime\prime})
  \nonumber\\ \label{eq:Pi(LL)Result} 
  & & \hspace{17ex}
  \Gamma^{(\lambda F^{\prime}G^{\prime}H^{\prime})}{}^{
           bc^{\prime}d^{\prime}e^{\prime}}_{jk^{\prime}l^{\prime}m^{\prime}}
  (k^{\prime}\! + k^{\prime\prime}\!-k, -k^{\prime}, -k^{\prime\prime})
\end{eqnarray} 
where we have used the symmetry of the vertex 
$V^{acde}_{iklm}$ in the last three pairs of indices to 
combine the first three terms arising from Eq.~(\ref{eq:4th-derivative-W}) 
into one. We have illustrated Eq.~(\ref{eq:Pi(LL)Result}) 
in Fig.~\ref{fig:DSE(LL)}.

%
%

\subsubsection[DSE for $\Pi^{(\lambda A)}$]{DSE for $\Pi^{(\lambda A)}(k)$}

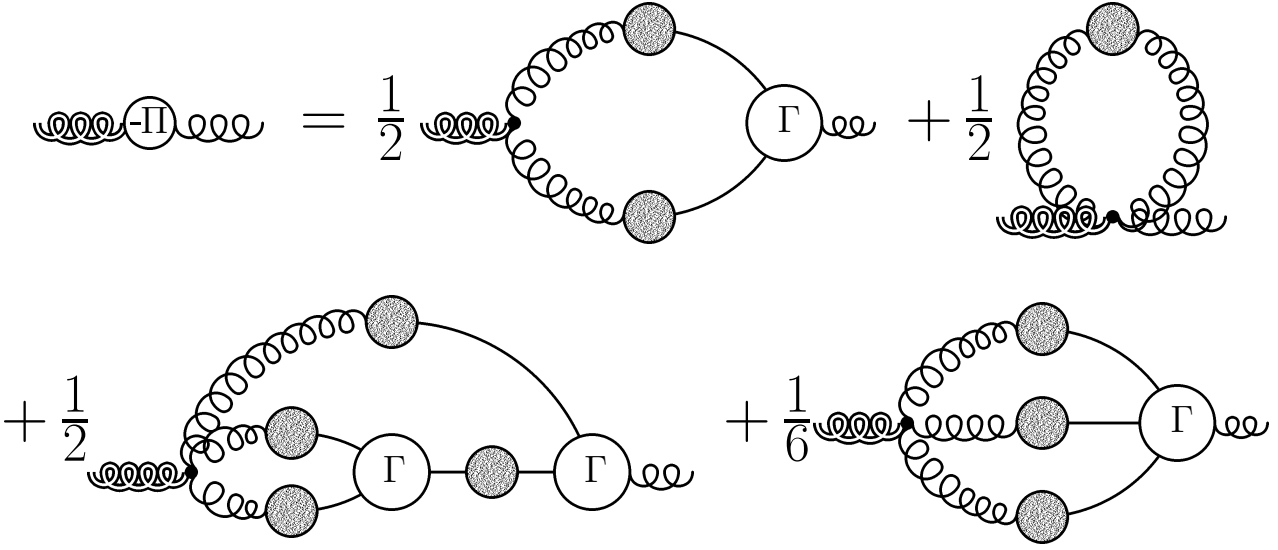
\begin{figure}[t]
\begin{center}
  \begin{picture}(180,70) 
  \put(-5.5,52.5){\usebox{\PiLA}}
  \put(29.5,54.1){\huge $=$}
  \put(39.5,54.1){\huge $\frac{1}{2}$}
  \put(46,40){\usebox{\DSELALP}}
  \put(110,54.1){\huge $+\,\frac{1}{2}$}
  \put(122.5,40){\usebox{\DSELATP}}
  \put(-10.25,14.1){\huge $+\,\frac{1}{2}$}
  \put(1.75,0){\usebox{\DSELAG}}
  \put(85.75,14.1){\huge $+\,\frac{1}{6}$}
  \put(98.25,0){\usebox{\DSELASS}}
  \end{picture}
  \caption{\label{fig:DSE(LA)}
    {
    \textsc{Dyson--Schwinger} equation of the $\Pi^{(\lambda A)}$ self-energy 
    component, Eq.~(\ref{eq:Pi(LA)Result}).
    }}
\end{center} 
\end{figure}

Taking the derivative of Eq.~(\ref{eq:DSE-lambda:Gamma}) with respect 
to $A^{bj}(y)$ instead of $\lambda^{bj}(y)$ and afterwards setting the 
sources to zero leads to the \textsc{Dyson--Schwinger} equation for the 
$\Pi^{(\lambda A)}$ self-energy component, namely
\begin{eqnarray}
  \frac{\delta^2\Gamma}{\delta\lambda^{ai}(x)\,\delta A^{bj}(y)}
  \biggr|_{J=I=0}
  \!\!\! &=& \!\!\!
  \overbrace{
    -i \,\delta^{ab}
    \left[
      (+\sigma\partial_t - \Delta) \,\delta_{ij} +
      (1 -{\textstyle\frac{\sigma}{\kappa}})\,\partial_i\partial_j
    \right] \delta(x-y)}^{\textstyle    
      (\Delta^{-1})^{(\lambda A)}{}^{ab}_{ij}(x,y)}
  \nonumber\\
  \!\!\! & & \!\!\!
  -\frac{ig}{2!}\, f^{acd}
  \Bigl[
    \left(1-{\textstyle\frac{\sigma}{\kappa}}\right) \!
    \bigl[
      \delta^{ik} \partial_l^{\,\prime} \!
     -\delta^{il} \partial_k
    \bigr]
    + \,2\,
    \bigl[
      \delta^{ik} \partial_l \!
     -\delta^{il} \partial_k^{\,\prime}
    \bigr]
  \nonumber\\
  \!\!\! & & \!\!\! \hspace{12ex}
    + \,
    \bigl[
      \delta^{kl} \partial_i^{\,\prime} \!
      -\delta^{lk} \partial_i
    \bigr]\,
  \Bigr]
  \frac{\delta}{\delta A^{bj}(y)}\,
  \frac{\delta^2 W}{\delta J_A^{ck}(x)\,\delta J_A^{dl}(x^{\prime})}
  \biggr|_{\!\!\!\begin{array}{lcl}
    \scriptstyle x^{\prime} = \,x \\[-1ex]
    \scriptstyle J \,=\, I \,=\, 0
  \end{array}}
  \nonumber\\
  \!\!\! & & \!\!\!
  - \frac{ig^2}{3!} V^{acde}_{iklm} 
  \frac{\delta}{\delta A^{bj}(y)}\,
  \frac{\delta^3 W}
       {\delta J_A^{ck}(x)\,\delta J_A^{dl}(x)\,\delta J_A^{em}(x)}
  \biggr|_{J=I=0}
  \nonumber\\
  \!\!\! & & \!\!\!
  -\frac{ig^2}{2!} V^{acdb}_{iklj}
  \,\delta(x-y)\,
  \frac{\delta^2 W}{\delta J_A^{ck}(x)\,\delta J_A^{dl}(x)}
  \biggr|_{J=I=0}
\end{eqnarray}
Reading off the self-energy component by comparing with
Eq.~(\ref{def:self-energy(AL)}), and
using Eqs.~(\ref{eq:3th-derivative-W})--(\ref{eq:4th-derivative-W}) one
arrives at
\begin{eqnarray} 
%
%
\lefteqn{
  \Pi^{(\lambda A)}{}^{ab}_{ij}(k) 
  =
  -\frac{1}{2}
  \intDkp (-g) V^{acd}_{i\,kl}
                 (\myvec{k}-\myvec{k}^{\prime}, \myvec{k}^{\prime})\,
  G^{(AG^{\prime})}{}^{cc^{\prime}}_{kk^{\prime}}
     (k-k^{\prime})\,
  G^{(AH^{\prime})}{}^{dd^{\prime}}_{l\,l^{\prime}}
     (k^{\prime})\,
  }
  \nonumber\\
  & & \hspace{55.0ex}
  \Gamma^{(A G^{\prime}H^{\prime})}{}^{
           bc^{\prime}d^{\prime}}_{jk^{\prime}l^{\prime}}
     (k^{\prime}\!-k, -k^{\prime})
  \nonumber\\ 
& &
  -\frac{1}{2}
  \int\!\frac{d^D\!k^{\prime}}{(2\pi)^D} 
        \frac{d^D\!k^{\prime\prime}}{(2\pi)^D}\;
  ig^2 V^{acde}_{iklm}\;
  G^{(AF^{\prime})}{}^{cc^{\prime}}_{kk^{\prime}}
     (k-k^{\prime})\,
  G^{(AG^{\prime})dd^{\prime}}_{l\,l^{\prime}}
     (k^{\prime}\!-k^{\prime\prime})\,
  G^{(AH^{\prime})}{}^{ee^{\prime}}_{mm^{\prime}}
     (k^{\prime\prime})
  \nonumber\\
  & & \hspace{17ex}
  \Gamma^{(L^{\prime}G^{\prime}H^{\prime})}{}^{
           h^{\prime}d^{\prime}e^{\prime}}_{s^{\prime}l^{\prime}m^{\prime}}
     (k^{\prime\prime}\!-k^{\prime}\!, -k^{\prime\prime})\,
  G^{(L^{\prime}\!K^{\prime})}{}^{h^{\prime}g^{\prime}}_{s^{\prime}r^{\prime}}
     (k^{\prime})\,
  \Gamma^{(A F^{\prime}K^{\prime})}{}^{
           bc^{\prime}g^{\prime}}_{jk^{\prime}r^{\prime}}
     (k^{\prime}\!-k, -k^{\prime})
  \nonumber\\ 
  & &
  -\frac{1}{6}
  \int\!\frac{d^D\!k^{\prime}}{(2\pi)^D} 
        \frac{d^D\!k^{\prime\prime}}{(2\pi)^D}\;
  ig^2 V^{acde}_{iklm}\;
  G^{(AF^{\prime})}{}^{cc^{\prime}}_{kk^{\prime}}
     (k-k^{\prime}\!-k^{\prime\prime})\,
  G^{(AG^{\prime})}{}^{dd^{\prime}}_{l\,l^{\prime}}
     (k^{\prime})\,
  G^{(AH^{\prime})}{}^{ee^{\prime}}_{mm^{\prime}}
     (k^{\prime\prime})
  \nonumber\\
  & & \hspace{17ex}
  \Gamma^{(A F^{\prime}G^{\prime}H^{\prime})}{}^{
           bc^{\prime}d^{\prime}e^{\prime}}_{jk^{\prime}l^{\prime}m^{\prime}}
  (k^{\prime}\! + k^{\prime\prime}\!-k, -k^{\prime}, -k^{\prime\prime})
  \nonumber\\ \label{eq:Pi(LA)Result} 
  & &
  -\frac{1}{2} \intDkp
  ig^2 V^{abcd}_{ijkl}\;
  G^{(AA)}{}^{cd}_{kl}(k^{\prime})
\end{eqnarray} 
which is depicted in Fig.~\ref{fig:DSE(LA)}.

%
%
\subsubsection[DSE for $\Pi^{(A\lambda)}$]{DSE for $\Pi^{(A\lambda)}(k)$}

From the gauge field equation (\ref{eq:DSE-A:Gamma}) one obtains 
by taking the derivative with respect to $\lambda^{bj}(y)$ for
vanishing sources
\vspace{-3ex}
\begin{eqnarray}\label{eq:Pi(Alambda)}
  \frac{\delta^2\Gamma}{\delta A^{ai}(x)\,\delta\lambda^{bj}(y)}
  \biggr|_{J=I=0}
  \!\!\! &=& \!\!\!
  \overbrace{
    -i \,\delta^{ab}
    \left[
      (-\sigma\partial_t - \Delta) \,\delta_{ij} +
      (1 -{\textstyle\frac{\sigma}{\kappa}})\,\partial_i\partial_j
    \right] \delta(x-y)}^{\textstyle    
      (\Delta^{-1})^{(A\lambda)}{}^{ab}_{ij}(x,y)}
  \nonumber\\
  \!\!\! & & \!\!\!
  -ig f^{acd}
  \Bigl[
    -\!\left(1-{\textstyle\frac{\sigma}{\kappa}}\right) \!
    \bigl[
      \delta^{ik} \partial_l^{\,\prime} \!
     +\delta^{kl} (\partial_i +  \partial_i^{\,\prime})
    \bigr]
    + \,2\,
    \bigl[
      \delta^{kl} \partial_i^{\,\prime} \!
     +\delta^{ik} (\partial_l + \partial_l^{\,\prime})
    \bigr]
  \nonumber\\
  & &  \hspace{10.0ex}
    - \,
    \bigl[
      \delta^{il} \partial_k^{\,\prime} \!
     +\delta^{il} (\partial_k + \partial_k^{\,\prime})
    \bigr]\,
  \Bigr]
  \frac{\delta}{\delta\lambda^{bj}(y)}\,
  \frac{\delta^2 W}{\delta J_{\lambda}^{ck}(x)\,\delta J_A^{dl}(x^{\prime})}
  \biggr|_{\!\!\!\begin{array}{lcl}
    \scriptstyle x^{\prime} = \,x \\[-1ex]
    \scriptstyle J \,=\, I \,=\, 0
  \end{array}}
  \nonumber\\[-3ex]
  & & \!\!\!
  - \frac{ig^2}{2!} V^{eacd}_{mikl} 
  \frac{\delta}{\delta\lambda^{bj}(y)}\,
  \frac{\delta^3 W}
       {\delta J_\lambda^{em}(x)\,\delta J_A^{ck}(x)\,\delta J_A^{dl}(x)}
  \biggr|_{J=I=0}
  \nonumber\\
  \!\!\! & & \!\!\!
  -\frac{ig^2}{2!} V^{bacd}_{jikl}
  \,\delta(x-y)\,
  \frac{\delta^2 W}{\delta J_A^{ck}(x)\,\delta J_A^{dl}(x)}
  \biggr|_{J=I=0}
  \nonumber\\
  \!\!\! & & \!\!\!
  -\frac{g}{\kappa} f^{acd} \,\partial_i \,
  \frac{\delta}{\delta\lambda^{bj}(y)}\,
  \frac{\delta^2 W}{\delta J_{\bar{\omega}}^{c}(x^{\prime})\,
                    \delta J_{\omega}^d(x)}
  \biggr|_{\!\!\!\begin{array}{lcl}
    \scriptstyle x^{\prime} = \,x \\[-1ex]
    \scriptstyle J \,=\, I \,=\, 0
  \end{array}}
\end{eqnarray}
As in Eq.~(\ref{eq:Pi(LA)Result}, we have again a self-energy, a tadpole,
and terms of the type in
Eqs.~(\ref{eq:3th-derivative-W})--(\ref{eq:4th-derivative-W}), 
but we also have a new term involving gauge ghosts. It can be calculated
in a similar way to the previous cases, and comes out to
\begin{eqnarray}
  \frac{\delta}{\delta E^{bj}(y)}\,
  \frac{\delta^2 W}{\delta J_{\bar{\omega}}^c(x^{\prime})\,
                    \delta J_{\omega}^d(x)}
  \biggr|_{J=I=0}
  \!\!\!\!\!\!\!\! &=& \!\!\!
  -\int\!\frac{d^D\!k}{(2\pi)^D} \frac{d^D\!k^{\prime}}{(2\pi)^D}\,
  e^{-ik(x-y)} e^{ik^{\prime}(x-x^{\prime})}\,
  G^{(\omega)\,c^{\prime}\!c}(-k^{\prime})\,
  \nonumber\\
  & & \hspace{7ex}
  G^{(\omega)\,dd^{\prime}}(k-k^{\prime})\,
  \Gamma^{(\bar{\omega}\omega E)}{}^{
     d^{\prime}\!c^{\prime}}{}^{b}_{j}
     (k^{\prime}\!-k,-k^{\prime})
\end{eqnarray}
With this, and the previous identities, Eq.~(\ref{eq:Pi(Alambda)}) can
be written as
\begin{eqnarray} 
%
%
\lefteqn{
  \Pi^{(A\lambda)}{}^{ab}_{ij}(k) 
  =
  -\intDkp (-g) V^{cda}_{kli}
                 (\myvec{k}^{\prime}, -\myvec{k})\,
  G^{(\lambda A)}{}^{cc^{\prime}}_{kk^{\prime}}
     (k-k^{\prime})\,
  G^{(AH^{\prime})}{}^{dd^{\prime}}_{l\,l^{\prime}}
     (k^{\prime})\,
  }
  \nonumber\\
  & & \hspace{60.2ex}
  \Gamma^{(\lambda AH^{\prime})}{}^{
           bc^{\prime}d^{\prime}}_{jk^{\prime}l^{\prime}}
     (k^{\prime}\!-k, -k^{\prime})
  \nonumber\\ 
  & &
  \hspace{1.7ex} -\int\!\frac{d^D\!k^{\prime}}{(2\pi)^D} 
        \frac{d^D\!k^{\prime\prime}}{(2\pi)^D}\;
  ig^2 V^{eacd}_{mikl}\;
  G^{(AF^{\prime})}{}^{cc^{\prime}}_{kk^{\prime}}
     (k-k^{\prime})\,
  G^{(AG^{\prime})}{}^{dd^{\prime}}_{l\,l^{\prime}}
     (k^{\prime}\!-k^{\prime\prime})\,
  G^{(\lambda A)}{}^{ee^{\prime}}_{mm^{\prime}}
     (k^{\prime\prime})
  \nonumber\\
  & & \hspace{17ex}
  \Gamma^{(L^{\prime}G^{\prime}A)}{}^{
           h^{\prime}d^{\prime}e^{\prime}}_{s^{\prime}l^{\prime}m^{\prime}}
     (k^{\prime\prime}\!-k^{\prime}\!, -k^{\prime\prime})\,
  G^{(L^{\prime}\!K^{\prime})}{}^{h^{\prime}g^{\prime}}_{s^{\prime}r^{\prime}}
     (k^{\prime})\,
  \Gamma^{(\lambda F^{\prime}K^{\prime})}{}^{
           bc^{\prime}g^{\prime}}_{jk^{\prime}r^{\prime}}
     (k^{\prime}\!-k, -k^{\prime})
  \nonumber\\ 
  & &
  -\frac{1}{2}
  \int\!\frac{d^D\!k^{\prime}}{(2\pi)^D} 
        \frac{d^D\!k^{\prime\prime}}{(2\pi)^D}\;
  ig^2 V^{eacd}_{mikl}\;
  G^{(\lambda A)}{}^{ec^{\prime}}_{mk^{\prime}}
     (k-k^{\prime})\,
  G^{(AG^{\prime})}{}^{cd^{\prime}}_{kl^{\prime}}
     (k^{\prime}\!-k^{\prime\prime})\,
  G^{(AH^{\prime})}{}^{de^{\prime}}_{lm^{\prime}}
     (k^{\prime\prime})
  \nonumber\\
  & & \hspace{16.7ex}
  \Gamma^{(L^{\prime}G^{\prime}H^{\prime})}{}^{
           h^{\prime}d^{\prime}e^{\prime}}_{s^{\prime}l^{\prime}m^{\prime}}
     (k^{\prime\prime}\!-k^{\prime}\!, -k^{\prime\prime})\,
  G^{(L^{\prime}\!K^{\prime})}{}^{h^{\prime}g^{\prime}}_{s^{\prime}r^{\prime}}
     (k^{\prime})\,
  \Gamma^{(\lambda AK^{\prime})}{}^{
           bc^{\prime}g^{\prime}}_{jk^{\prime}r^{\prime}}
     (k^{\prime}\!-k, -k^{\prime})
  \nonumber\\ 
  & &
  -\frac{1}{2}
  \int\!\frac{d^D\!k^{\prime}}{(2\pi)^D} 
        \frac{d^D\!k^{\prime\prime}}{(2\pi)^D}\;
  ig^2 V^{eacd}_{mikl}\;
  G^{(AF^{\prime})}{}^{cc^{\prime}}_{kk^{\prime}}
     (k-k^{\prime}\!-k^{\prime\prime})\,
  G^{(AG^{\prime})}{}^{dd^{\prime}}_{l\,l^{\prime}}
     (k^{\prime})\,
  G^{(\lambda A)}{}^{ee^{\prime}}_{mm^{\prime}}
     (k^{\prime\prime})
  \nonumber\\
  & & \hspace{46.3ex}
  \Gamma^{(\lambda F^{\prime}G^{\prime}\!A)}{}^{
           bc^{\prime}d^{\prime}e^{\prime}}_{jk^{\prime}l^{\prime}m^{\prime}}
  (k^{\prime}\! + k^{\prime\prime}\!-k, -k^{\prime}, -k^{\prime\prime})
  \nonumber\\ 
  & &
  \hspace{1.7ex}+ \intDkp
  \frac{ig}{\kappa}\,f^{cda} (k-k^{\prime})^i
  G^{(\omega)\,c^{\prime}\!c}(-k^{\prime})\,
  G^{(\omega)\,dd^{\prime}}(k-k^{\prime})\,
  \Gamma^{(\bar{\omega}\omega\lambda)}{}^{d^{\prime}c^{\prime}}{}^{b}_{j}
     (k^{\prime}\!-k,-k^{\prime})
  \nonumber\\ \label{eq:Pi(AL)Result} 
  & &
  -\frac{1}{2} \intDkp
  ig^2 V^{bacd}_{jikl}\;
  G^{(AA)}{}^{cd}_{kl}(k^{\prime})
\end{eqnarray} 
A graphical representation of this identity can be found in 
Fig.~\ref{fig:DSE(AL)}.

%
%
\begin{figure}[t]
\begin{center}
  \begin{picture}(180,110) 
  \put(-0.5,92.5){\usebox{\PiAL}}
  \put(34.5,94.1){\huge $=$}
  \put(44.5,80){\usebox{\DSEALLP}}
  \put(108.5,94.1){\huge $+\,\frac{1}{2}$}
  \put(121,80){\usebox{\DSEALTP}}
  \put(-16,54.1){\huge $+$}
  \put(-10,40){\usebox{\DSEALG}}
  \put(72,54.1){\huge $+\,\frac{1}{2}$}
  \put(84,40){\usebox{\DSEALH}}
  \put(0,14.1){\huge $+\,\frac{1}{2}$}
  \put(12.5,0){\usebox{\DSEALSS}}
  \put(80,14.1){\huge $-$}
  \put(87.5,0){\usebox{\DSEALLG}}
  \end{picture}
  \caption{\label{fig:DSE(AL)}
    {
    \textsc{Dyson--Schwinger} equation of the $\Pi^{(A\lambda)}$ self-energy 
    component, Eq.~(\ref{eq:Pi(AL)Result}).
    }}
\end{center} 
\end{figure} 

\subsubsection[DSE for $\Pi^{(AA)}$]{DSE for $\Pi^{(AA)}(k)$}

Finally, we come to the pure gauge field component $\Pi^{(AA)}$. 
Because 
\begin{math}
  (\Delta^{-1})^{(AA)}{}^{ab}_{ij}=0
\end{math},
one has in this case
\begin{equation}
  \Pi^{(AA)}{}^{ab}_{ij}(x,y) =
  \frac{\delta^2\Gamma}{\delta A^{ai}(x)\,\delta A^{bj}(y)}
  \biggr|_{J=I=0}
\end{equation}
and thus one obtains from Eq.~(\ref{eq:DSE-A:Gamma}) the final identity
\begin{eqnarray} 
%
%
\lefteqn{
  \Pi^{(AA)}{}^{ab}_{ij}(k) 
  =
  -\intDkp (-g) V^{cda}_{kli}
                 (\myvec{k}^{\prime}, -\myvec{k})\,
  G^{(\lambda A)}{}^{cc^{\prime}}_{kk^{\prime}}
     (k-k^{\prime})\,
  G^{(A\lambda)}{}^{dd^{\prime}}_{l\,l^{\prime}}
     (k^{\prime})\,
  }
  \nonumber\\
  & & \hspace{61.4ex}
  \Gamma^{(AA\lambda)}{}^{
           bc^{\prime}d^{\prime}}_{jk^{\prime}l^{\prime}}
     (k^{\prime}\!-k, -k^{\prime})
  \nonumber\\ 
  & &
  \hspace{1.7ex} -\int\!\frac{d^D\!k^{\prime}}{(2\pi)^D} 
        \frac{d^D\!k^{\prime\prime}}{(2\pi)^D}\;
  ig^2 V^{eacd}_{mikl}\;
  G^{(AF^{\prime})}{}^{cc^{\prime}}_{kk^{\prime}}
     (k-k^{\prime})\,
  G^{(AG^{\prime})}{}^{dd^{\prime}}_{l\,l^{\prime}}
     (k^{\prime}\!-k^{\prime\prime})\,
  G^{(\lambda A)}{}^{ee^{\prime}}_{mm^{\prime}}
     (k^{\prime\prime})
  \nonumber\\
  & & \hspace{17ex}
  \Gamma^{(L^{\prime}G^{\prime}A)}{}^{
           h^{\prime}d^{\prime}e^{\prime}}_{s^{\prime}l^{\prime}m^{\prime}}
     (k^{\prime\prime}\!-k^{\prime}\!, -k^{\prime\prime})\,
  G^{(L^{\prime}\!K^{\prime})}{}^{h^{\prime}g^{\prime}}_{s^{\prime}r^{\prime}}
     (k^{\prime})\,
  \Gamma^{(AF^{\prime}K^{\prime})}{}^{
           bc^{\prime}g^{\prime}}_{jk^{\prime}r^{\prime}}
     (k^{\prime}\!-k, -k^{\prime})
  \nonumber\\ 
& &
  -\frac{1}{2}
  \int\!\frac{d^D\!k^{\prime}}{(2\pi)^D} 
        \frac{d^D\!k^{\prime\prime}}{(2\pi)^D}\;
  ig^2 V^{eacd}_{mikl}\;
  G^{(\lambda A)}{}^{ec^{\prime}}_{mk^{\prime}}
     (k-k^{\prime})\,
  G^{(AG^{\prime})}{}^{cd^{\prime}}_{kl^{\prime}}
     (k^{\prime}\!-k^{\prime\prime})\,
  G^{(AH^{\prime})}{}^{de^{\prime}}_{lm^{\prime}}
     (k^{\prime\prime})
  \nonumber\\
  & & \hspace{19.8ex}
  \Gamma^{(AG^{\prime}H^{\prime})}{}^{h^{\prime}d^{\prime}
      e^{\prime}}_{s^{\prime}l^{\prime}m^{\prime}}
     (k^{\prime\prime}\!-k^{\prime}\!, -k^{\prime\prime})\,
  G^{(A\lambda)}{}^{h^{\prime}g^{\prime}}_{s^{\prime}r^{\prime}}
     (k^{\prime})\,
  \Gamma^{(AA\lambda)}{}^{
           bc^{\prime}g^{\prime}}_{jk^{\prime}r^{\prime}}
     (k^{\prime}\!-k, -k^{\prime})
  \nonumber\\ 
  & &
  -\frac{1}{2}
  \int\!\frac{d^D\!k^{\prime}}{(2\pi)^D} 
        \frac{d^D\!k^{\prime\prime}}{(2\pi)^D}\;
  ig^2 V^{eacd}_{mikl}\;
  G^{(AF^{\prime})}{}^{cc^{\prime}}_{kk^{\prime}}
     (k-k^{\prime}\!-k^{\prime\prime})\,
  G^{(AG^{\prime})}{}^{dd^{\prime}}_{l\,l^{\prime}}
     (k^{\prime})\,
  G^{(\lambda A)}{}^{ee^{\prime}}_{mm^{\prime}}
     (k^{\prime\prime})
  \nonumber\\
  & & \hspace{46.5ex}
  \Gamma^{(AF^{\prime}G^{\prime}\!A)}{}^{
           bc^{\prime}d^{\prime}e^{\prime}}_{jk^{\prime}l^{\prime}m^{\prime}}
  (k^{\prime}\! + k^{\prime\prime}\!-k, -k^{\prime}, -k^{\prime\prime})
  \nonumber\\ 
  & &
  \hspace{1.7ex}+ \intDkp
  \frac{ig}{\kappa}\,f^{cda} (k-k^{\prime})^i
  G^{(\omega)\,c^{\prime}\!c}(-k^{\prime})\,
  G^{(\omega)\,dd^{\prime}}(k-k^{\prime})\,
  \Gamma^{(\bar{\omega}\omega A)}{}^{
     d^{\prime}\!c^{\prime}}{}^{b}_{j}
     (k^{\prime}\!-k,-k^{\prime})
  \nonumber\\ \label{eq:Pi(AA)Result}
  & &
  \hspace{1.7ex}- \intDkp
  ig^2 V^{eacb}_{mikj}\;
  G^{(\lambda A)}{}^{ec}_{mk}(k^{\prime})
\end{eqnarray} 

%
%
\begin{figure}[t]
\begin{center}
  \begin{picture}(180,110) 
  \put(-0.5,92.5){\usebox{\PiAA}}
  \put(34.5,94.1){\huge $=$}
  \put(44.5,80){\usebox{\DSEAALP}}
  \put(108.5,94.1){\huge $+$}
  \put(116,80){\usebox{\DSEAATP}}
  \put(-16,54.1){\huge $+$}
  \put(-10,40){\usebox{\DSEAAG}}
  \put(72,54.1){\huge $+\,\frac{1}{2}$}
  \put(84,40){\usebox{\DSEAAH}}
  \put(0,14.1){\huge $+\,\frac{1}{2}$}
  \put(12.5,0){\usebox{\DSEAASS}}
  \put(80,14.1){\huge $-$}
  \put(87.5,0){\usebox{\DSEAALG}}
  \end{picture}
  \caption{\label{fig:DSE(AA)}
    {
    \textsc{Dyson--Schwinger} equation of the $\Pi^{(AA)}$ self-energy 
    component, Eq.~(\ref{eq:Pi(AA)Result}).
    }}
\end{center} 
\end{figure}
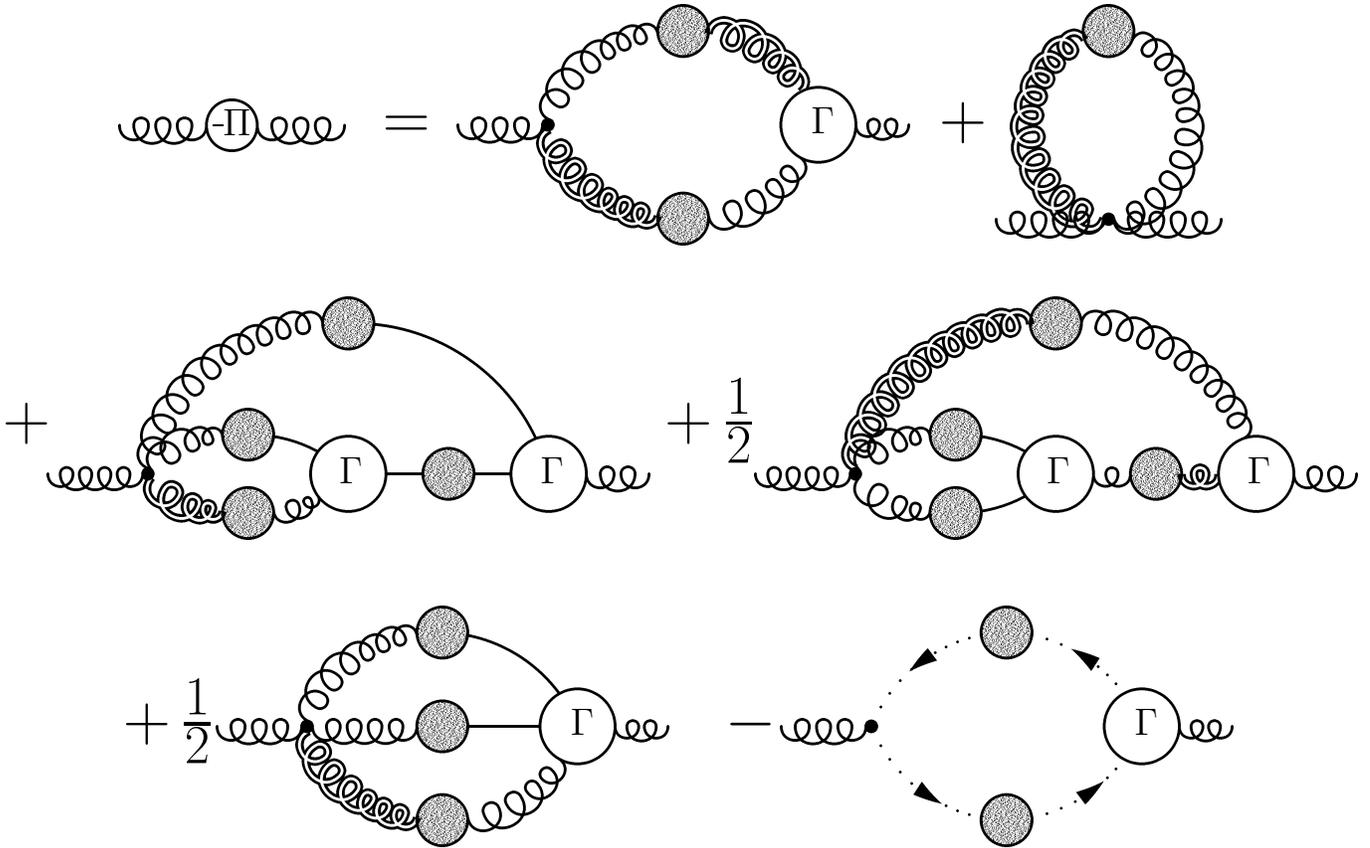 

\noindent
which completes our derivation of the \textsc{Dyson--Schwinger} equations 
in B\"odeker's effective theory.
\end{fmffile}

\section{Discussion and Outlook}
\label{Outlook}

In this work we have constructed an analytic approach to the non-perturbative
physics encoded in B\"odeker's effective theory \cite{Boedeker:1,Boedeker:2}.
Our approach is based on \textsc{Dyson--Schwinger} equations and allows for
an investigation of the non-perturbative dynamics of soft, non-abelian hot
gauge fields 
that is independent of the existing lattice studies 
of B\"odeker's theory \cite{Moore:SphaleronLLO,Moore:SphaleronSym:00}.

The basic starting point is to transform B\"odeker's \textsc{Langevin}
equation into a path integral. From this path integral, in principle, one
could deduce the \textsc{Dyson--Schwinger} equations. However, it would
hardly be avoidable to introduce an uncontrolled gauge dependence when
finally truncating these equations. To control this gauge dependence,
we therefore enlarged the system by the introduction of gauge ghosts 
(which is optional in stochastic quantisation). This enlarged system 
is endowed with a~BRST symmetry reflecting the gauge invariance; 
and we have derived the corresponding \textsc{Ward--Takahashi} identities. 
A consistent truncation of the \textsc{Dyson--Schwinger} equations is 
achieved if the gauge and ghost sectors are truncated in accordance with
these identities.

We also derived a second class of restrictions, so-called stochastic 
Ward identities known from stochastic quantisation
\cite{ZinnJustinZwanziger:WardIdforSQ}. These reflect the characteristic
structure of the path integral action induced by its origin in a stochastic
differential equation.

Finally, we have deduced the \textsc{Dyson--Schwinger} equations of the
theory.
They contain, in principle, the possibility of (finite!)~vertices 
coupling auxiliary fields to gauge ghosts or gauge field/auxiliary field 
vertices with more than one auxiliary field, both of which are not present 
at tree level. Whether these vertices are really non-zero, will be an
interesting question to be decided by an implementation of our formalism.

In combination with the gauge and stochastic Ward identities given in
Eqs.~(\ref{eq:StochasticWID-zero}) -- (\ref{eq:StochasticWID-2nd}),
the \textsc{Dyson--Schwinger} equations 
(\ref{eq:DSE:Pi(omega)(k)}), (\ref{eq:DSE:Pi(omega)(k)II}),
(\ref{eq:Pi(LL)Result}), (\ref{eq:Pi(LA)Result}), 
(\ref{eq:Pi(AL)Result}) and (\ref{eq:Pi(AA)Result})
provide all the necessary tools for an analytic study of the
non-perturbative physics encoded in B\"odeker's effective theory. In particular, 
it can be used to study the sphaleron rate Eq.~(\ref{sphaleron}), where
$N_{\rm CS}$ in terms of the gauge field takes the form
\begin{equation}\label{Ncs}
N_{\rm CS}(t_2) - N_{\rm CS}(t_1) = \int_{t_1}^{t_2} dt \int d^3 x 
	\frac{g^2}{8 \pi^2} E_i^a B_i^a(x) \,
\end{equation}
Restricting to the lowest correlators, we are then interested in the
unequal time correlators $\langle E_i^a(x_1) E_j^b (x_2)\rangle$, $\langle E_i^a(x_1) B_j^b (x_2)\rangle$, and 
$\langle B_i^a(x_1) B_j^b (x_2)\rangle$. The first one should approach a delta function, while the second one should be subleading \cite{Moore:SphaleronTalk:00}. It would be a good test of our ansatz if we could (roughly) reproduce the factor in front of the sphaleron rate \cite{Moore:SphaleronLLO,Moore:SphaleronSym:00}.

We close this discussion with a few comments on what is to come. Since for
the hot sphaleron rate we are interested primarily in the infrared behaviour
of the theory, the first thing to be done moving forward is determining the
appropriate relation between the anomalous dimensions for $k_0$ and
$\vert\kvec\vert^2$. This can be done by investigating the limit when
$k_0\rightarrow0$, and comparing with the anomalous dimension in
\textsc{Yang-Mills} theory in three dimensions \cite{Lerche}. One can also
analyse the importance of the ansatz for the vertex functions by comparing
with time-independent stochastic quantisation \cite{Zwanziger2003}. 

\section*{Acknowledgments}
\noindent
A.H. is supported by CONACYT/DAAD, Contract No. A/05/12566. M.G.S. gratefully acknowledges very 
useful discussions with Tomislav Prokopec during the first part of this project. 

\begin{appendix}
\appendix
\setcounter{section}{0}
\setcounter{equation}{0}
\renewcommand{\theequation}{\thesection.\arabic{equation}}
\section{Calculation of Jacobians} \label{Jacobian}
Throughout this work, there appear several times Jacobians as products
of change of variables. As is well known from the
literature \cite{ZinnJustin}, we have claimed that they are constants
and have generally absorbed them in the measure. To make this work more
self-contained, we provide here a derivation of this claim.

In order to simplify the expressions, we will suppress the colour and
space indices until it becomes necessary. The first Jacobian that we
encountered was in Eq.~(\ref{eq:1byA}), where the following expression
appears
\begin{equation}\label{eq:Det1}
\mathrm{Det}\!\left(\frac{\delta \Eb}{\delta\Ab}\right)
\end{equation}
with
\begin{equation}
\frac{\delta\Eb[\Ab]}{\delta\Ab}=\frac{\partial}{\partial t}+\frac{1}{2}
\frac{\delta\Kb[\Ab]}{\delta\Ab}
=\frac{\partial}{\partial t}\left(\mathbbm{1}+\frac{1}{2}\left(\frac{\partial}{\partial
t}\right)^{-1} \frac{\delta\Kb[\Ab]}{\delta\Ab}\right)
\end{equation}
where $\Kb$ contains all the terms in the left-hand side of
Eq.(\ref{eq:Boedeker+v}) without time derivatives. The kernel of the
operator $(\partial/\partial t)^{-1}$ is constrained by causality to
be $\Theta(t_2-t_1)$. We then have
\begin{equation}\label{eq:DetWithConstPulledOut}
        \mathrm{Det}\!\left(\frac{\delta \Eb}{\delta\Ab}\right)        =
        \;\mathrm{const.}
\cdot\mathrm{Det}\left(\mathbbm{1}+\frac{1}{2}\left(\frac{\partial}{\partial
t}\right)^{-1} \frac{\delta\Kb[\Ab]}{\delta\Ab}\right)
\end{equation}
with
\begin{equation} \label{eq:PartOfDet}
  \biggl[
  \frac{1}{2}\!
  \left(\frac{\partial}{\partial t}\right)^{\!\!-1}
  \!\frac{\delta \Kb[\Ab]}{\delta\Ab}
  \biggr]^{ab}_{ij}\!\!\!(t,\myvec{x};t^{\prime},\myvec{x}^{\prime})
  \;=\;
  \frac{1}{2}\int\!\!dt^{\prime\prime}\;
  \Theta(t-t^{\prime\prime})\,
  \frac{\delta K^a_i[\Ab](t^{\prime\prime}\!,\myvec{x})}
       {\delta A^b_j(t^{\prime},\myvec{x}^{\prime})}
\end{equation}
Since $\Kb$ contains no time derivatives, the functional derivative
produces a delta function in the time variable i.e.
\begin{equation}
  \frac{\delta K^a_i[\Ab](t^{\prime\prime}\!,\myvec{x})}
       {\delta A^b_j(t^{\prime},\myvec{x}^{\prime})}
  = \delta(t^{\prime\prime}\! - t^{\prime})\,
     \frac{\delta_x K^a_i[\Ab](t^{\prime},\myvec{x})}
          {\delta_x A^b_j(t^{\prime},\myvec{x}^{\prime})}
\end{equation}
where we have introduced the symbol $\delta_x$ to denote a variation
with respect to the $\myvec{x}$ dependence only. Hence, we find
\begin{equation} \label{eq:PartOfDetFinal}
  \biggl[
  \frac{1}{2}\!
  \left(\frac{\partial}{\partial t}\right)^{\!\!-1}
  \!\frac{\delta K[\phi]}{\delta\phi}
  \biggr]_{\alpha\beta}\!\!\!(t,\myvec{x};t^{\prime},\myvec{x}^{\prime})
  \;=\;
  \frac{1}{2}\,
  \Theta(t-t^{\prime})\,
     \frac{\delta_x K^a_i[\phi](t^{\prime},\myvec{x})}
          {\delta_x A^b_j(t^{\prime},\myvec{x}^{\prime})}
\end{equation}
Coming back to Eq.~(\ref{eq:DetWithConstPulledOut}) and using
$\mathrm{Tr}\ln (\dots) = \ln\mathrm{Det} (\dots)$ in addition to
the series expansion of the logarithm, the determinant takes the form
\begin{equation} \label{eq:DetWithSum}
  \mathrm{Det}\!\left(\frac{\delta \Eb[\Ab]}{\delta\Ab}\right)
  = \;\mathrm{const.} \cdot
    \exp
    \biggl\{\,
      \sum_{n=1}^{\infty} \frac{(-1)^{n+1}}{n}
      \,\frac{1}{2^n}
      \mathrm{Tr}
      \biggl[
      \left(\frac{\partial}{\partial t}\right)^{\!\!-1}
      \!\frac{\delta \Kb[\Ab]}{\delta\Ab}
      \biggr]^n
    \biggr\}
\end{equation}
The trace in this expression can be evaluated with the help of
Eq.~(\ref{eq:PartOfDetFinal}). One obtains
\begin{eqnarray} \hspace{-5.5ex}
  \mathrm{Tr}
    \biggl[
      \left(\frac{\partial}{\partial t}\right)^{\!\!-1}
      \!\frac{\delta \Kb[\Ab]}{\delta\Ab}
    \biggr]^n
  \!\!\!& = &\!\!\!
  \int\!\!dt_1\cdots dt_n\, d^{D-1}\!x_1\cdots d^{D-1}\!x_n\;
  \Theta(t_1-t_2)\,
  \frac{\delta_x K^{a_1}_{i_1}[\Ab](t_2,\myvec{x}_1)}
          {\delta_x A^{a_2}_{i_2}(t_2,\myvec{x}_2)}
  \nonumber\\
  & & \qquad\!\!\!\!\!
  \Theta(t_2-t_3)\,
  \frac{\delta_x K^{a_2}_{i_2}[\Ab](t_3,\myvec{x}_2)}
          {\delta_x A^{a_3}_{i_3}(t_3,\myvec{x}_3)}
  \;\cdots\;
  \Theta(t_n-t_1)\,
  \frac{\delta_x K^{a_n}_{i_n}[\Ab](t_1,\myvec{x}_n)}
          {\delta_x A^{a_1}_{i_1}(t_1,\myvec{x}_1)}
\end{eqnarray}
and thus
\begin{equation} \label{eq:TimeTrace}
  \mathrm{Tr}
    \biggl[
      \left(\frac{\partial}{\partial t}\right)^{\!\!-1}
      \!\frac{\delta K[\phi]}{\delta\phi}
    \biggr]^n
  =
  \int\!\!dt_1\cdots dt_n\;
  \Theta(t_1-t_2)\,\Theta(t_2-t_3)\,\cdots\,\Theta(t_n-t_1)\,
  f_n(t_1,t_2,\dots,t_n)
\end{equation}
if we set
\begin{displaymath}
  f_n(t_1,t_2,\dots,t_n)
  =
  \int\!\!d^{D-1}\!x_1\cdots d^{D-1}\!x_n\;
  \frac{\delta_x K^{a_1}_{i_1}[\Ab](t_2,\myvec{x}_1)}
          {\delta_x A^{a_2}_{i_2}(t_2,\myvec{x}_2)}
  \frac{\delta_x K^{a_2}_{i_2}[\Ab](t_3,\myvec{x}_2)}
          {\delta_x A^{a_3}_{i_3}(t_3,\myvec{x}_3)}
  \;\cdots\;
  \frac{\delta_x K^{a_n}_{i_n}[\Ab](t_1,\myvec{x}_n)}
          {\delta_x A^{a_1}_{i_1}(t_1,\myvec{x}_1)}
\end{displaymath}
for abbreviation. Unless $n=1$, however, the
expression (\ref{eq:TimeTrace}) vanishes
for any function $f_n$.
Therefore, only the first term of the sum
in Eq.~(\ref{eq:DetWithSum}) survives and we finally arrive at
\begin{equation} \label{eq:DetWithoutGhosts}
  \mathrm{Det}\!\left(\frac{\delta \Eb[\Ab]}{\delta\Ab}\right)
  = \;\mathrm{const.} \cdot
    \exp
    \biggl\{\,
      \frac{1}{2}\,\Theta(0)\!
      \int\!\!dt\,d^{D-1}\!x
      \left.
      \frac{\delta_x K^a_i[\Ab](t,\myvec{x})}
           {\delta_x A^a_b(t,\myvec{x}^{\prime})}
      \right|_{\myvec{x}^{\prime}=\myvec{x}}
    \biggr\}
\end{equation}
Our next task is to calculate the functional derivative of
$K^a_i[\myvec{A}]$. To this end, it is easiest to write it down
in components which clarifies the structure
\begin{eqnarray} \label{eq:K[A]inComponents}
  \frac{1}{2}\,K^a_i[\myvec{A}]
  &=&
  g f^{abc}
  \left[
    \left(1-\frac{\sigma}{\kappa}\right) A^b_i \partial_j A^c_j
    +
    2 A^c_j \partial_j A^b_i
    +
    A^b_j \partial_i A^c_j
  \right] \nonumber\\[1ex]
  & &  \quad + \;\,
  \left[
    \left(1-\frac{\sigma}{\kappa}\right) \partial_i\partial_j
    - \delta_{ij}\Delta
  \right] A^a_j
  \;+\;
  g^2 f^{abc} f^{bde} A^c_j A^d_j A^e_i
\end{eqnarray}
Obviously, the first term, i.e.~the term quadratic in the gauge field,
does not contribute to the functional derivative with respect to
$A^a_i$ because it always produces a $\delta^{ab}$
or $\delta^{ac}$ that is contracted with the structure constants
$f^{abc}$ in front of the square bracket. The linear term, on the
other hand, only contributes a constant that can be absorbed
into the constant in Eq.~(\ref{eq:DetWithoutGhosts}). Thus, we only
have to take care of the third order term which leads to
\begin{equation} \label{eq:FuncDetGaugeResult}
  \mathrm{Det}\left(
    \frac{\delta\myvec{E}[\myvec{A}]}{\delta\myvec{A}}\right)
  = \;\mathrm{const.}^{\prime} \cdot
    \exp
    \biggl\{\,
       C_A\,(D-2)\,\Theta(0)\,\delta^{D-1}(0)\;
      \frac{g^2}{\sigma}
      \int\!\!dt\,d^{D-1}\!x\;
      \myvec{A}^{\!a}(t,\myvec{x}) \cdot \myvec{A}^{\!a}(t,\myvec{x})
    \biggr\}
\end{equation}
where $f^{acd} f^{bcd} = C_A\,\delta^{ab}$ as usual.
However, in dimensional regularisation $\delta^{D-1}(0)$ gives zero as a
consequence of the general rules of D--dimensional
integration, and the determinant is simply a constant.

We came across another determinant
in Eq.~(\ref{eq:F[A0,A]_ChangeofVariables})
\begin{equation}
  \mathrm{Det}
  \left(
    \frac{\delta{}^{\omega}\!\gvec{\zeta}}{\delta\gvec{\zeta}}
  \right)
\end{equation}
One finds
\begin{equation}
  \frac{\delta{}^{\omega}\!\zeta^{ai}(t\,,\myvec{x}\,)}
       {\delta\zeta^{bj}(t^{\prime},\myvec{x}^{\prime})}
  =
  \delta^{ab}\delta_{ij} \,
  \delta(t-t^{\prime}) \,
  \delta^{D-1}(\myvec{x} - \myvec{x}^{\prime})
  \,+\,
  g f^{acd}\,\frac{\delta}{\delta\zeta^{bj}(t^{\prime},\myvec{x}^{\prime})}
  \left[
    \omega^d[\gvec{\zeta}](t,\myvec{x})\,\zeta^{ci}(t,\myvec{x})
  \right]
\end{equation}
and thus because $\omega$ is infinitesimal
\begin{equation} \label{eq:DetZetaP/Zeta}
  \mathrm{Det}
  \left(
    \frac{\delta{}^{\omega}\!\gvec{\zeta}}{\delta\gvec{\zeta}}
  \right)
  =
  1 + \int\!\!dt\,d^{D-1}\!x\;
  g f^{acd}
  \left[
  \frac{\delta}{\delta\zeta^{ai}(t^{\prime},\myvec{x}^{\prime})}
  \left[
    \omega^d[\gvec{\zeta}](t,\myvec{x})\,\zeta^{ci}(t,\myvec{x})
  \right]
  \right]_{t^{\prime} \,=\,t \atop \myvec{x}^{\prime} =\,\myvec{x}}
\end{equation}
The functional derivative acting on $\zeta^{ci}$ produces a 
$\delta^{ac}$ and therefore does not contribute because the
Kronecker delta is contracted with the structure constants.
To determine the remaining functional derivative of 
$\omega^d[\gvec{\zeta}]$, let us formally integrate
Eq.~(\ref{eq:CondOnOmega})
\begin{equation} \label{eq:OmegaFormalInt}
  \omega^a(t,\myvec{x}) =
  \omega^a(-\infty,\myvec{x})
  - \int\limits_{-\infty}^t\!\!dt^{\prime\prime}\,
    \left[ H[\myvec{A}]\omega\right]^a\!(t^{\prime\prime},\myvec{x})
  - \int\limits_{-\infty}^t\!\!dt^{\prime\prime}\,
    \delta v^a[\myvec{A}](t^{\prime\prime},\myvec{x})
\end{equation}
Since $H[\myvec{A}]$ and $\delta v^a[\myvec{A}]$ are local functionals
in time, this equation for $\omega$ has a causal character, 
i.e.~$\omega(t,\myvec{x})$ does only depend on the values of
the gauge field $\myvec{A}(t^{\prime\prime},\myvec{x})$ 
at times $t^{\prime\prime}\!< t$. On the other hand,
Eq.~(\ref{eq:Original}) leads to
\begin{equation} \label{eq:AformalInt}
  \sigma\myvec{A}^{\!a}(t,\myvec{x}) =
  \sigma\myvec{A}^{\!a}(-\infty,\myvec{x})
  - \int\limits_{-\infty}^t\!\!dt^{\prime\prime}\,
  \left[ \,
    \myvec{D}^{ab} \!\times \myvec{B}^b
    + \sigma\myvec{D}^{ab} v^b[\myvec{A}]\,
  \right]\!(t^{\prime\prime},\myvec{x})
  + \int\limits_{-\infty}^t\!\!dt^{\prime\prime}\,
  \gvec{\zeta}^a(t^{\prime\prime},\myvec{x})
\end{equation}
and $\myvec{A}(t,\myvec{x})$ itself only depends on the 
stochastic force $\gvec{\zeta}(t^{\prime\prime},\myvec{x})$
for $t^{\prime\prime}\!< t$. Hence, neither 
$\myvec{A}(t,\myvec{x})$ nor $\omega(t,\myvec{x})$ have
a dependence on $\gvec{\zeta}(t^{\prime\prime},\myvec{x})$
unless $t^{\prime\prime}\!< t$ and in taking the functional 
derivative of Eq.~(\ref{eq:OmegaFormalInt}), we can restrict
the integration range accordingly
\begin{equation} \label{eq:delOmega/delZeta_IntEq}
  \frac{\delta\omega^a[\gvec{\zeta}](t,\myvec{x})}
       {\delta\zeta^{bi}(t^{\prime},\myvec{x}^{\prime})} =
  - \int\limits_{t^{\prime}}^t\!\!dt^{\prime\prime}\,
    \frac{\delta
    \left[ H[\myvec{A}]\omega\right]^a\!(t^{\prime\prime},\myvec{x})
         }{\delta\zeta^{bi}(t^{\prime},\myvec{x}^{\prime})} \;
  - \int\limits_{t^{\prime}}^t\!\!dt^{\prime\prime}\,
    \frac{\delta\,
    \delta v^a[\myvec{A}](t^{\prime\prime},\myvec{x})
         }{\delta\zeta^{bi}(t^{\prime},\myvec{x}^{\prime})}
\end{equation}
Evaluating this relation for $t=t^{\prime}$ as in 
Eq.~(\ref{eq:DetZetaP/Zeta}) leads to
\begin{equation}
  \left.
  \frac{\delta\omega^a[\gvec{\zeta}](t,\myvec{x})}
       {\delta\zeta^{bi}(t^{\prime},\myvec{x}^{\prime})}
  \right|_{t=t^{\prime}} = 0
\end{equation}
The only way to escape this conclusion would be an integrand that is
singular in time. However, if $\delta\omega/\delta\zeta$ 
appearing under the integral in Eq.~(\ref{eq:delOmega/delZeta_IntEq}) 
was singular, the integrated expression would be finite which
again is $\delta\omega/\delta\zeta$. Therefore, $\delta\omega/\delta\zeta$
can not be singular. $\delta\myvec{A}/\delta\zeta$ on the other hand
can not be singular neither because of the same argument applied to
the functional derivative of Eq.~(\ref{eq:AformalInt}) with 
respect to $\zeta$. Thus, we conclude
\begin{equation}
  \mathrm{Det}
  \left(
    \frac{\delta{}^{\omega}\!\gvec{\zeta}}{\delta\gvec{\zeta}}
  \right)
  = 1
\end{equation}
which completes the proof.

During the introduction of gauge ghosts to the path integral,
Eq.~(\ref{eq:UnityByOmegas}), there appears in our work another
Jacobian. We can see that it has the same form as the one we have 
already calculated, but with 
\begin{equation}
  \frac{1}{2}\,K^a[\omega,\myvec{A}](t,\myvec{x})
  =
  -\frac{1}{\kappa}\,(\myvec{D}^{ab}\!\cdot\nabla\omega^b)(t,\myvec{x})
\end{equation}
Hence, we can rely on our general result for the determinant,
Eq.~(\ref{eq:DetWithoutGhosts}), 
\begin{equation}
  \mathrm{Det}\!\left(\frac{\delta \gamma[\omega,\myvec{A}]}
                           {\delta\omega}\right)  
  = \;\mathrm{const.} \cdot
    \exp
    \biggl\{
      -\frac{1}{\kappa}\,\Theta(0)\!
        \int\!\!dt\,d^{D-1}\!x
        \left.
         \frac{\delta_x (\myvec{D}^{ab}\!\cdot\nabla\omega^b)(t,\myvec{x})}
              {\delta_x\omega^a(t,\myvec{x}^{\prime})}
        \right|_{\myvec{x}^{\prime}=\myvec{x}}
    \biggr\}
\end{equation}
The functional derivative with respect to spacial variations is given
by
\begin{equation}
  \frac{\delta_x (\myvec{D}^{ab}\!\cdot\nabla\omega^b)(t,\myvec{x})}
       {\delta_x\omega^d(t,\myvec{x}^{\prime})}
  =
  (\delta^{ab}\nabla -g f^{abc} \myvec{A}^{\!c}) \cdot
  \nabla\delta^D(\myvec{x} - \myvec{x}^{\prime}) \,\delta^{bd}
\end{equation}
and thus, evaluated for $d=a$, gives a constant because
the $\myvec{A}$ dependent contribution is set to zero due to the 
antisymmetry of the structure constants. Note that,
this time, we did not have to rely on dimensional regularisation 
to proof the constancy of the determinant as we had to in the case
of $\mathrm{Det}(\delta\myvec{E[\myvec{A}]}/\delta\myvec{A})$.

When we performed the BRST transformation  in our derivation of the
Ward identities, Eq.~(\ref{eq:GaugeBRST}), one more type of determinant
appeared. In general, if $x_a$ are Grassmann even and $\vartheta_i$
Grassmann odd quantities, a mixed change of variables of the form
\begin{equation} \label{eq:mixedChangeStart}
\begin{array}{rcl}
  x_a & \!\!=\!\! & x_a^{\prime} + 
            \varepsilon \,f_a(x^{\prime},\vartheta^{\prime})\\[1ex]
  \vartheta_i & \!\!=\!\! & \vartheta_i^{\prime} \,+
            \varepsilon \,\phi_i(x^{\prime},\vartheta^{\prime})
\end{array}
\end{equation}
with $\varepsilon$ being a Grassmann odd parameter leads to a Jacobian
\begin{equation}
  J = 1 + \varepsilon\,\textrm{str}(M)
\end{equation}
In this expression, the matrix $M$ under the super trace is given by
\begin{equation}
  M = 
  \left(
  \begin{array}{cc}
    A & B \\[1ex] C & D
  \end{array}
  \right)
  =
  \left(
  \begin{array}{cc}
    \frac{\partial f_a}{\partial x_b^{\prime}} &
    -\frac{\partial f_a}{\partial \vartheta_i^{\prime}} \\[1ex]
    \frac{\partial \phi_i}{\partial x_a^{\prime}} &
    - \frac{\partial \phi_i}{\partial\vartheta_j^{\prime}}
  \end{array}
  \right)
\end{equation}
and hence
\begin{equation} \label{eq:mixedChangeEnd}
  \textrm{str}(M) = \textrm{tr}(A) - \textrm{tr}(D)
  =
  \frac{\partial f_a}{\partial x_a^{\prime}} +
  \frac{\partial \phi_i}{\partial\vartheta_i^{\prime}}
\end{equation}
(See e.g.~\cite{ZinnJustin}, Section 1.8.2. Note, however, that in our 
case $\varepsilon$ is Grassmann odd which leads to the additional minus 
signs in the matrix $M$ when $\varepsilon$ is commuted with the
derivative $\partial/\partial\vartheta$).

In our case, we have two sets of commuting variables, 
$\myvec{A}^{\!ai}(x)$ and $\gvec{\lambda}^{\!ai}(x)$, and 
two sets of anti-commuting ones, $\omega^a(x)$ and
$\bar{\omega}^a(x)$. Therefore, the Jacobian is given by
\begin{equation} \label{eq:JacobianGaugeWard}
  J = 1 + \varepsilon \!\int\!\!dx \left[
      \frac{\delta \,sA^{\prime ai}(x)}{\delta\!A^{\prime ai}(x)}
     +\frac{\delta \,s\lambda^{\prime ai}(x)}{\delta\lambda^{\prime ai}(x)}
     +\frac{\delta \,s\omega^{\prime a}(x)}{\delta\omega^{\prime a}(x)}
     +\frac{\delta \,s\bar{\omega}^{\prime a}(x)}
           {\delta\bar{\omega}^{\prime a}(x)}
      \right]
\end{equation}
However, any of these functional derivatives vanishes as a short glance
at the BRST transformed fields in Eq.~(\ref{eq:finiteBRST}) makes 
obvious: The derivative always produces a Kronecker delta 
that is to be contracted with the structure constants. Consequently,
the Jacobian of the change of variables 
(\ref{eq:GaugeBRST}) is unity.



\appendix
\setcounter{section}{1}
\setcounter{equation}{0}
\renewcommand{\theequation}{\thesection.\arabic{equation}}
\section{Feynman Rules} \label{FeynmanRules}
\begin{fmffile}{fmFeynmanRules}
The action, as given by Eq.~(\ref{eq:S[A,Lambda,Omega,OmegaBar]}), is
\begin{equation}
  S[\myvec{A},\gvec{\lambda},\omega,\bar{\omega}] =
  S^{\scriptscriptstyle\mathrm{(D)}}[\myvec{A},\gvec{\lambda}]
  \,+\, S^{\scriptscriptstyle\mathrm{(GG)}}
  [\myvec{A},\omega,\bar{\omega}],
\end{equation}
with
\begin{eqnarray}\label{eq:SDynamic}
  S^{\scriptscriptstyle\mathrm{(D)}}[\myvec{A},\gvec{\lambda}] &=&
  \int\!\!dx\,
  \Bigl[
    \sigma T \,\gvec{\lambda}^a\!\cdot\!\gvec{\lambda}^a
    -i \gvec{\lambda}^a \!\cdot\!
    \left(
      \myvec{D}^{ab} \!\times \myvec{B}^b
      + \sigma (\dot{\myvec{A}}^{\!a}
                -{\textstyle\frac{1}{\kappa}\,}
                 \myvec{D}^{ab}\,\nabla\!\cdot\!\myvec{A}^{\!b}
               )
    \right)
  \Bigr]\\
\label{eq:Sgg}
  S^{\scriptscriptstyle\mathrm{(GG)}}[\myvec{A},\omega,\bar{\omega}]
  &=&
  \int\!\!dx\,
  \Bigl[
    -\,\bar{\omega}^a \dot{\omega}^a
    +\frac{1}{\kappa}\,
      \bar{\omega}^a \myvec{D}^{ab}\!\cdot\nabla \omega^b
  \Bigr]
\end{eqnarray}
\subsection{The Propagators}
The free, quadratic part of the dynamical action
$S^{\scriptscriptstyle\mathrm{(D)}}[\myvec{A},\gvec{\lambda}]$ can be cast
into the following symmetric form reflecting the mixing that occur between
the gauge field $\Av$ and the auxiliary field $\Lv$
\begin{equation}
  S^{\scriptscriptstyle\mathrm{(D)}}_0[\myvec{A},\gvec{\lambda}] = 
  \int\!\!dx dy\;
  \frac{1}{2}\;
  (\lambda^{ai}(x), A^{ai}(x))
  \;(\hat{\Delta}^{-1})^{\,ab}_{\,ij}\,(x,y)\,
  \left(
  \begin{array}{c}
    \!\! \lambda^{bj}(y) \!\! \\
    \!\! A^{bj}(y)       \!\!
  \end{array}
  \right)
\end{equation}
with the matrix
\begin{equation} \label{eq:InvPropagatorMatrix}
  (\hat{\Delta}^{-1})^{\,ab}_{\,ij}\,(x,y) =
  \left(
  \begin{array}{cc}
  \!\! (\Delta^{-1})^{(\lambda\lambda)}{}^{ab}_{ij}(x,y)   & 
       (\Delta^{-1})^{(\lambda A)}{}^{ab}_{ij}(x,y) \!\! 
  \\[2ex] 
  \!\! (\Delta^{-1})^{(A      \lambda)}{}^{ab}_{ij}(x,y) & 
       (\Delta^{-1})^{(A       A)}{}^{ab}_{ij}(x,y) \!\!
  \end{array}
  \right)
\end{equation}
and
\begin{eqnarray} \label{eq:Delta^-1:LL}
  (\Delta^{-1})^{(\lambda\lambda)}{}^{ab}_{ij}(x,y) & = &
  2 \sigma T \,\delta^{ab}\delta_{ij} \,\delta(x-y) \\ \label{eq:Delta^-1:LA}
  (\Delta^{-1})^{(\lambda A)}{}^{ab}_{ij}(x,y) & = &
  -i \,\delta^{ab}
  \left[
    (+\sigma\partial_t - \Delta) \,\delta_{ij} +
    (1 -{\textstyle\frac{\sigma}{\kappa}})\,\partial_i\partial_j
  \right] \delta(x-y) \\ \label{eq:Delta^-1:AL}
  (\Delta^{-1})^{(A \lambda)}{}^{ab}_{ij}(x,y) & = &
  -i \,\delta^{ab}
  \left[
    (-\sigma\partial_t - \Delta) \,\delta_{ij} +
    (1 -{\textstyle\frac{\sigma}{\kappa}})\,\partial_i\partial_j
  \right] \delta(x-y) \\ \label{eq:Delta^-1:AA}
  (\Delta^{-1})^{(AA)}{}^{ab}_{ij}(x,y) & = & 0
\end{eqnarray}
We denote by non-bold symbols combinations of time and space variables, 
e.g.~$\delta(x-y) = \delta(t_x\!-t_y)\,\delta^{D-1}(\myvec{x}-\myvec{y})$.
The matrix $\hat{\Delta}^{-1}$ 
is symmetric in the following sense
\begin{equation} \label{eq:SymDeltaInv:x-space}
  (\Delta^{-1})^{(FG)}{}^{ab}_{ij}(x,y) =
  (\Delta^{-1})^{(GF)}{}^{ba}_{ji}(y,x)
\end{equation}
Hence, the matrix propagator
\begin{samepage}
\begin{displaymath} \addtocounter{equation}{1}
  \hat{\Delta}^{\,ab}_{\,ij}\,(x,y) =
  \left(
  \begin{array}{cc}
  \!\! \Delta^{(\lambda\lambda)}{}^{ab}_{ij}(x,y)   & 
       \Delta^{(\lambda A)}{}^{ab}_{ij}(x,y) \!\! 
  \\[2ex] 
  \!\! \Delta^{(A      \lambda)}{}^{ab}_{ij}(x,y) & 
       \Delta^{(A       A)ab}_{}{}^{ij}(x,y) \!\!
  \end{array}
  \right)
  =
  \left(
  \begin{array}{cc}
  \!\! \left\langle \lambda^{ai}(x) \,\lambda^{bj}(y) \right\rangle_0 
       \rule{0pt}{10pt} & 
       \left\langle \lambda^{ai}(x) \,      A^{bj}(y) \right\rangle_0 \!\! 
  \\[2ex] 
  \!\! \left\langle       A^{ai}(x) \,\lambda^{bj}(y) \right\rangle_0 & 
       \left\langle       A^{ai}(x) \,      A^{bj}(y) \right\rangle_0 \!\!
  \end{array}
  \right)
\end{displaymath}
is given by its inverse 
\hspace{\fill}(\theequation)\\
\end{samepage}
\begin{equation}
  \int\!\!d^D\!y\;
  \Delta^{(FG)}{}^{ab}_{ij}(x,y) \,
  (\Delta^{-1})^{(GH)}{}^{bc}_{jk}(y,z)
  = \delta^{ac} \, \delta_{ik} \, \delta^{F\!H} \,\delta^D(x-z)
\end{equation}
or equivalently
\begin{equation}
  \Delta^{(FG)}{}^{ab}_{ij}(k) \,
  (\Delta^{-1})^{(GH)}{}^{bc}_{jk}(k)
  = \delta^{ac} \, \delta_{ik} \, \delta^{F\!H}
\end{equation}
for the momentum space functions
\begin{eqnarray} \label{eq:Fourier}
  \Delta^{(FG)}{}^{ab}_{ij}(x,y) & = &
  \intDk e^{-ik(x-y)}
  \Delta^{(FG)}{}^{ab}_{ij}(k)\\
  (\Delta^{-1})^{(FG)}{}^{ab}_{ij}(x,y) & = &
  \intDk e^{-ik(x-y)}
  (\Delta^{-1})^{(FG)}{}^{ab}_{ij}(k)
\end{eqnarray}
Note again that though we are most of the time dealing with three-vectors,  
in the \textsc{Fourier} transform we use four-vector notation,
i.e.~$e^{-ik(x-y)}=e^{-ik_0 (x_0-y_0) 
+ i\myvec{k}\cdot(\myvec{x}-\myvec{y})}$
leading to
\begin{equation} \label{eq:InvProp(AL)(k)}
  (\hat{\Delta}^{-1})^{\,ab}_{\,ij}\,(k) =
  \left(
  \begin{array}{cc}
  \!\! 2 \sigma T \,\delta^{ab}\delta_{ij} \rule{0pt}{10pt} & 
  \!\!\!\!\!\!\!\!\!\!\!
  -i \,\delta^{ab}
  \left[
    (-i\sigma k_0 \!+ \myvec{k}^2) \,\delta_{ij}\! -
    (1\! -\!{\textstyle\frac{\sigma}{\kappa}})\,k_i k_j
  \right] \!\! 
  \\[2ex] 
  \!\! -i \,\delta^{ab}
  \left[
    (+i\sigma k_0 \!+ \myvec{k}^2) \,\delta_{ij}\! -
    (1\! -\!{\textstyle\frac{\sigma}{\kappa}})\,k_i k_j
  \right] & 0 \!\!
  \end{array} 
  \right).
\end{equation}
In momentum space, the gauge/auxiliary field propagators are given by:

\vspace{4ex}
\unitlength=1mm
\noindent 
\begin{minipage}{147mm}
\parbox{50mm}{
\begin{picture}(50,25) 
  \makebox(50,25){
    \put(16,17){\vector(-1,0){4.0}} 
    \put(16.5,16){$k$}
    \put(-8,11.75){$a,\,i$}
    \put(33,11.75){$b,\,j$}
    \begin{fmfgraph*}(30,25)
      \fmfkeep{LLProp}
      \fmfstraight
      \fmfleft{o}
      \fmfright{i}
      \fmf{dbl_curly}{o,i}
      \fmfdot{i,o}
    \end{fmfgraph*}
  }
\end{picture} 
}
\begin{minipage}{91mm}
\begin{displaymath}
  \Delta^{(\lambda\lambda)}{}^{ab}_{ij}(k) = 0
  \hspace{44.0ex}
\end{displaymath}
\vspace{-1ex}
\end{minipage}
\end{minipage} 

\unitlength=1mm
\noindent 
\begin{minipage}{147mm}
\parbox{50mm}{
\begin{picture}(50,25) 
  \makebox(50,25){
    \put(16,17){\vector(-1,0){4.0}} 
    \put(16.5,16){$k$}
    \put(-8,11.75){$a,\,i$}
    \put(33,11.75){$b,\,j$}
    \begin{fmfgraph*}(30,25)
      \fmfkeep{LAProp}
      \fmfstraight
      \fmfleft{o}
      \fmfright{i}
      \fmf{dbl_curly}{o,v}
      \fmf{gluon}{v,i}
      \fmfdot{i,o}
    \end{fmfgraph*}
  }
\end{picture} 
}
\begin{minipage}{91mm}
\begin{displaymath}
  \Delta^{(\lambda A)}{}^{ab}_{ij}(k) =
  \frac{i \delta^{ab}}{+i\sigma k_0 + |\kvec|^2}
  \left[
    \delta_{ij} + 
    \left(1-{\textstyle\frac{\sigma}{\kappa}}\right)
    \frac{k_i k_j}{+i\sigma k_0 + 
    {\textstyle\frac{\sigma}{\kappa}} |\kvec|^2}
  \right]
\end{displaymath}
\vspace{0.05ex}
\end{minipage}
\end{minipage} 

\unitlength=1mm
\noindent 
\begin{minipage}{147mm}
\parbox{50mm}{
\begin{picture}(50,25) 
  \makebox(50,25){
    \put(16,17){\vector(-1,0){4.0}} 
    \put(16.5,16){$k$}
    \put(-8,11.75){$a,\,i$}
    \put(33,11.75){$b,\,j$}
    \begin{fmfgraph*}(30,25)
      \fmfkeep{ALProp}
      \fmfstraight
      \fmfleft{o}
      \fmfright{i}
      \fmf{dbl_curly}{v,i}
      \fmf{gluon}{o,v}
      \fmfdot{i,o}
    \end{fmfgraph*}
  }
\end{picture} 
}
\begin{minipage}{91mm}
\begin{displaymath}
  \Delta^{(A \lambda)}{}^{ab}_{ij}(k) =
  \frac{i \delta^{ab}}{-i\sigma k_0 + |\kvec|^2}
  \left[
    \delta_{ij} + 
    \left(1-{\textstyle\frac{\sigma}{\kappa}}\right)
    \frac{k_i k_j}{-i\sigma k_0 + 
    {\textstyle\frac{\sigma}{\kappa}} |\kvec|^2}
  \right]
\end{displaymath}
\vspace{0.05ex}
\end{minipage}
\end{minipage} 

\unitlength=1mm
\noindent 
\begin{minipage}{147mm}
\parbox{50mm}{
\begin{picture}(50,25) 
  \makebox(50,25){
    \put(16,17){\vector(-1,0){4.0}} 
    \put(16.5,16){$k$}
    \put(-8,11.75){$a,\,i$}
    \put(33,11.75){$b,\,j$}
    \begin{fmfgraph*}(30,25)
      \fmfkeep{AAProp}
      \fmfstraight
      \fmfleft{o}
      \fmfright{i}
      \fmf{gluon}{o,i}
      \fmfdot{i,o}
    \end{fmfgraph*}
  }
\end{picture} 
}
\begin{minipage}{91mm}
\begin{displaymath}
  \Delta^{(AA)}{}^{ab}_{ij}(k) =
  \frac{2\sigma T \delta^{ab}}{\sigma^2 k_0^2 + |\kvec|^4}\,
  \biggl[
    \delta_{ij} +
    \bigl(1-{\textstyle\frac{\sigma^2}{\kappa^2}}\bigr) \,
    \frac{k_i k_j |\kvec|^2 }
         {\sigma^2 k_0^2 + {\textstyle\frac{\sigma^2}{\kappa^2}}
                   |\kvec|^4}
  \biggr]
  \hspace{1.0ex}
\end{displaymath}
\vspace{-0.5ex}
\end{minipage}
\end{minipage} 

\vspace{2ex}
\noindent
For the gauge ghosts, we have the corresponding contribution to the
action, Eq.~(\ref{eq:Sgg}), comprises the free part
\begin{equation}
  S^{\scriptscriptstyle\mathrm{(GG)}}_0[\omega,\bar{\omega}] =
  \int\!\!dx\,
  \bar{\omega}^a \!
  \left(
    -\partial_t + {\textstyle\frac{1}{\kappa}}\,\Delta
  \right)
  \omega^a
\end{equation}
and therefore
\begin{equation} \label{eq:Delta^-1:omega}
  (\Delta^{-1})^{(\omega)\,ab}(x,y) =
  \delta^{ab} \!
  \left(
    -\partial_t + {\textstyle\frac{1}{\kappa}}\,\Delta
  \right)
  \delta(x-y)
\end{equation}
or in momentum space
\begin{equation} \label{eq:InvPropOmega(k)}
  (\Delta^{-1})^{(\omega)\,ab}(k) =
  \delta^{ab} \!
  \left(
    ik_0 - {\textstyle\frac{1}{\kappa}} |\myvec{k}|^2
  \right)
\end{equation}
Hence the gauge ghost propagator is given by 

\vspace{4ex}
\unitlength=1mm
\noindent 
\begin{minipage}{147mm}
\parbox{50mm}{
\begin{picture}(50,25) 
  \makebox(50,25){
    \put(16,17){\vector(-1,0){4.0}} 
    \put(16.5,16){$k$}
    \put(-4.7,11.75){$a$}
    \put(33,11.75){$b$}
    \begin{fmfgraph*}(30,25)
      \fmfkeep{GProp}
      \fmfstraight
      \fmfleft{o}
      \fmfright{i}
      \fmf{dots_arrow}{i,o}
      \fmfv{label=$a$}{o}
      \fmfv{label=$b$}{i}
      \fmfdot{i,o}
    \end{fmfgraph*}
  }
\end{picture} 
}
\begin{minipage}{91mm}
\begin{displaymath}
  \Delta^{(\omega)\,ab}(k) =
  \frac{\kappa\,\delta^{ab}}{i\kappa k_0 -|\myvec{k}|^2}
  \hspace{35ex}
\end{displaymath}
\vspace{0.05ex}
\end{minipage}
\end{minipage} 
\subsection{The Vertices}
For the interacting part of the dynamical action (\ref{eq:SDynamic}) we have
\begin{eqnarray}
  S^{\scriptscriptstyle\mathrm{(D)}}_{\,\mathrm{int}}
  [\myvec{A},\gvec{\lambda}]
  &=&
  \int\!\!dx\,
  \Bigl\{
  -ig f^{abc} \lambda^{ai}
  \Bigl[
    \left(1-{\textstyle\frac{\sigma}{\kappa}}\right) \!
    A^{bi} \partial_j A^{cj}
    + 
    2 A^{cj} \partial_j A^{bi}
    +
    A^{bj} \partial_i A^{cj}
  \Bigr] \nonumber\\[1ex]
  & &  \hspace{6.3ex}
  -ig^2 f^{abc} f^{bde} \lambda^{ai} A^{cj} A^{dj} A^{ei}
  \Bigr\}
\end{eqnarray}
Thus, the theory provides a 3--point vertex containing
one auxiliary and two gauge fields and a 4--point vertex of three 
gauge fields and one auxiliary field. To simplify explicit 
calculations, it is useful to symmetrise the vertices with respect to 
the two and three gauge fields in either case. Splitting 
\begin{math}
  S^{\scriptscriptstyle\mathrm{(D)}}_{\,\mathrm{int}}
  [\myvec{A},\gvec{\lambda}]
\end{math}
into the contributions corresponding to the 3-- and 4--point vertex
\begin{equation}
  S^{\scriptscriptstyle\mathrm{(D)}}_{\,\mathrm{int}}
  [\myvec{A},\gvec{\lambda}] =
  S^{\scriptscriptstyle\mathrm{(D)}}_{\,\mathrm{int},3}
  [\myvec{A},\gvec{\lambda}] +
  S^{\scriptscriptstyle\mathrm{(D)}}_{\,\mathrm{int},4}
  [\myvec{A},\gvec{\lambda}]  
\end{equation}
one obtains
\begin{eqnarray} \label{eq:S(D)int,3-sym}
  S^{\scriptscriptstyle\mathrm{(D)}}_{\,\mathrm{int},3}
  [\myvec{A},\gvec{\lambda}]
  \!\!\! &=& \!\!\!\!
  \int\!\!dx\,
  \frac{1}{2!}\,(-ig) f^{abc} \lambda^{ai}
  \Bigl\{
    \left(1-{\textstyle\frac{\sigma}{\kappa}}\right) \!
    \bigl[
      \delta^{ij} A^{bj} \partial_k A^{ck} \!
     -\delta^{ik} A^{ck} \partial_j A^{bj}
    \bigr] \nonumber\\[1ex]
  & &  \hspace{24.6ex}
    + \,2\,
    \bigl[
      \delta^{ij} A^{ck} \partial_k A^{bj} \!
     -\delta^{ik} A^{bj} \partial_j A^{ck}
    \bigr] \nonumber\\[1ex]
  & &  \hspace{24.6ex}
    + \hspace{2ex}
    \bigl[
      \delta^{jk} A^{bj} \partial_i A^{ck} \!
     -\delta^{kj} A^{ck} \partial_i A^{bj}
    \bigr]
  \Bigr\} \\[2ex] \label{eq:S(D)int,4-sym}
  S^{\scriptscriptstyle\mathrm{(D)}}_{\,\mathrm{int},4}
  [\myvec{A},\gvec{\lambda}]
  \!\!\! &=& \!\!\!\!
  \int\!\!dx\,
  \frac{1}{3!}\,(-ig^2) \, V^{abcd}_{ijkl} \,
  \lambda^{ai} A^{bj} A^{ck} A^{dl}
\end{eqnarray}
where
\vspace{-3ex}
\begin{eqnarray} \label{def:V4}
  V^{abcd}_{ijkl} &=& \hspace{1ex}
  f^{ace} f^{bde} (\delta^{ij}\delta^{kl} \!- \delta^{il}\delta^{kj})
  \nonumber\\
  & & \!\!\!+\,
  f^{abe} f^{cde} (\delta^{ik}\delta^{jl} \!- \delta^{il}\delta^{jk})
  \nonumber\\
  & & \!\!\!+\,
  f^{ade} f^{bce} (\delta^{ij}\delta^{kl} \!- \delta^{ik}\delta^{jl}) 
\end{eqnarray}
Observing that there is an additional minus sign because we 
have $-S^{\scriptscriptstyle\mathrm{(D)}}[\myvec{A},\gvec{\lambda}]$ in 
the exponent of the generating functional and noting our conventions
of the \textsc{Fourier} transform (\ref{eq:Fourier}) of the propagators,
we find for the 3--point vertex in momentum space
that is symmetrised with respect to the two $\myvec{A}$ fields

\vspace{3ex}
\unitlength=1mm
\noindent 
\begin{minipage}{147mm}
\underline{$\gvec{\lambda} \myvec{A}^{\!2}$ vertex}\\
\parbox{60mm}
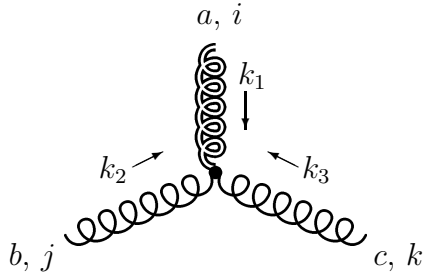
{
\begin{picture}(60,40) 
  \makebox(60,40){
    \put( 9 ,9){\vector( 2, 1){4.0}} 
    \put(23,20){$k_1$}
    \put(17.5,28){$a,\,i$} 
    \put(31 ,9){\vector(-2, 1){4.0}}
    \put( 4.5, 7.5){$k_2$}
    \put(-7.5,-4){$b,\,j$} 
    \put(24,19){\vector( 0,-1){5.0}}
    \put(31.5, 7.5){$k_3$}
    \put(41,-4){$c,\,k$}
    \begin{fmfgraph*}(40,25)
      \fmfkeep{LAA}
      \fmfstraight
      \fmftop{i1}
      \fmfbottom{i2,i3}
      \fmf{dbl_curly}{i1,v}
      \fmf{gluon}{i2,v}
      \fmf{gluon}{v,i3}
      \fmfdot{v}
    \end{fmfgraph*}
  }
\end{picture} 
}
\begin{minipage}{81mm}
\begin{eqnarray*}
  -g\,V^{abc}_{ijk}(\myvec{k}_2,\myvec{k}_3) 
  \!\!\! &=& \!\!\!
  -g f^{abc}
  \Bigr\{\!
    \left(1-{\textstyle\frac{\sigma}{\kappa}}\right) \!
    \bigl(
      \delta^{ij} k_3^k - \delta^{ik} k_2^j
    \bigr) \\ 
    && \hspace{12.3ex} \!
    + \,2
    \bigl(
      \delta^{ij} k_2^k - \delta^{ik} k_3^j
    \bigr) \\
    &&  \hspace{16.2ex} \!
    + \,\delta^{jk}
    \bigl(
      k_3^i - k_2^i
    \bigr)
  \Bigl\}
\end{eqnarray*}
\end{minipage}
\end{minipage} 

\vspace{2ex}
\noindent
Momentum conservation is thereby to be understood. By construction, 
the object $V^{abc}_{ijk}(\myvec{k}_2,\myvec{k}_3)$ is symmetric in 
the last two pairs of indices (and corresponding momenta), i.e.
\begin{equation}
  V^{abc}_{ijk}(\myvec{k}_2,\myvec{k}_3) =
  V^{acb}_{ikj}(\myvec{k}_3,\myvec{k}_2)
\end{equation}

\noindent
Analogously, the symmetrised 4--point vertex is found to be

\nopagebreak
\vspace{3ex}
\noindent 
\begin{minipage}{147mm}
\underline{$\gvec{\lambda} \myvec{A}^{\!3}$ vertex}\\
\parbox{60mm}{
\begin{picture}(60,40) 
  \makebox(60,40){
    \put( 7,24){$k_1$}
    \put(-7.5,27){$a,\,i$} 
    \put(11,23.5){\vector( 3,-2){4.0}}
    \put(31,24){$k_2$}
    \put(41,27){$b,\,j$}
    \put(30,23.5){\vector(-3,-2){4.0}}
    \put( 2, 6){$k_3$}
    \put(-7.5,-4){$c,\,k$} 
    \put( 6, 8){\vector( 3, 2){4.0}}
    \put(35, 6){$k_4$}
    \put(41,-4){$d,\,l$}
    \put(34, 8){\vector(-3, 2){4.0}}
    \begin{fmfgraph*}(40,25)
      \fmfkeep{LAAA}
      \fmfstraight
      \fmftop{i1,i2}
      \fmfbottom{i3,i4}
      \fmf{dbl_curly}{i1,v}
      \fmf{gluon}{v,i2}
      \fmf{gluon}{i3,v}
      \fmf{gluon}{v,i4}
      \fmfdot{v}
    \end{fmfgraph*}
  }
\end{picture} 
}
\begin{minipage}{81mm}
\begin{eqnarray*}
  ig^2\,V^{abcd}_{ijkl}
  \!\!\! &=& \!\!\!
  ig^2 \Bigl\{\ \;
  f^{ace} f^{bde} (\delta^{ij}\delta^{kl} \!- \delta^{il}\delta^{kj})
  \nonumber\\
  & & \hspace{2.7ex} +\,
  f^{abe} f^{cde} (\delta^{ik}\delta^{jl} \!- \delta^{il}\delta^{jk})
  \nonumber\\[0.2ex]
  & & \hspace{2.7ex} +\,
  f^{ade} f^{bce} (\delta^{ij}\delta^{kl} \!- \delta^{ik}\delta^{jl}) 
  \Bigr\}
\end{eqnarray*}
\end{minipage}
\end{minipage} 

\vspace{2ex}
\noindent
where $V^{abcd}_{ijkl}$ was already introduced in 
Eq.~(\ref{def:V4}) and is symmetric in the last three pairs of indices
\begin{equation}
  V^{abcd}_{ijkl} =
  V^{abdc}_{ijlk} =
  V^{acbd}_{ikjl} =
  V^{acdb}_{iklj} =
  V^{adbc}_{iljk} =
  V^{adcb}_{ilkj}
\end{equation}

\noindent
For the ghost sector, the corresponding interaction term extracted 
from Eq.~(\ref{eq:Sgg}) is given by 
\begin{equation}
  S^{\scriptscriptstyle\mathrm{(GG)}}_{\,\mathrm{int}}
  [\myvec{A},\omega,\bar{\omega}] =
  \int\!\!dx\,
  {\textstyle\frac{(-g)}{\kappa}} f^{abc}
  \bar{\omega}^a \!\left(\myvec{A}^{\!c} \cdot \nabla\right) \omega^b
  =
  \int\!\!dx\,
  {\textstyle\frac{(-g)}{\kappa}} f^{abc}
  \bar{\omega}^a {A}^{ck} \partial_k \omega^b
\end{equation}
and leads to the momentum space vertex

\nopagebreak
\vspace{3ex}
\noindent 
\begin{minipage}{147mm}
\underline{$\gvec{\omega} \bar{\gvec{\omega}} \myvec{A}$ vertex}\\
\parbox{60mm}{
\begin{picture}(60,40) 
  \makebox(60,40){
    \put(23,20){$k_1$} 
    \put(19.3,28){$a$} 
    \put(24,19){\vector( 0,-1){5.0}}
    \put( 2, 5){$k_2$}
    \put(-3.5,-4){$b$} 
    \put( 6,6.6){\vector( 3, 1){4.0}}
    \put(35, 5){$k_3$}
    \put(41,-4){$c,\,k$}
    \put(34.5,6.6){\vector(-2, 1){4.0}}
    \begin{fmfgraph*}(40,25)
      \fmfkeep{ooA}
      \fmfstraight
      \fmftop{i1}
      \fmfbottom{i2,i3}
      \fmf{dots_arrow}{v,i1}
      \fmf{dots_arrow}{i2,v}
      \fmf{gluon}{v,i3}
      \fmfdot{v}
      \fmfv{label=$a$}{i1}
      \fmfv{label=$b$}{i2}
      \fmfv{label=$c,,k$}{i3}
    \end{fmfgraph*}
  }
\end{picture} 
}
\begin{minipage}{81mm}
\begin{displaymath}
  \frac{ig}{\kappa} \, f^{abc} k_2^k
  \hspace{33.5ex}
\end{displaymath}
\vspace{1ex}
\end{minipage}
\end{minipage} 

\vspace{2ex}
\end{fmffile}

\appendix
\setcounter{section}{2}
\setcounter{equation}{0}
\renewcommand{\theequation}{\thesection.\arabic{equation}}
\section{Explicit Consequences of Identities to Lower N-Point Functions}
\label{NPointFunctions}
We will now find explicit identities for the lower n-point function from the
identities obtained in Section \ref{WardIDs}.
\subsection{1-point Functions}
Let us start by explicitly writing down the consequences of Ghost number
conservation, Eqs.~(\ref{eq:GhostNumber-Z}) -- (\ref{eq:GhostNumber-Gamma}), 
to the one-point functions of the theory. Taking the functional derivative
of Eq.~(\ref{eq:GhostNumber-W}) with respect to one of the sources
$J_{\omega}$, $J_{\bar{\omega}}$, $I_{s\!A}$, $I_{s\!\lambda}$
or $I_{s\omega}$  and evaluating for $J=I=0$ yields
\begin{equation} \label{eq:OnePont-W}
\begin{array}{r@{\!\!}cl@{\ \qquad}r@{\!\!}cl@{\ \qquad}r@{\!\!}cl}
  \displaystyle
  \left.
  \frac{\delta W[J,I]}{\delta J_{\omega}^{a}(x)}
  \right|_{J=I=0}
  &=& \!\!\!0 &
  \displaystyle
  \left.
  \frac{\delta W[J,I]}{\delta J_{\bar{\omega}}^{a}(x)}
  \right|_{J=I=0}
  &=& \!\!\!0 &
  \displaystyle
  \left.
  \frac{\delta W[J,I]}{\delta I_{s\!A}^{ai}(x)}
  \right|_{J=I=0}
  &=& \!\!\!0 \\[4ex]
  \displaystyle
  \left.
  \frac{\delta W[J,I]}{\delta I_{s\!\lambda}^{ai}(x)}
  \right|_{J=I=0}
  &=& \!\!\!0 &
  \displaystyle
  \left.
  \frac{\delta W[J,I]}{\delta I_{s\omega}^a(x)}\,
  \right|_{J=I=0}
  &=& \!\!\!0 &
  \displaystyle
  & &
\end{array}
\end{equation}
The same relations follow for the derivatives of $Z[J,I]$ from 
Eq.~(\ref{eq:GhostNumber-Z}). On the other hand, 
Eq.~(\ref{eq:DerivativesOfGamma}) implies
\begin{equation} \label{eq:OnePont-Gamma-Fields}
\begin{array}{r@{\!\!}cl@{\quad}r@{\!\!}cl@{\quad}r@{\!\!}cl@{\quad}r@{\!\!}cl}
  \displaystyle
  \left.
  \frac{\delta \Gamma}{\delta A^{ai}(x)}
  \right|_{J=I=0}
  &=& \!\!\!0 &
  \displaystyle
  \left.
  \frac{\delta \Gamma}{\delta \lambda^{ai}(x)}
  \right|_{J=I=0}
  &=& \!\!\!0 &
  \displaystyle
  \left.
  \frac{\delta \Gamma}{\delta \omega^a(x)}
  \right|_{J=I=0}
  &=& \!\!\!0 &
  \displaystyle
  \left.
  \frac{\delta \Gamma}{\delta \bar{\omega}^a(x)}
  \right|_{J=I=0}
  &=& \!\!\!0
\end{array}
\end{equation}
and the combination of Eq.~(\ref{eq:MoreDerivativesOfGamma})
and (\ref{eq:OnePont-W}) gives
\begin{equation} \label{eq:OnePont-Gamma-I}
\begin{array}{r@{\!\!}cl@{\ \qquad}r@{\!\!}cl@{\ \qquad}r@{\!\!}cl}
  \displaystyle
  \left.
  \frac{\delta \Gamma}{\delta I_{s\!A}^{ai}(x)}
  \right|_{J=I=0}
  &=& \!\!\!0 &
  \displaystyle
  \left.
  \frac{\delta \Gamma}{\delta I_{s\!\lambda}^{ai}(x)}
  \right|_{J=I=0}
  &=& \!\!\!0 &
  \displaystyle
  \left.
  \frac{\delta \Gamma}{\delta I_{s\omega}^a(x)}\,
  \right|_{J=I=0}
  &=& \!\!\!0
\end{array}
\end{equation}
The last first derivative of $\Gamma$ can be computed from the stochastic
Ward identities, Eq.~(\ref{eq:GammastochasticWI}) 
\begin{equation}\label{eq:OnePont-Gamma-II}
  \left.\frac{\delta \Gamma}{\delta I_{s\bar\omega}^a(x)}\,
  \right|_{J=I=0}
  = \!\!\!0
\end{equation}
Thus, all first derivatives of $\Gamma$ have to vanish. 
\subsection{2-point Functions}
The consequences of ghost number conservation to the second derivatives
of $Z[J,I]$ and $W[J,I]$ are summarised in the following table,
indicating for any pair of sources whether the corresponding second
derivative (evaluated for $J=I=0$) is restricted to vanish or not by
ghost number conservation
\begin{equation} \label{eq:TwoPont-W}
\begin{array}{c|c|c|c|c|c|c|c|c}
  & J_A & J_{\lambda} & J_{\omega} & J_{\bar{\omega}}
  & I_{s\!A} & I_{s\!\lambda} & I_{s\omega} & I_{s\bar{\omega}} \\ \hline
  J_A               &   &   & 0 & 0 & 0 & 0 & 0 &   \\ \hline
  J_{\lambda}       &   &   & 0 & 0 & 0 & 0 & 0 &   \\ \hline
  J_{\omega}        & 0 & 0 & 0 &   & 0 & 0 & 0 & 0 \\ \hline
  J_{\bar{\omega}}  & 0 & 0 &   & 0 &   &   & 0 & 0 \\ \hline
  I_{s\!A}          & 0 & 0 & 0 &   & 0 & 0 & 0 & 0 \\ \hline
  I_{s\!\lambda}    & 0 & 0 & 0 &   & 0 & 0 & 0 & 0 \\ \hline
  I_{s\omega}       & \makebox[3.0ex]{0}
                    & \makebox[3.0ex]{0}
                    & \makebox[3.0ex]{0}
                    & \makebox[3.0ex]{0}
                    & \makebox[3.0ex]{0}
                    & \makebox[3.0ex]{0}
                    & \makebox[3.0ex]{0}
                    & \makebox[3.0ex]{0} \\ \hline
  I_{s\bar{\omega}} &   &   & 0 & 0 & 0 & 0 & 0 &   \\ \hline
\end{array}
\end{equation}
\noindent
The analogous result for the second derivatives of the 1PI generating
functional $\Gamma[\myvec{A},\gvec{\lambda},\omega,\bar{\omega};I]$
(as well evaluated for vanishing sources $J=I=0$) is
\begin{equation} \label{eq:TwoPont-Gamma}
\begin{array}{c|c|c|c|c|c|c|c|c}
  & A & \lambda & \omega & \bar{\omega}
  & I_{s\!A} & I_{s\!\lambda} & I_{s\omega} & I_{s\bar{\omega}} \\ \hline
  A                 &   &   & 0 & 0 & 0 & 0 & 0 &   \\ \hline
  \lambda           &   &   & 0 & 0 & 0 & 0 & 0 &   \\ \hline
  \omega            & 0 & 0 & 0 &   &   &   & 0 & 0 \\ \hline
  \bar{\omega}      & 0 & 0 &   & 0 & 0 & 0 & 0 & 0 \\ \hline
  I_{s\!A}          & 0 & 0 &   & 0 & 0 & 0 & 0 & 0 \\ \hline
  I_{s\!\lambda}    & 0 & 0 &   & 0 & 0 & 0 & 0 & 0 \\ \hline
  I_{s\omega}       & \makebox[3.0ex]{0}
                    & \makebox[3.0ex]{0}
                    & \makebox[3.0ex]{0}
                    & \makebox[3.0ex]{0}
                    & \makebox[3.0ex]{0}
                    & \makebox[3.0ex]{0}
                    & \makebox[3.0ex]{0}
                    & \makebox[3.0ex]{0} \\ \hline
  I_{s\bar{\omega}} &   &   & 0 & 0 & 0 & 0 & 0 &   \\ \hline
\end{array}
\end{equation}
\noindent
In the following,
we will often rely on the information summarised in these tables
dropping certain terms that are bound to zero by ghost number conservation
from our calculations without further notice.
To start with, let us recall the gauge Ward identity in terms of the
generating functional of connected correlation functions $W[J,I]$.
It was found in Eq.~(\ref{eq:ST-W}) to read
\begin{equation}
  \int\!\!dx\,
  \Bigl[
    J_{\!A}^{ai}(x) \,\frac{\delta W[J,I]}
                           {\delta I_{s\!A}^{ai}(x)}
  + J_{\!\lambda}^{ai}(x) \,\frac{\delta W[J,I]}
                                 {\delta I_{s\!\lambda}^{ai}(x)}
  + J_{\omega}^a(x) \,\frac{\delta W[J,I]}
                           {\delta I_{s\omega}^a(x)}
  + J_{\bar{\omega}}^a(x) \,\frac{\delta W[J,I]}
                                 {\delta I_{s\bar{\omega}}^a(x)}
  \Bigr]
   = 0
\end{equation}
Taking second derivatives, a variety of possibilities arise.
For instance, choosing $\delta/\delta J_A^{ai}(x)$ and 
$\delta/\delta J_A^{bj}(y)$  yields after setting sources to zero
\begin{equation}
  \frac{\delta^2 W[J,I]}{\delta J_A^{bj}(y) \,\delta I_{s\!A}^{ai}(x)}
  \biggr|_{J=I=0}
  +\hspace{3.5ex}
  \frac{\delta^2 W[J,I]}{\delta J_A^{ai}(x) \,\delta I_{s\!A}^{bj}(y)}
  \biggr|_{J=I=0} = 0
\end{equation}
However, due to ghost number conservation both of these terms are zero
by themselves. Likewise, the combination of $\delta/\delta J_A^{ai}(x)$
with $\delta/\delta J_{\lambda}^{bj}(y)$ or 
$\delta/\delta J_{\omega}^{b}(y)$ does not lead to any new relation 
when ghost number conservation is taken into account.
The fourth possibility however, combining $\delta/\delta J_A^{ai}(x)$
and a derivative with respect to $J_{\bar{\omega}}^b(y)$, results in the
identity
\begin{equation} \label{eq:GaugeWI:RelationForW}
  \frac{\delta^2 W[J,I]}{\delta J_{\bar{\omega}}^{b}(y) 
                         \,\delta I_{s\!A}^{ai}(x)}
  \biggr|_{J=I=0}
  +\hspace{3.5ex}
  \frac{\delta^2 W[J,I]}{\delta J_A^{ai}(x)
                         \,\delta I_{s\bar{\omega}}^{b}(y)}
  \biggr|_{J=I=0} = 0
\end{equation}
that will be further exploited in a moment. Considering the 
combinations of $\delta/\delta J_{\lambda}^{ai}(x)$
with one of the derivatives $\delta/\delta J_{\lambda}^{bj}(y)$ or
$\delta/\delta J_{\omega}^{b}(y)$ again only leads to trivial 
relations in view of ghost number conservation. 
The pairing of $\delta/\delta J_{\lambda}^{ai}(x)$ with
$\delta/\delta J_{\bar{\omega}}^{b}(y)$ yields
\begin{equation}
  \frac{\delta^2 W[J,I]}{\delta J_{\bar{\omega}}^{b}(y) 
                         \,\delta I_{s\!\lambda}^{ai}(x)}
  \biggr|_{J=I=0}
  +\hspace{3.5ex}
  \frac{\delta^2 W[J,I]}{\delta J_{\lambda}^{ai}(x)
                         \,\delta I_{s\bar{\omega}}^{b}(y)}
  \biggr|_{J=I=0} = 0
\end{equation}
However, this relation is a consequence of the two simpler identities
\begin{equation} \label{eq:someConseqOfSDI}
  \frac{\delta^2 W[J,I]}{\delta J_{\bar{\omega}}^{b}(y) 
                         \,\delta I_{s\!\lambda}^{ai}(x)}
  \biggr|_{J=I=0} \!\!\!\!= 0 
  \hspace{5ex}
  \frac{\delta^2 W[J,I]}{\delta J_{\lambda}^{ai}(x)
                         \,\delta I_{s\bar{\omega}}^{b}(y)}
  \biggr|_{J=I=0} \!\!\!\!= 0
\end{equation}
induced by the stochastic Ward identity (\ref{eq:stochasticWI}).
The remaining possibilities finally, choosing two derivatives
with respect to $\omega$, two derivatives with respect to $\bar{\omega}$,
or one with respect to $\omega$, one to $\bar{\omega}$
again express ghost number conservation only.
Hence, up to the level of second derivatives 
Eq.~(\ref{eq:GaugeWI:RelationForW}) is the only restriction
imposed by the gauge BRST symmetry beyond relations that
already follow from the stochastic Ward identity or
simply are a consequence of ghost number conservation.
Implications of the stochastic Ward identities (\ref{eq:stochasticWI})
and (\ref{eq:GammastochasticWI}) are most importantly the vanishing
of the auxiliary field propagator to all orders
\begin{equation}
  G^{(\lambda\lambda)}{}^{ab}_{ij}(x,y) =
  \frac{\delta^2 W[J,I]}{\delta J_{\lambda}^{ai}(x) \, 
        \delta J_{\lambda}^{bj}(y)}
  \biggr|_{J=I=0} = 0
\end{equation}
or, equivalently, of the $(AA)$ self-energy component
\begin{equation} \label{eq:Pi(AA)=0}
  \Pi^{(AA)}{}^{ab}_{ij}(x,y) =
  \frac{\delta^2 \Gamma}{\delta A^{ai}(x) \, 
        \delta A^{bj}(y)}
  \biggr|_{J=I=0} -
  (\Delta^{-1})^{(AA)}{}^{ab}_{ij}(x,y) = 0
\end{equation}
where in addition to Eq.~(\ref{eq:GammastochasticWI}) it was used 
that the $(AA)$ component of the inverse free propagator is zero too
(cf.~Eq.~(\ref{eq:Delta^-1:AA})). Note that
Eq.~(\ref{eq:Pi(AA)=0}) is a special case of the general statement
that there are no pure gauge field vertices in the theory: All proper
vertex functions of the form
\begin{equation}
  \Gamma^{(AA\dots A)}{}^{ab\dots c}_{ij\dots k}
  (x,y,\dots,z) =
  \frac{\delta^n \Gamma}{\delta A^{ai}(x) \, 
        \delta A^{bj}(y) \cdots \delta A^{ck}(z)}
  \biggr|_{J=I=0} 
\end{equation}
vanish as an immediate consequence of the stochastic Ward identity
(\ref{eq:GammastochasticWI}).
Further implications up to second derivatives (neglecting those
only expressing ghost number conservation) are
\begin{equation} \label{eq:StochasticWID-Main}
  \frac{\delta^2 \Gamma}{\delta {\omega}^{b}(y) 
                         \,\delta I_{s\!\lambda}^{ai}(x)}
  \biggr|_{J=I=0} \!\!\!\!= 0 
  \hspace{5ex}
  \frac{\delta^2 \Gamma}{\delta A^{ai}(x)
                         \,\delta I_{s\bar{\omega}}^{b}(y)}
  \biggr|_{J=I=0} \!\!\!\!= 0
\end{equation}
together with the equivalent identities (\ref{eq:someConseqOfSDI}),
\begin{equation} \label{eq:StochasticWID-Norm}
  \left.
  \frac{\delta \Gamma}{\delta I_{s\bar{\omega}}^a(x)}
  \right|_{J=I=0} \!\!\!\!= 
  -\left.
  \frac{\delta W}{\delta I_{s\bar{\omega}}^a(x)}
  \right|_{J=I=0} \!\!\!\!= 0  
\end{equation}
and for completeness finally
\begin{equation}
  \frac{\delta^2 \Gamma}{\delta I_{s\bar{\omega}}^{a}(x)
                         \,\delta I_{s\bar{\omega}}^{b}(y)}
  \biggr|_{J=I=0} \!\!\!\!= -\,
  \frac{\delta^2 W}{\delta I_{s\bar{\omega}}^{a}(x)
                         \,\delta I_{s\bar{\omega}}^{b}(y)}
  \biggr|_{J=I=0} \!\!\!\!= 0
\end{equation}
This last identity, however, does not lead to a simple relation 
among the lower n-point functions because both of the derivatives act on
sources of the BRST transformed fields.
In general, to make sense of the above identities we will
have to translate the derivatives of the $I$-type to such
with respect to sources of the fundamental fields.
For instance, one has
\begin{eqnarray}
  \frac{\delta Z}{\delta I_{s\!A}^{ai}(x)} &=&
  \int\!\mathcal{D}\!\myvec{A}\mathcal{D}\!\gvec{\lambda}
        \mathcal{D}\omega\mathcal{D}\bar{\omega}
        \left(
          D^{ab}_i(x)\,\omega^b(x)
        \right)
        \exp\Bigl\{ (\dots) \Bigr\}
        \nonumber\\
  &=&
  \int\!\mathcal{D}\!\myvec{A}\mathcal{D}\!\gvec{\lambda}
        \mathcal{D}\omega\mathcal{D}\bar{\omega}
        \Bigl(
          \partial_i 
          \Bigl(
            -\frac{\delta}{\delta J_{\omega}^a(x)}
          \Bigr)
          -g f^{abc}
           \frac{\delta}{\delta J_A^{ci}(x)}
          \Bigl(
            -\frac{\delta}{\delta J_{\omega}^b(x)}
          \Bigr)
        \Bigr)
        \exp\Bigl\{ (\dots) \Bigr\}
        \nonumber\\
  &=&
  - \partial_i \frac{\delta Z}{\delta J_{\omega}^a(x)}
  + g f^{abc} 
  \frac{\delta^2 Z}{\delta J_A^{ci}(x)\,\delta J_{\omega}^b(x)}
\end{eqnarray}
where the dots abbreviate the usual exponent of the generating
functional as given in Eq.~(\ref{eq:Z[J,I]}). Expressing this identity 
in terms of $W=\ln Z$ yields
\begin{equation}
  \frac{\delta W}{\delta I_{s\!A}^{ai}(x)} =
  - \partial_i \frac{\delta W}{\delta J_{\omega}^a(x)}
  + g f^{abc}
    \left(
      \frac{\delta^2 W}{\delta J_A^{ci}(x)\,\delta J_{\omega}^b(x)}
      +
      \frac{\delta W}{\delta J_A^{ci}(x)}\,
      \frac{\delta W}{\delta J_{\omega}^b(x)}
    \right)
\end{equation}
Analogously, one obtains after some algebra
\begin{eqnarray}
  \frac{\delta W}{\delta I_{s\!\lambda}^{ai}(x)} \!\!\!&=&\!\!\!
  \hspace{1ex}g f^{abc}
    \left(
      \frac{\delta^2 W}{\delta J_{\lambda}^{ci}(x)\,\delta J_{\omega}^b(x)}
      +
      \frac{\delta W}{\delta J_{\lambda}^{ci}(x)}\,
      \frac{\delta W}{\delta J_{\omega}^b(x)}
    \right) \\
  \frac{\delta W}{\delta I_{s\omega}^{a}(x)} \!\!\!&=&\!\!\!
  \frac{1}{2} g f^{abc}
    \left(
      \frac{\delta^2 W}{\delta J_{\omega}^{c}(x)\,\delta J_{\omega}^b(x)}
      +
      \frac{\delta W}{\delta J_{\omega}^{c}(x)}\,
      \frac{\delta W}{\delta J_{\omega}^b(x)}
    \right) \\
  \frac{\delta W}{\delta I_{s\bar{\omega}}^{a}(x)} \!\!\!&=&\!\!\!
  i\sigma\partial_i \frac{\delta W}{\delta J_{\lambda}^{ai}(x)}
  -i\sigma g f^{abc}
    \left(
      \frac{\delta^2 W}{\delta J_A^{ci}(x)\,\delta J_{\lambda}^{bi}(x)}
      +
      \frac{\delta W}{\delta J_A^{ci}(x)}\,
      \frac{\delta W}{\delta J_{\lambda}^{bi}(x)}
    \right) \nonumber\\
  \!\!\!& &\!\!\!
  + g f^{abc}
    \left(
      \frac{\delta^2 W}{\delta J_{\omega}^{c}(x)\,
                                  \delta J_{\bar{\omega}}^b(x)}
      +
      \frac{\delta W}{\delta J_{\omega}^{c}(x)}\,
      \frac{\delta W}{\delta J_{\bar{\omega}}^b(x)}
    \right)  
\end{eqnarray}
With these substitutions Eq.~(\ref{eq:GaugeWI:RelationForW})
translates to
\begin{eqnarray} \label{eq:GaugeWardwithW}
\lefteqn{
  \partial_i G^{(\omega) \,ab}(x,y)
  -i\sigma\partial_j G^{(A\lambda)}{}^{ab}_{ij}(x,y) =}
  \\[1ex]
  & & 
  -g f^{acd} 
     W^{(\bar{\omega}\omega A)}{}^{bc}{}^{d}_{i}(y,x,x)
  -i\sigma g f^{bcd} 
     W^{(AA\lambda)}{}^{acd}_{ijj}(x,y,y)
  +g f^{bcd}
     W^{(\bar{\omega}\omega A)}{}^{cd}{}^{a}_{i}(y,y,x)
  \nonumber
\end{eqnarray}
To further proceed, we express the connected three-point functions by 
their 1PI counterparts and transform into momentum space. 
Especially note that we pull out the momentum conserving delta 
function from the definition of our proper vertices. 
Hence, only $N-1$ momentum variables appear in the argument
of a $N$-point vertex. 
For instance, we use
\begin{math}
  \Gamma^{(\bar{\omega}\omega G)\,abc}_{\hspace{7.5ex}j}(k_1,k_2)
\end{math}
where the superscript $G$ is either the gauge field $A$ or the 
auxiliary field $\lambda$ 
and $k_1$ and $k_2$ refer to the (incoming) momenta along the ghost 
lines leaving and entering the vertex in this order.
Accordingly, in
\begin{math}
  \Gamma^{(FGH)\,abc}_{\hspace{6.35ex}ijk}(k_2,k_3)
\end{math}
with $F,G,H \in \{A,\lambda\}$
the two arguments $k_2$ and $k_3$ refer to the incoming 
momenta along the $G$ and $H$ line respectively.
\noindent
With these definitions, Eq.~(\ref{eq:GaugeWardwithW}) takes the form
\begin{eqnarray}
\lefteqn{ \hspace{-7ex}
  i k^i G^{(\omega) \,ab}(k) 
  + \sigma k^j G^{(A\lambda)}{}^{ab}_{i\,j}(k)
  =} \nonumber\\
  & & \hspace{-8ex}
  +\hspace{1.5ex}G^{(\omega) \,b^{\prime}b}(k) \intDkp g f^{acd}\,
  G^{(\omega) \,cc^{\prime}}\!(k^{\prime})\,
  G^{(AF)}{}^{dd^{\prime}}_{i\,i^{\prime}}
    \!(k-k^{\prime}) \,
  \Gamma^{(\bar{\omega}\omega F)}{}^{
           c^{\prime} b^{\prime}}{}^{d^{\prime}}_{i^{\prime}}
           \!(-k^{\prime}\!,k)
  \nonumber\\
  & & \hspace{-8ex}
  -\,G^{(AF)}{}^{aa^{\prime}}_{i\,i^{\prime}}\!(k)
  \!\intDkp 
  g f^{bcd}
  \Bigl[
  G^{(\omega) \,c^{\prime}c}(k^{\prime})\,
  G^{(\omega) \,dd^{\prime}}\!(k^{\prime}\!-k)\,
  \Gamma^{(\bar{\omega}\omega F)}{}^{
           d^{\prime}c^{\prime}}{}^{a^{\prime}}_{i^{\prime}}
           \!(k-k^{\prime}\!,k^{\prime})
  \nonumber\\ \label{eq:GaugeWardBeforeCancel}
  & & \hspace{15ex} - i\sigma\,
  G^{(A\lambda)}{}^{c^{\prime}c}_{j^{\prime}j}(k^{\prime})\,
  G^{(AG)}{}^{dd^{\prime}}_{jk^{\prime}}
    \!(k^{\prime}\!-k) \,
  \Gamma^{(FGA)}{}^{
           a^{\prime} d^{\prime} c^{\prime}}_{
             i^{\prime} k^{\prime} j^{\prime}}
           (k-k^{\prime}\!, k^{\prime})
  \Bigr]
\end{eqnarray}
The indices $F$ and $G$ in this equation are summation indices taking
the two values $A$ and $\lambda$. However, as we will show now, the
stochastic Ward identity leads to a cancellation among some of the 
terms involved.
To this end, let us express also the identities derived from 
the stochastic Ward identity in the language of full propagators 
and proper vertex functions.
As mentioned above, identity (\ref{eq:StochasticWID-Norm})
relates the normalisations of the gauge ghost and mixed
auxiliary/gauge field propagator
\begin{equation} \label{eq:StochasticWID-zero}
  g f^{abc} \! \intDk \!
  \left[
    G^{(\omega) \,cb}(k)
    -i\sigma G^{(A\lambda)}{}^{cb}_{i\,i}(k)
  \right] = 0
\end{equation}
From the first of the Eqs.~(\ref{eq:StochasticWID-Main}) one obtains
after some relabelling
\begin{equation} \label{eq:StochasticWID-1st}
  \intDkp g f^{bcd}\,
  G^{(A\lambda)}{}^{c^{\prime}c}_{i^{\prime}i}(k^{\prime})\,
  G^{(\omega) \,dd^{\prime}}\!(k^{\prime}\!-k)\,
  \Gamma^{(\bar{\omega}\omega A)}{}^{
           d^{\prime} a}{}^{c^{\prime}}_{i^{\prime}}
           \!(k-k^{\prime}\!,-k)
  = 0
\end{equation}
from the second equation
\begin{eqnarray}
  & & \hspace{4.9ex}
  \intDkp g f^{bcd} \,
  G^{(\omega) \,c^{\prime}c}(k^{\prime})\,
  G^{(\omega) \,dd^{\prime}}\!(k^{\prime}\!-k)\,
  \Gamma^{(\bar{\omega}\omega A)}{}^{
           d^{\prime} c^{\prime}}{}^{a}_{i}
           (k-k^{\prime}\!,k^{\prime})
  \nonumber\\
  & & -\,i\sigma \intDkp g f^{bcd}\,
  G^{(A\lambda)}{}^{c^{\prime}c}_{j^{\prime}j}(k^{\prime})\,
  G^{(A \lambda)}{}^{dd^{\prime}}_{jk^{\prime}}
    \!(k^{\prime}\!-k) \,
  \Gamma^{(A\lambda A)}{}^{
           a d^{\prime} c^{\prime}}_{i\,k^{\prime}j^{\prime}}
           (k-k^{\prime}\!, k^{\prime})
  = 0 \hspace{9ex} \label{eq:StochasticWID-2nd}
\end{eqnarray}
Here we have used
\begin{math}
  \Gamma^{(FGH)}{}^{abc}_{ijk}(k_2, k_3)
  =
  \Gamma^{(GHF)}{}^{bca}_{jki}(k_3, -k_2-k_3)  
\end{math}
in accordance with our definition of the vertex functions.
Let us now come back to Eq.~(\ref{eq:GaugeWardBeforeCancel}), that was found
to be the expression of the gauge Ward identity on the level of second
derivatives. With the summation index $F$ taking the value $A$, the second
integral in Eq.~(\ref{eq:GaugeWardBeforeCancel}) consists of three terms:
the one with the two ghost propagators and two copies of the second term
corresponding to the two possible values $G=\lambda$ and $G=A$. The last of
these terms is zero because it contains $\Gamma^{(AAA)}$. Moreover, the
remaining two terms cancel each other due to 
Eq.~(\ref{eq:StochasticWID-2nd}) as 
a consequence of the stochastic Ward identity. Hence, there is only a
contribution of the second integral in Eq.~(\ref{eq:GaugeWardBeforeCancel})
for $F=\lambda$. The first integral, however, contributes for both choices
$F=\lambda$ and $F=A$ (and likewise if $F$ is set to $\lambda$ in the 
second integral, $G$ can still take both values $G=\lambda,A$).

\noindent
The gauge BRST symmetry therefore leads to the
following identity to be obeyed by the full propagators and proper 
vertex functions of the theory
\begin{eqnarray}
\lefteqn{ \hspace{-7ex}
  i k^i G^{(\omega) \,ab}(k) 
  + \sigma k^j G^{(A\lambda)}{}^{ab}_{i\,j}(k)
  =} \nonumber\\
  & & \hspace{-8ex}
  +\hspace{1.5ex}G^{(\omega) \,b^{\prime}b}(k) \intDkp g f^{acd}\,
  G^{(\omega) \,cc^{\prime}}\!(k^{\prime})\,
  G^{(AA)}{}^{dd^{\prime}}_{i\,i^{\prime}}
    \!(k-k^{\prime}) \,
  \Gamma^{(\bar{\omega}\omega A)}{}^{c^{\prime}
           b^{\prime}}{}^{d^{\prime}}_{i^{\prime}}
           \!(-k^{\prime}\!,k)
  \nonumber\\
  & & \hspace{-8ex}
  +\hspace{1.5ex}G^{(\omega) \,b^{\prime}b}(k) \intDkp g f^{acd}\,
  G^{(\omega) \,cc^{\prime}}\!(k^{\prime})\,
  G^{(A\lambda)}{}^{dd^{\prime}}_{i\,i^{\prime}}
    \!(k-k^{\prime}) \,
  \Gamma^{(\bar{\omega}\omega\lambda)}{}^{c^{\prime}
           b^{\prime}}{}^{d^{\prime}}_{i^{\prime}}
           \!(-k^{\prime}\!,k)
  \nonumber\\
  & & \hspace{-8ex} 
  -\,G^{(A\lambda)}{}^{aa^{\prime}}_{i\,i^{\prime}}\!(k)
  \!\intDkp 
  g f^{bcd}
  \Bigl[
  G^{(\omega) \,c^{\prime}c}(k^{\prime})\,
  G^{(\omega) \,dd^{\prime}}\!(k^{\prime}\!-k)\,
  \Gamma^{(\bar{\omega}\omega\lambda)}{}^{
           d^{\prime} c^{\prime}}{}^{a^{\prime}}_{i^{\prime}}
           \!(k-k^{\prime}\!,k^{\prime})
  \nonumber\\
  & & \hspace{15ex} - i\sigma\,
  G^{(A\lambda)}{}^{c^{\prime}c}_{j^{\prime}j}(k^{\prime})\,
  G^{(AA)}{}^{dd^{\prime}}_{jk^{\prime}}
    \!(k^{\prime}\!-k) \,
  \Gamma^{(\lambda AA)}{}^{a^{\prime} d^{\prime}
           c^{\prime}}_{i^{\prime} k^{\prime} j^{\prime}}
           (k-k^{\prime}\!, k^{\prime})
  \nonumber\\[1.0ex] \label{eq:GaugeWID}
  & & \hspace{15ex} - i\sigma\,
  G^{(A\lambda)}{}^{c^{\prime}c}_{j^{\prime}j}(k^{\prime})\,
  G^{(A\lambda)}{}^{dd^{\prime}}_{jk^{\prime}}
    \!(k^{\prime}\!-k) \;
  \Gamma^{(\lambda\lambda A)}{}^{a^{\prime} d^{\prime}
           c^{\prime}}_{i^{\prime} k^{\prime} j^{\prime}}
           (k-k^{\prime}\!, k^{\prime})
  \,\Bigr]
\end{eqnarray}

\end{appendix}

\end{document}